\documentclass[useAMS]{gGAF2e}

\usepackage{epstopdf}
\usepackage{oubraces}

\begin{document}
\doi{10.1080/03091920xxxxxxxxx}
 \issn{1029-0419} \issnp{0309-1929} \jvol{00} \jnum{00} \jyear{2011} 

\markboth{N. Yokoi}{Cross helicity and related dynamo}

\title{{\textit{Cross helicity and related dynamo}}}

\author{
N. YOKOI${\dag}$$\ddag$$\S$
\thanks{$\ddag$Guest Researcher at the National Astronomical Observatory of Japan (NAOJ)} \thanks{\S Guest Researcher  at the Nordic Institute for Theoretical Physics (NORDITA)}
$^{\ast}$\thanks{$^\ast$Corresponding author. Email: nobyokoi@iis.u-tokyo.ac.jp}
\vspace{6pt}\\
${\dag}$Institute of Industrial Science, University of Tokyo, 4-6-1, Komaba, Meguro,
Tokyo 153-8505, Japan\\ 
\vspace{6pt}
\received{received: Dec.\ 8, 2011, revised: May 12, 2012, 2nd revised: Aug.\ 5, 2012, finally revised: Oct.\ 11, 2012} 
}

\maketitle

\begin{abstract}
The turbulent cross helicity is directly related to the coupling coefficients for the mean vorticity in the electromotive force and for the mean magnetic-field strain in the Reynolds stress tensor. This suggests that the cross-helicity effects are important in the cases where global inhomogeneous flow and magnetic-field structures are present. Since such large-scale structures are ubiquitous in geo/astrophysical phenomena, the cross-helicity effect is expected to play an important role in geo/astrophysical flows. In the presence of turbulent cross helicity, the mean vortical motion contributes to the turbulent electromotive force. Magnetic-field generation due to this effect is called the cross-helicity dynamo. Several features of the cross-helicity dynamo are introduced. Alignment of the mean electric-current density ${\bf{J}}$ with the mean vorticity $\mbox{\boldmath$\Omega$}$, as well as the alignment between the mean magnetic field ${\bf{B}}$ and velocity ${\bf{U}}$, is supposed to be one of the characteristic features of the dynamo. Unlike the case in the helicity or $\alpha$ effect, where ${\bf{J}}$ is aligned with ${\bf{B}}$ in the turbulent electromotive force, we in general have a finite mean-field Lorentz force ${\bf{J}} \times {\bf{B}}$ in the cross-helicity dynamo. This gives a distinguished feature of the cross-helicity effect. By considering the effects of cross helicity in the momentum equation, we see several interesting consequences of the effect. Turbulent cross helicity coupled with the mean magnetic shear reduces the effect of turbulent or eddy viscosity. Flow induction is an important consequence of this effect. One key issue in the cross-helicity dynamo is to examine how and how much cross helicity can be present in turbulence. On the basis of the cross-helicity transport equation, its production mechanisms are discussed. Some recent developments in numerical validation of the basic notion of the cross-helicity dynamo are also presented.
\bigskip

\bigskip

\centerline{\bfseries Index}\medskip

\noindent\hbox to
\textwidth{\hsize\textwidth\vbox{\noindent\hsize20pc
{1.}    Introduction\\
{2.}    Cross helicity\\
{3.}    Turbulent electromotive force\\
\hspace*{10pt}{3.1.}  Mean and fluctuation\\
\hspace*{10pt}{3.2.}  Reynolds stress and turbulent electromotive force\\
\hspace*{10pt}{3.3.}  Physical interpretation of each effect\\
\hspace*{24pt} {3.3.1.}   Electromotive force due to turbulent motion;\\  \hspace*{50pt}$\beta$-related terms\\
\hspace*{24pt} {3.3.2.}  Electromotive force due to helicity;\\  \hspace*{50pt}$\alpha$-related terms\\
\hspace*{24pt} {3.3.3.}   Electromotive force due to cross helicity;\\  \hspace*{50pt}$\gamma$-related term\\
{4.}    Cross-helicity dynamo\\
{5.}    Cross-helicity generation mechanisms\\
\hspace*{10pt}{5.1.}   Turbulence modelling and statistical quantities\\
\hspace*{10pt}{5.2.}   Transport equation of turbulent cross helicity\\
\hspace*{26pt} (incompressible case)}
\vbox{\noindent\hsize20pc
\hspace*{10pt}{5.3.}   Cross-helicity production mechanisms\\
\hspace*{10pt}{5.4.}   Transport equation of turbulent cross helicity\\
\hspace*{26pt} (compressible case)\\
{6.}   Illustrative examples \\
\hspace*{10pt}{6.1.}   Galactic magnetic field\\
\hspace*{10pt}{6.2.}   Accretion disks\\
\hspace*{10pt}{6.3.}   Solar dynamos\\
{7.}   Flow generation\\
\hspace*{10pt}{7.1.}   Plasma rotation in internal-transport-barrier mode\\  
\hspace*{26pt}in tokamaks\\
\hspace*{10pt}{7.2.}   Torsional oscillation inside the Sun\\
\hspace*{10pt}{7.3.}   Flow--turbulence interaction in magnetic reconnection\\
\noindent   {8.}   Numerical tests\\
{9.}   Concluding remarks\\
{}   References\\
{}   Appendix
      }}
\end{abstract}

\section{Introduction\label{sec:intro}}
	The primary effect of turbulence is enhancing the effective transport. The rates of transport enhancement as compared with the molecular viscosity $\nu$ and the magnetic diffusivity $\eta$ are approximately expressed by the turbulent Reynolds and magnetic Reynolds numbers, $Re^{(\rm{T})}$ and $Rm^{(\rm{T})}$, respectively. They are Reynolds numbers defined using the characteristic velocity of turbulence, $v$. If we adopt the mixing length $\ell$ as the characteristic length scale of turbulence, the turbulent or eddy viscosity $\nu_{\rm{T}}$ is estimated as $\nu_{\rm{T}} \sim v \ell$. Hence, $\nu_{\rm{T}} / \nu = Re^{(\rm{T})}$ and for the turbulent magnetic diffusivity or anomalous resistivity $\beta$, we have $\beta / \eta = Rm^{(\rm{T})}$. In geophysical and astrophysical phenomena, the Reynolds and magnetic Reynolds numbers, and consequently the turbulent counterparts, are usually huge, so the transport enhancement is expected to be very large. These transport enhancements by turbulence often play an essential role in the dynamics of geophysical and astrophysical bodies.

	If we have some symmetry breakage in turbulence, even in the presence of strong fluctuations, the transport enhancement may be effectively suppressed or balanced by some other turbulence effects. In this situation, large-scale or mean-field structures such as the large-scale vorticity, global magnetic field, etc.\ are generated and sustained persistently in turbulence. Turbulent dynamos, in which global magnetic fields are generated and sustained by fluctuation motion, is one of the most interesting and important physical processes in turbulence.
	
	Here in this paper, dynamo is considered in the broadest sense. Of course, one of the most important aspects of the dynamo is instability problem: how weak seed fields can be amplified to strong fields. At the same time, however, the dynamo has the aspect of transport suppression. The sustainment of the magnetic configuration in the presence of strong turbulent magnetic diffusivity is also very important topic in dynamo theory. This is because enhancement of magnetic diffusivity is the primary effect of turbulence in magnetic-field evolution. Without the strong effective resistivity, a large-scale magnetic configuration cannot be ever formulated from the original or previous configurations. One of the obvious challenges for dynamo is to elucidate and predict the solar cycle. In order to elucidate the internal rotation of the Sun and the Maunder Minimum-like ``anomaly'' of the solar activity cycle, we have to consider the dynamical balance between the field generation and destruction mechanisms, which is beyond the instability.

	Starting from 1950's, the mean-field dynamo theory made a great achievement in understanding physics of magnetic-field generation and sustainment in highly turbulent electrically conducting media \citep{mof1978,par1979,kra1980}. First of all, we should point out that it is fabulous to derive, explain, and predict the basic behaviours of the magnetic fields in the geo/astrophysical bodies on the basis of a very simple system of equations, magnetohydrodynamics (MHD). At the same time, several criticisms have been made against the mean-field dynamo theory with several connotations, which include (i) kinematic approach; (ii) transport coefficients as parameters; (iii) ``generic'' form of the turbulent electromotive force; (iv) physical interpretations of main processes; (v) azimuthal averaging; (vi) incompressible treatment.

\begin{quote}
{\it{(i) Kinematic approach:}} In the kinematic dynamo approach, with expectation that the Lorentz back-reaction force to the flow can be small enough to be neglected, we assume that the velocity does not depend on the magnetic field. With this prescribed velocity, evolution of magnetic field is examined. However, as the magneto-rotational instability (MRI) studies have shown, even small magnetic field (much less than the equipartition field) will affect the dynamic evolution of turbulent motion \citep{bal1998}. Also it has been argued that even with a very small magnetic field, the Lorentz back-reaction will restrict dynamo action through the suppression of the generation and diffusion of the magnetic field (quenching) \citep{vai1992} In these senses, the central assumption of the kinematic approach completely failed. 
\end{quote}

\begin{quote}
{\it{(ii) Transport coefficients as parameters:}} In some mean-field dynamo models, the transport coefficients appearing in the turbulent electromotive force are treated as adjustable parameters with or without a prescribed spatial distribution \citep{dik1999}. However, from the viewpoint of turbulence theory, this is quite questionable. The transport coefficients should be determined by statistical properties of turbulence, which in general depends on the spatiotemporal evolution of turbulent flow.
\end{quote}

\begin{quote}
{\it{(iii) ``Generic'' form of the turbulent electromotive force:}} In some mean-field dynamo theory, the ``generic'' form of turbulent electromotive force $\langle {\bf{u}}' \times {\bf{b}}' \rangle$ is assumed to be a linear functional of the mean magnetic field and its derivatives (${\bf{u}}'$: velocity fluctuation, ${\bf{b}}'$: magnetic-field fluctuation, $\langle \cdots \rangle$: ensemble average). Even if the proportional coefficients $\mbox{\boldmath$\cal{\alpha}$}$, $\mbox{\boldmath$\cal{\beta}$}$ are treated as tensors, the assumption of such expansion with respect to the mean magnetic field may not be sufficient \citep{rae2010}.
\end{quote}

\begin{quote}
{\it{(iv) Physics of main processes:}} In order to explain magnetic-field evolution intuitively, in some mean-field models, a combination of the turbulent helicity effect ($\alpha$ effect) and the differential rotation effect ($\Omega$ effect) is employed. However, each process contains several assumptions. For example, the so-called $\Omega$ effect contains, at least, magnetic flux freezing in highly turbulent medium, favorable differential rotations, and magnetic reconnection at a particular location. Analysis of all these processes is not so simple as some mean-field model explanation naively assumes \citep{yok2011c}.
\end{quote}

\begin{quote}
{\it{(v) Azimuthal averaging:}} In some mean-field theory, the azimuthal average along the rotation axis is adopted as the ensemble average. However, non-axisymmetric properties are expected to be essential in some magnetic-field generation processes. Azimuthal averaging procedure will delete possibility of such non-axisymmetric effects. So, mean-field theory with azimuthal averaging is nothing to do with the magnetic-field evolution associated with non-axisymmetric behaviour of the field \citep{sch2003}.
\end{quote}

\begin{quote}
{\it{(vi) Incompressible treatment:}} In some mean-field theory, key notions of the dynamo, such as the turbulent electromotive force, are derived under the assumption of the incompressibility. However, in realistic geo/astrophysical situations, the compressibility or at least the mean-density stratification plays an essential role in magnetic-field generation processes. In this sense, the mean-field theory under the incompressible assumption is not an appropriate approach to the realistic dynamo phenomena.
\end{quote}

	Depending on what kind of phenomenon we are interested in, some mean-field theory employs some of these assumptions or approximations listed above as connotations. However, none of these connotations are essential ingredients of the mean-field dynamo theory. 

	On the criticism related to point (i), kinematic approach, it is worth while to point out the following point. The velocity field is certainly influenced by the magnetic field. But the degree of influence depends on the stage of turbulence (or instabilities). At the fully developed turbulence stage (or fully saturated stage of relevant instabilities), the influence of the magnetic field is entirely different from the one at the developing stage of turbulence or instabilities \citep{mat1995}. In this sense, oversimplified argument against the kinematic approach is sometimes misleading. Of course, it is true that the kinematic approach has its limitation. The force-free field configuration argument in the $\alpha$ dynamo may support such approach to some extent. However, the most interesting aspect of dynamo action lies in the dynamical interaction between the flow and magnetic field, as will be stressed later in the context of the cross-helicity dynamo. 

	Point (ii), transport coefficients as parameters, will be considered in this paper. Transport coefficients should be determined from the statistical properties of turbulence. And statistical properties change depending on the evolution of turbulent flow. Only in the case of homogeneous turbulence, these transport coefficients can be treated as constants. But still they are not adjustable parameters. In this paper, we stress the importance of self-consistent turbulence modelling, where transport coefficients are determined by solving transport equations for the coefficients. In other words, if the mean-field dynamo theory is accompanied by some closure scheme that determines the transport coefficients in a nonlinear and self-consistent manner, the mean-field dynamo approach is very strong and useful in realistic applications to the geo/astrophysical phenomena.

	Points (iii)-(iv) are directly related to the subject of this article. If we take pseudoscalars other than the helicity into account, the ``generic'' form of the turbulent electromotive force should be changed. If we have a third party who participates in the dynamo game, the physics of magnetic-field generation and sustainment may change drastically.

	As for the questions on (v) azimuthal averaging and (vi) incompressible treatment, again we stress these treatment is not the essential ingredients of the mean-field dynamo theory. For the latter, we point out the fact that the magnetic induction equation does not contain the density. So, as far as the formal expression of the turbulent electromotive force is concerned, the incompressible treatment is expected to give a good result. However, transport coefficients appearing in the turbulent electromotive force, $\alpha$, $\beta$, etc., depend on the compressibility.
	
	The mean-field dynamo approach, in particular after clearing all the arguments listed above, is very strong and useful. However, it is also true the term ``mean-field'' has several historical connotations. So, we prefer the term ``turbulent dynamo'' to mean-field dynamo, and hereafter denote this approach without connotations as turbulent dynamo.
	
	In the study of turbulent dynamos, pseudoscalars play an important role in the generation and sustainment of the large-scale structures in turbulence. One of the representative pseudoscalars is the turbulent kinetic helicity $\langle {{\bf u'}\cdot \mbox{\boldmath$\omega$}' }\rangle$, which characterizes the helical property of the turbulent motion [$\mbox{\boldmath$\omega$}' (= \nabla \times {\bf u'})$: vorticity fluctuation]. The generation of the large-scale magnetic fields has long been studied with special attention focussed on the helicity or $\alpha$ effect \citep{mof1978,par1979,kra1980}. In the context of the inverse cascade of the energy from small scales to large scales, \citet{pou1976} showed that it is not the kinetic helicity $\langle {{\bf u'} \cdot \mbox{\boldmath$\omega$}'} \rangle$ or the current helicity $\langle {{\bf b'} \cdot {\bf{j}}' }\rangle$ but the difference of them that induces the growth of the large-scale magnetic-field energy [${\bf{j}}' (= \nabla \times {\bf{b}}')$: electric-current density fluctuation]. The difference defined by $\langle {- {\bf{u}'} \cdot \mbox{\boldmath$\omega$}' + {\bf b'}\cdot {\bf{j}}'} \rangle$ is called the turbulent residual helicity. 

	A pseudoscalar defined by the cross-correlation between the turbulent velocity and magnetic field, $\langle {{\bf{u}}' \cdot {\bf{b}}'} \rangle$, is called the turbulent cross helicity.  In contrast to the helicity or $\alpha$ effect, not so much attention has been paid to the cross-helicity effect in the turbulent dynamo studies. The cross helicity itself has been investigated extensively in particular in the relation with solar-wind turbulence. From the pioneering work by \citet{dob1980a,dob1980b} not a few works have been done in the study of relaxation properties of the magnetohydrodynamic turbulence with cross helicity. \citet{gra1982,gra1983} and \citet{pou1988} worked on the energy transfer in MHD turbulence with the cross correlation between the velocity and magnetic fields. In addition, the velocity--magnetic-fields alignment itself is ubiquitous in geo/astrophysical flow phenomena such as solar winds, and is often called the dynamic alignment. Dynamic alignment has been discussed in relation to the Alfv\'{e}n wave and Alfv\'{e}n effect \citep{rob1967,pou1993}, and the notion of dynamic alignment has been confirmed through numerical simulations of the two-dimensional MHD turbulence \citep{bis1989,bis1993}. However, in this paper, we confine ourselves to the cross-helicity effects in the dynamo action and turbulent transport. Those who are interested in other aspects of cross helicity are referred to \citet{yok2011a} and works cited therein.

	As was mentioned, in the context of dynamos, the cross-helicity effect has not drawn so much attention as compared with the helicity or $\alpha$ effect. We can point out several reasons why people have considered the cross-helicity dynamo would not be so much relevant.
\begin{enumerate}

\item[(Q-i)] Due to the Galilean invariance of the fluid equation, we may put ${\bf{U}} = 0$ in the equation of the fluctuation velocity. As this result, the large-scale fluid motion represented by ${\bf{U}}$ is excluded from the expression of ${\bf{E}}_{\rm{M}}$ in the mean induction equation.

\item[(Q-ii)] The inner and outer products of ${\bf{u}}'$ and ${\bf{b}}'$ are related to each other as $({\bf{u}}' \cdot {\bf{b}}')^2 + ({\bf{u}}' \times {\bf{b}}')^2 = {|{\bf{u}}'|^2 |{\bf{b}}'|^2}$.
This relation suggests that a large turbulent cross helicity $\langle {{\bf{u}}' \cdot {\bf{b}}'} \rangle$ corresponds to a small turbulent electromotive force $\langle {{\bf{u}}' \times {\bf{b}}'} \rangle$. In other words, in the situation where the turbulent electromotive force plays an essential role, the turbulent cross helicity is expected to be very small or negligible. Thus there is no need for us to take the cross-helicity effect into account in the turbulent dynamo process.

\item[(Q-iii)] Turbulent cross helicity is the transport coefficient that couples with the large-scale vorticity $\mbox{\boldmath$\Omega$}$. The large-scale vorticity is locally equivalent to the system rotation. Since the system rotation will not directly affect the magnetic field, the large-scale vorticity is not expected to enter in the expression for the turbulent electromotive force. In this sense, the cross helicity must be irrelevant to the turbulent dynamo process.

\item[(Q-iv)] Even if the cross helicity can be related to the turbulent dynamo process, it is difficult for the cross helicity to be present in turbulence. Turbulent cross helicity represents breakage of symmetry between the directions parallel and antiparallel to the magnetic field. It is unlikely for large amount of cross helicity to exist in usual turbulent situation. Namely, the cross-helicity effect is too weak to play an important role in the real dynamo process in turbulence.
\end{enumerate}

	We shall answer these arguments as follows.

\begin{enumerate}
\item[(A-i)] By the Galilean invariance of the governing equation, we eliminate only the translational motion from the equation, but not the rotation or strained motion. However, if we naively drop the mean velocity by putting ${\bf{U}} = 0$, we also delete the possibility that the inhomogeneity of the mean velocity ${\bf{U}}$ may work. Namely, the effects of the mean vortical motion $\mbox{\boldmath$\Omega$} (= \nabla \times {\bf{U}})$ and the  mean velocity strain $\nabla {\bf{U}}$. Only in the homogeneous turbulence, such treatment can be allowed. In this sense, we should be careful to treat the mean velocity.

\item[(A-ii)] First, as we will see from applications of the cross-helicity effects to several geo/astrophysical phenomena in \S\ref{sec:ch_examples} and \S\ref{sec:flow_generation}, the magnitude of the scaled turbulent cross helicity (the turbulent cross helicity normalized by the turbulent MHD energy) we need for the effect to be relevant is $|W/K| = O(10^{-2})-O(10^{-1})$. Not so strong correlation such as $0.1-1$ in most cases. 

	Secondly, the relationship between the turbulent electromotive force $|\langle {{\bf{u}}' \times {\bf{b}}'} \rangle|$ and the turbulent cross helicity $|\langle {{\bf{u}}' \cdot {\bf{b}}'} \rangle|$ is not so simple. For instance, let us consider the case with fully aligned ${\bf{u}}'$ and ${\bf{b}}'$ as ${\bf{u}}' = \pm {\bf{b}}'$. If the number of parallel and antiparallel ones is almost the same, we have very small $|\langle {{\bf{u}}' \cdot {\bf{b}}'} \rangle|$. At the same time, due to the alignment, $|\langle {{\bf{u}}' \times {\bf{b}}'} \rangle| = 0$.
	
	Thirdly, the turbulent electromotive force $\langle {{\bf{u}}' \times {\bf{b}}'} \rangle$ cannot be estimated only by one term of $\alpha {\bf{B}}$, $- \beta {\bf{J}}$, and $\gamma \mbox{\boldmath$\Omega$}$. The balance of these three terms should be important. In reality, we may have a situation such as
	\begin{equation}
		\underbrace{\langle {{\bf{u}}' \times {\bf{b}}'} \rangle}_{\mbox{small or large}}
		= \underbrace{\alpha {\bf{B}}}_{\mbox{large}} 
		- \underbrace{\beta {\bf{J}}}_{\mbox{large}} 
		+ \underbrace{\gamma \mbox{\boldmath$\Omega$}}_{\mbox{large}}.
	\end{equation}

\item[(A-iii)] Properties of turbulence will be changed by the rotation effect. Actually, as we will see in \S\ref{sec:gamma_term}, the velocity under the rotation or vortical motion is subject to the Coriolis-like force due to the local angular momentum conservation.

\item[(A-iv)] How and how much cross helicity can exist in turbulence is a problem of substantial importance. As will be suggested by the estimates of the galactic magnetic field, the period of magnetic activity, the torsional oscillation inside the Sun, etc., the turbulent cross helicity scaled by the turbulent MHD energy, $|W/K| = O(10^{-2})$ seems to be large enough for the cross-helicity effect to be relevant for several phenomena. Further information through the experiments, observations and numerical simulations is needed on the estimate of $|W/K|$.

\end{enumerate}

	These considerations suggest that there is no definite reason why we can deny the possibility of the cross-helicity-related dynamo. Such a dynamo other than the usual helicity or $\alpha$ dynamo may serve itself as a supplementary player in the dynamo process. How much cross-helicity is relevant depends on how much cross helicity we have in turbulence.
	
	As will be seen in \S\ref{sec:turb_emf}, if we retain the inhomogeneous mean velocity ${\bf{U}}$ in the fluctuation equations, we do have a cross-helicity contribution to the turbulent electromotive force ${\bf{E}}_{\rm{M}} = \langle {{\bf{u}}' \times {\bf{b}}'} \rangle$. This contribution was first calculated by \citet{yos1990} with the aid of an analytical statistical theory of inhomogeneous turbulence. Physical interpretations of this effect have been proposed with the aid of the stationary dynamo solution consisting of the alignment of the mean magnetic field $\bf B$ and the velocity field $\bf U$ \citep{yos1993b,yok1996a}. Also the physical origin of the cross-helicity effect has been clarified by \citet{yok1999}. The importance of the cross-helicity effects has been pointed out in the context of the mean-field dynamo theory, with special emphasis on the magnetic-field generation in astrophysical phenomena such as accretion disks \citep{yos1993b,nis1998}, the Sun and the Earth \citep{yos1993a, yos1996b}, galaxies \citep{yok1996a,bra1998}, and on the turbulence suppression in fusion devices such as the improved confinement mode in tokamaks \citep{yos1991,yok1996b,yos1999}. What has been lacking in this dynamo study is a numerical test of the basic notion of the idea.

	As we see in the following sections, the cross-helicity dynamo shows features different from the usual helicity or $\alpha$ dynamo. One of such features is the configuration of the mean electric-current density ${\bf{J}}$. As we show in \S \ref{sec:phys_interprt}, the ${\bf{E}}_{\rm{M}}$ expression itself does not tell us any alignments between the mean electric-current density ${\bf{J}}$, the mean magnetic field ${\bf{B}}$, or the mean vorticity $\mbox{\boldmath$\Omega$}$. The direction of the mean fields is determined by the spatial distribution of several turbulent quantities with the boundary conditions. However there is alignment tendency between the corresponding parts of the turbulent electromotive force. As the celebrated figure of the $\alpha$ dynamo indicates (later in Figure~\ref{fig:alpha_dynamo}), an essential ingredient of the $\alpha$ effect lies in its ability to produce a mean-field configuration with the mean magnetic field ${\bf{B}}$ which has a component parallel or anti-parallel to the mean electric-current density ${\bf{J}}$. If the main balancer against the turbulent-magnetic-diffusivity-related term ($\beta {\bf{J}}$) is the $\alpha$-related term ($\alpha {\bf{B}}$), the essential feature of the $\alpha$ effect is the alignment of ${\bf{B}}$ and ${\bf{J}}$. This point would be clearer if we consider the mean Ohm's law as is shown later in Eq.~(\ref{eq:mean_Ohms_law}). With the ${\bf{E}}_{\rm{M}}$ expression, the mean electric-current density is expressed as in Eq.~(\ref{eq:mean_Lorentz_force_exp}). This clearly shows that the $\alpha {\bf{B}}$ term never enters into the mean-field Lorentz force ${\bf{J}} \times {\bf{B}}$. In this sense, irrespective of the boundary conditions, the magnetic field induced by the $\alpha$ effect never contributes to the ${\bf{J}} \times {\bf{B}}$ back reaction.
	
	Alignment of the mean electric-current density ${\bf{J}}$ with the mean vorticity $\mbox{\boldmath$\Omega$}$ in the ${\bf{E}}_{\rm{M}}$ expression is one of the characteristics of the cross-helicity dynamo. This configuration may naturally lead to the alignment of the mean magnetic field ${\bf{B}}$ and the mean velocity ${\bf{U}}$. Unlike the $\alpha$ effect, the mean-field configuration in the cross-helicity dynamo allows a non-vanishing mean-field Lorentz force $({\bf{J}} \times {\bf{B}} \neq 0)$. This is a distinct difference from the mean-field configuration in the $\alpha$ dynamo. This feature is fully utilized when we investigate the flow generation or flow dynamo by considering the cross-helicity effects in the momentum equation in \S\ref{sec:flow_generation}. Related to the Lorentz force, we should note the following point. So far we have argued only the mean-field Lorentz force, ${\bf{J}} \times {\bf{B}}$. The mean of the Lorentz force, $\langle {{\bf{j}} \times {\bf{b}}} \rangle$, contains the other part expressed by $\langle {{\bf{j}}' \times {\bf{b}}'} \rangle$. The latter is directly related to the turbulent Maxwell stress, which is included in the definition of the MHD Reynolds stress [Eq.~(\ref{eq:Re_strss_def})] in this work. This certainly gives an important contribution of turbulence to the mean momentum equation. Actually, inclusion of both ${\bf{J}} \times {\bf{B}}$ and $\langle {{\bf{j}}' \times {\bf{b}}'} \rangle$ is an essential point when we consider the flow generation related to the turbulent dynamo (\S \ref{sec:flow_generation}).

	In the general situation of dynamo process, both the helicity and cross-helicity effects would play a certain role in generating and sustaining the mean magnetic field against the enhanced magnetic diffusion due to turbulence. Corresponding to this, as will be referred to later, the turbulent electromotive force in its generic form is a functional of $\bf B$, $\bf J$, and $\mbox{\boldmath$\Omega$}$. This fact implies that the alignment of $\bf J$ and $\mbox{\boldmath$\Omega$}$ is a direct consequence of the turbulent electromotive force that lacks the $\bf B$-related term but consists of the $\bf J$- and $\bf\Omega$-related terms only. On the other hand, the alignment of $\bf J$ and $\bf B$, which is realized in the helicity or $\alpha$ dynamo, might be an immediate consequence of the turbulent electromotive force with the $\mbox{\boldmath$\Omega$}$-related term dropped. These points lead us to the questions: What physical process underlies in each term of the turbulent electromotive force that is originated from the presence of the mean magnetic field $\bf B$, the mean electric-current density $\bf J$, and the mean vorticity $\bf\Omega$? Which effect is dominant under what conditions? These questions are addressed in the following sections where physical origin of each term of the turbulent electromotive force is discussed, and where production mechanisms of the turbulent cross helicity are examined.
	
	The organization of this paper is as follows. In \S\ref{sec:ch_properties}, we introduce the turbulent cross helicity and present its properties that are relevant to the dynamo process. In \S\ref{sec:turb_emf}, the turbulent electromotive force is viewed from several aspects. By considering the evolution equation of the turbulent electromotive force, the physical origin of each term of the electromotive force is shown. In \S\ref{sec:ch_effect}, basic properties of the cross-helicity dynamo are explained. In comparison with the $\alpha$ or helicity dynamo, the main features of the field configuration in the cross-helicity dynamo are stressed. In the applications of the cross-helicity dynamo to real phenomena, how and how much cross helicity exists in turbulence are very important issues. By considering the transport equation of the turbulent cross helicity, we examine the production mechanisms of turbulent cross helicity in \S\ref{sec:ch_generation}. In \S\ref{sec:ch_examples}, some illustrative applications of the cross-helicity dynamo to real phenomena are presented, which include galactic magnetic field (\S\ref{sec:galaxy_dynamo}), accretion disk (\S\ref{sec:accretion_dynamo}), solar dynamos (\S\ref{sec:solar_dynamo}). Another interesting feature of the cross-helicity effect is flow generation. Some examples of this flow dynamo are presented in \S\ref{sec:flow_generation}. Recently several numerical tests on the cross-helicity effect have been performed or in progress. Some of these numerical results are presented in \S\ref{sec:numerical_tests}. Concluding remarks are given in \S\ref{sec:conclusions}

\section{Cross helicity\label{sec:ch_properties}}
	In order to examine the turbulence effects on the evolution of the mean velocity and magnetic field, we adopt an ensemble average $\langle \cdots \rangle$ and divide a field quantity $f$ into the mean $F$ and the fluctuation around it, $f'$:
\begin{equation}
	f = F + f',\;\; F = \langle{f}\rangle
	\label{eq:Rey_decomp}
\end{equation}
with
\begin{subequations}
\begin{equation}
	f = \left( {
		{\bf{u}}, \mbox{\boldmath$\omega$}, {\bf{b}}, {\bf{j}}, {\bf{e}}, 
		{\bf{a}}, \rho, p, p_{\rm{M}}, q, \theta
	} \right),
	\label{eq:instant_f}
\end{equation}
\begin{equation}
	F = \left( {
		{\bf{U}}, \mbox{\boldmath$\Omega$}, {\bf{B}}, {\bf{J}}, {\bf{E}}, 
		{\bf{A}}, \overline{\rho}, P, P_{\rm{M}}, Q, \Theta}
	\right),
	\label{eq:mean_f}
\end{equation}
\begin{equation}
	f = \left( {
		{\bf{u}}', \mbox{\boldmath$\omega$}', {\bf{b}}', {\bf{j}}', {\bf{e}}', 
		{\bf{a}}', \rho', p', p'_{\rm{M}}, q', \theta'}
	\right),
	\label{eq:fluct_f}
\end{equation}
\end{subequations}
where ${\bf{u}}$ is the velocity, $\mbox{\boldmath$\omega$} (= \nabla \times {\bf{u}})$ the vorticity, ${\bf{b}}$ the Alfv\'{e}n velocity (magnetic field measured in the Alfv\'{e}n speed unit), ${\bf{j}} (= \nabla \times {\bf{b}})$ the electric-current density counterpart, ${\bf{e}}$ the electric field counterpart, ${\bf{a}}$ the vector potential, $\rho$ the density, $p$ the pressure, $p_{\rm{M}} (= p + {\bf{b}}^2/2)$ the MHD pressure, $q$ the internal energy, $\theta$ the temperature. Note that the decomposition [Eq.~(\ref{eq:Rey_decomp})] itself does not require any scale separation between the mean and fluctuation quantities. Here, the magnetic field etc.\ are expressed using the Alfv\'{e}n-speed unit, and are related to the ones with the original or physical unit denoted with $_\ast$ as
\begin{equation}
	{\bf{b}} = \frac{{\bf{b}}_\ast}{\sqrt{\mu_0 \rho}},\;\;
	{\bf{j}} = \frac{{\bf{j}}_\ast}{\sqrt{\rho / \mu_0}},\;\;
	{\bf{e}} = \frac{{\bf{e}}_\ast}{\sqrt{\mu_0 \rho}},\;\;
	p = \frac{p_\ast}{\rho}
	\label{eq:alfven_unit}
\end{equation}
($\mu_0$: magnetic permeability).

	The main reason of introducing the Alfv\'{e}n-speed formulations is to make the system of equations more symmetric and the treatment of nonlinearity simpler. This is along the same line with introducing the pressure function $\tilde{p} = \int(p/\rho) dp$ in the barotropic fluid analysis, and with the Favr\'{e} or mass-weighted average of the velocity $\{ {\bf{u}} \}_{\rm{m}} = \langle {\rho {\bf{u}}} \rangle / \langle {\rho} \rangle$ in the compressible turbulence analysis. In order to tackle a strongly nonlinear problem, typically represented by the response-function equation [Eq.~(\ref{eq:GreenFnDef})], it is very useful to introduce some variables that make the governing equations simpler or more symmetric.
	
	Another (physically more important) reason to adopt the magnetic field in Alfv\'{e}n-speed units is related to its usefulness in the physical arguments. It is not the turbulent MHD energy (density) $\overline{\rho} \langle {{\bf{u}}'{}^2} \rangle /2 + \langle {{\bf{b}}'_\ast} \rangle / (2 \mu_0)$ or the turbulent cross helicity (density) $\langle {{\bf{u}}' \cdot {\bf{b}}'_\ast} \rangle$ themselves but a dimensionless quantities constructed by the energy and the cross helicity that represents the dynamic properties of turbulence transports. In the later sections we see that the turbulent cross helicity $\langle {{\bf{u}}' \cdot {\bf{b}}'} \rangle$ normalized by the turbulent MHD energy (per unit mass) $\langle {{\bf{u}}'{}^2 + {\bf{b}}'{}^2} \rangle /2$ is such a measure. This is in contrast to the geometrical or topological measure $\langle {{\bf{u}}' \cdot {\bf{b}}'_\ast} \rangle / (\sqrt{{\bf{u}}'{}^2} \sqrt{{\bf{b}}'_\ast{}^2})$, which just represents the alignment angle between the velocity and magnetic field.

	Cross helicity is the correlation between the velocity and magnetic field defined by
\begin{equation}
	{\cal{W}}_{\rm{tot}} \equiv \int_V {{\bf{u}} \cdot {\bf{b}}_\ast} dV,
	\label{eq:tot_cross_hel_def}
\end{equation}
where $V$ is the volume of the system considered. In this paper, we mainly consider its local density ${\bf{u}} \cdot {\bf{b}}$ with the magnetic field measured in Alfv\'{e}n-speed units, and denote it as cross helicty. Then the turbulent cross helicity (density) is defined as
\begin{equation}
	W \equiv \langle {{\bf{u}}' \cdot {\bf{b}}'} \rangle,
	\label{eq:turb_cross_hel_def}
\end{equation}
while the mean-field cross helicity (density) is defined as
\begin{equation}
	W_{\rm{M}} \equiv {\bf{U}} \cdot {\bf{B}}.
	\label{eq:mean_cross_hel_def}
\end{equation}
These definitions are adopted on the basis that we mostly treat turbulence in an incompressible framework. In the case of incompressible turbulence, whether we define the cross helicity by the magnetic field in physical units, ${\bf{b}}_\ast$, or by the counterpart in Alfv\'{e}n-speed units, ${\bf{b}}$, makes no substantial difference. It is quite common in the literature of turbulence studies to express the magnetic field by the one measured in Alfv\'{e}n-speed units. An even more symmetrical form of magnetohydrodynamic (MHD) equations; the Elsasser-variable formulation is often adopted in the analysis of incompressible MHD turbulence. 

	Treating fully nonlinear compressible MHD turbulence is a very difficult task. There are several strategies for it. One approach is a hybrid treatment: turbulence is treated as incompressible but the compressibility effects are taken into account for the estimate of the effective transport coefficients \citep{can1991}. The assumption that the turbulence is incompressible is made because of the simplicity of the mathematical treatment. We should note that even in the compressible turbulence case if we can neglect the density fluctuations ($\rho' = 0$), the expressions for the Reynolds stress $\mbox{\boldmath$\cal{R}$}$ and the turbulent electromotive force ${\bf{E}}_{\rm{M}}$ are written in a form similar to the counterpart in the incompressible case. We also note that the Elsasser formulation can be extended to the compressible MHD cases [\citet{mar1987}, see also \citet{yok2007} for compressible MHD turbulence].
	
	Cross helicity possesses several important features, which include (i) conservation, (ii) topological interpretation, (iii) pseudo-scalar, (iv) transport suppression, (v) boundedness, (vi) relation to Alfv\'{e}n wave. Since these features are related to the properties of cross-helicity-related dynamo, we briefly explain them.

\paragraph{(i)  Conservation}
	Cross helicity, as well as the magnetohydrodynamic (MHD) energy $\int_V {({\bf{u}}^2 +{\bf{b}}^2)/2} dV$ and magnetic helicity $\int_V {{\bf{a}} \cdot {\bf{b}}} dV$, is an inviscid invariant of the MHD equations. It is conserved in the absence of the molecular viscosity and magnetic diffusivity ($\nu = \eta =0$). This can be easily shown as
\begin{subequations}
\begin{eqnarray}
	\frac{d{\cal{W}}_{\rm{tot}}}{dt}
&=& \int_V {\left( {
\frac{\partial {\bf{u}}}{\partial t} \cdot {\bf{b}}_\ast
+ {\bf{u}} \cdot \frac{\partial{\bf{b}}_\ast}{\partial t} 
} \right) } dV
	\nonumber\\
&=& \int_V \left\{ {
	\left[ {
	- \left( {{\bf{u}} \cdot \nabla} \right) {\bf{u}}
	- \frac{1}{\rho} \nabla p_\ast
	+ \frac{1}{\rho} {\bf{j}}_\ast \times {\bf{b}}_\ast} \right] \cdot {\bf{b}}_\ast
	+ {\bf{u}} \cdot 
	\left[ {\nabla \times \left( { {\bf{u}} \times {\bf{b}}_\ast } \right)} \right] 
	} \right\}  dV
	\nonumber\\
	&=& \int_V \nabla \cdot \left[ {
		\left( {
		\frac{1}{2} {\bf{u}}^2
		- \frac{\gamma_0}{\gamma_0 - 1} \frac{p_\ast}{\rho}
	} \right) {{\bf{b}}_\ast}
	- \left( {{\bf{u}} \cdot {\bf{b}}_\ast} \right) {\bf{u}}
	} \right]\ dV\\
	\label{eq:total_ch_eq_div}
	&=& \int_S \left[ {
	 \left( {
	\frac{1}{2}{\bf{u}}^2 - \frac{\gamma_0}{\gamma_0 - 1} \frac{p_\ast}{\rho}
	} \right) {\bf{b}}_\ast
	- \left( {{\bf{u}} \cdot {\bf{b}}_\ast} \right) {\bf{u}}
	} \right] \cdot {\bf{n}}\ dS,
	\label{eq:total_ch_eq}
\end{eqnarray}
\end{subequations}
where $\gamma_0$ is the ratio of the pressure and volume specific heats, and ${\bf{n}}$ is the outward normal unit vector. A polytropic relation between the pressure and density $p = \rho^{\gamma_0}$ is assumed. Here use has been made of vector identities:
\begin{equation}
	\left( {{\bf{u}} \cdot \nabla} \right) {\bf{u}}
	= \nabla \left( {\frac{{\bf{u}}^2}{2}} \right)
	- {\bf{u}} \times \mbox{\boldmath$\omega$},
\end{equation}
\begin{equation}
	{\bf{u}} \cdot \left[ {
	\nabla \times \left( {{\bf{u}} \times {\bf{b}}_\ast} \right)
	} \right]
	= - \nabla \cdot \left[ {
	{\bf{u}} \times \left( {{\bf{u}} \times {\bf{b}}_\ast} \right)
	} \right]
	+ \left( {
	{\bf{u}} \times {\bf{b}}_\ast
	} \right) \cdot \mbox{\boldmath$\omega$}
\end{equation}
[$\mbox{\boldmath$\omega$} (= \nabla \times {\bf{u}})$: vorticity]. Equation~(\ref{eq:total_ch_eq}) shows that if we have no velocity nor magnetic field at the boundary surface (${\bf{u}} = {\bf{b}}_\ast = 0$), the total amount of cross helicity is conserved. The last line of Eq.~(\ref{eq:total_ch_eq}) shows that in addition to the cross-helicity influx:
\begin{equation}
	\int_S {\left( {
	{\bf{u}} \cdot {\bf{b}}_\ast
	} \right) {\bf{u}} \cdot (-{\bf{n}})}\ dS,
\end{equation}
if we have a sort of energy inhomogeneity along the magnetic field:
\begin{equation}
	\int_S {\left( {
		\frac{1}{2} {\bf{u}}^2
		- \frac{\gamma_0}{\gamma_0 - 1} \frac{p_\ast}{\rho}
	} \right) {\bf{b}}_\ast \cdot {\bf{n}}}\ dS
	= \int_V {{\bf{b}}_\ast} \cdot \nabla \left( {
		\frac{1}{2} {\bf{u}}^2
		- \frac{\gamma_0}{\gamma_0 - 1} \frac{p_\ast}{\rho}
	} \right)\ dV,
	\label{eq:ch_gene_en_inhomo}
\end{equation}
the cross helicity is supplied to the system. As the divergence form [Eq.~(\ref{eq:total_ch_eq_div})] shows, this just expresses the transport effect. However, it may play important role in the local cross-helicity generation because the cross helicity is not positive definite. This is a very important point for cross helicity generation mechanism and in strong contrast with the positive definite quantities like the energy. We return to this point later in \S\ref{sec:ch_generation}.

\begin{figure}[htpb]
\begin{center}
\includegraphics[width=.30\textwidth]{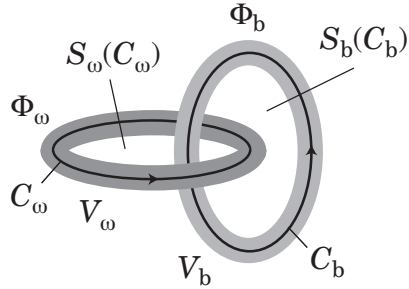}%
\caption{Topological interpretation of cross helicity.}%
\label{fig:topology_ch}
\end{center}
\end{figure}

\paragraph{(ii) Topological interpretation}
	Like the kinetic helicity ${\bf{u}} \cdot \mbox{\boldmath$\omega$}$ ($\mbox{\boldmath$\omega$}$: vorticity, $\mbox{\boldmath$\omega$} = \nabla \times {\bf{u}}$) and the magnetic helicity ${\bf{a}} \cdot {\bf{b}}$ (${\bf{a}}$: magnetic vector potential, ${\bf{b}} = \nabla \times {\bf{a}}$), the cross helicity can be topologically interpreted. The cross helicity provides a measure of degree of linkage of the vortex tubes of the velocity field with the flux tubes of the magnetic-field. To see this, we consider a special situation where the magnetic field is ${\bf{b}} = 0$ except in a single flux-tube volume $V_{b}$ in the neighborhood of the closed line $C_b$ and the vorticity is $\mbox{\boldmath$\omega$} = 0$ except in a single vortex-tube volume $V_\omega$ in the neighborhood of the closed line $C_\omega$ as in Figure~\ref{fig:topology_ch}. 

	In this case, the cross helicity is expressed as
\begin{equation}
	\int_V {{\bf{u}} \cdot {\bf{b}}}\ dV
	= \int_{V_{b}} {{\bf{u}} \cdot {\bf{b}}}\ dV_{b}
	+ \int_{V_{\omega}} {{\bf{u}} \cdot {\bf{b}}}\ dV_{\omega}.
	\label{eq:topol_int_1}
\end{equation}
If the magnetic field is homogeneous in the cross section of the volume $V_b$, the first term in Eq.~(\ref{eq:topol_int_1}) is expressed as
\begin{eqnarray}
	\int_{V_{b}} {{\bf{u}} \cdot {\bf{b}}}\ dV_{b}
	&=& \Phi_{b} \int_{C_{b}} {\bf{u}} \cdot d{\bf{s}}
	= \Phi_{b} \int_{S_{b}(C_{b})} 
		(\nabla \times {\bf{u}}) \cdot {\bf{n}}\ dS_{b}
	\nonumber\\
	&=& \Phi_{b} \int_{S_{b}(C_{b})} 
		\mbox{\boldmath$\omega$} \cdot {\bf{n}}\ dS_{b}
	= \Phi_{b} \Phi_{\omega},
\end{eqnarray}
where 
\begin{equation}
	\Phi_b = \int_{S(V_b)} {\bf{b}} \cdot {\bf{n}}\ dS
	= \int_{S_{\omega}(C_\omega)} {\bf{b}} \cdot {\bf{n}}\ dS_{\omega}
	\label{eq:topol_int_2}
\end{equation}
is the magnetic flux through the cross section of the volume $V_b$, which is equal to the magnetic flux through the surface $S(C_\omega)$ spanned by the loop $C_\omega$. And $\Phi_\omega$, the vortex flux through the surface $S_b$ spanned by the loop $C_b$, is equal to the vortex flux through the cross section of volume $V_\omega$. A similar argument is applicable to the second term in Eq.~(\ref{eq:topol_int_1}). Finally, the total amount of cross helicity is expressed
\begin{equation}
	\int_V {{\bf{u}} \cdot {\bf{b}}}\ dV = 2n \Phi_{\omega} \Phi_{b},
	\label{eq:topol_int_3}
\end{equation}
where $n$ shows how many times the vortex (or magnetic flux) tube thread through the surface spanned by the magnetic flux (or vortex) tube. Equation~(\ref{eq:topol_int_3}) shows that the cross helicity is equivalent to the knottedness of the vortex tube with the magnetic flux tube. The conservation of cross helicity topologically means the number of knottedness is conserved.

\paragraph{(iii) Pseudoscalar}
	Unlike the energy, the cross helicity is not positive definite. Inversion of the coordinate system is equivalent to a combination of rotation and reflection (mirror) transformations. It corresponds to the change of the coordinate system from right-handed into left-handed. Velocity is a polar vector whose components change their sign under inversion ($x^i \mapsto \hat{x}^i = - x^i$) as $u^i \mapsto \hat{u}^i = - u^i$, whereas magnetic field is an axial vector whose components do not change their sign as $b^i \mapsto \hat{b}^i = b^i$ ($\hat{\cdot}$ denotes a quantity under inversion). Defined as the inner product of the velocity and magnetic field, the cross helicity (density) changes its sign under inversion as $W \mapsto \hat{W} = - W$.
	
	In a mirror or reflectional symmetric system, all statistical quantities show $f({\bf{r}}) \mapsto \hat{f}(\hat{\bf{r}}) = \hat{f}(-{\bf{r}}) = f({\bf{r}})$. At the same time, by definition, a pseudoscalar quantity changes its sign under the inversion as $f({\bf{r}}) \mapsto \hat{f}(\hat{\bf{r}}) = \hat{f}(-{\bf{r}}) = - f({\bf{r}})$. Thus a pseudoscalar in a mirrorsymmetric system obeys $f({\bf{r}}) = - f({\bf{r}})$. Namely, a pseudoscalar in a mirrorsymmetric system always vanishes: $f({\bf{r}}) = 0$. To put it other way, a finite value of pseudoscalar appears only in non-mirrorsymmetric systems. In this sense, pseudoscalar is a measure for representing the breakage of mirrorsymmetry. Pseudoscalar nature of cross helicity is of fundamental importance in dynamo.

\paragraph{(iv) Transport suppression}
	In the magnetic induction equation: 
\begin{equation}
	\frac{\partial {\bf{b}}}{\partial t}
	= \nabla \times \left( {{\bf{u}} \times {\bf{b}}} \right)
	+\eta  \nabla^2 {\bf{b}},
	\label{eq:mag_ind_eq}
\end{equation}
the nonlinear mixing is represented by the first or ${\bf{u}} \times {\bf{b}}$-related term. Note that in general the velocity ${\bf{u}}$ depends on the magnetic field ${\bf{b}}$. If the velocity and the magnetic field are aligned, ${\bf{u}} \parallel {\bf{b}}$, the mixing term vanishes. In such a case, we have no nonlinear mixing and magnetic-field evolution obeys just a diffusion equation:
\begin{equation}
	\frac{\partial {\bf{b}}}{\partial t}
	= \eta  \nabla^2 {\bf{b}}.
	\label{eq:mag_diff_eq}
\end{equation}
Since the magnitude of inner and outer products of velocity and magnetic-field vectors are related as
\begin{equation}
\frac{\left( {{\bf{u}} \cdot {\bf{b}}} \right)^2}{|{\bf{u}}|^2 |{\bf{b}}|^2}
	+ \frac{\left( {{\bf{u}} \times {\bf{b}}} \right)^2}{|{\bf{u}}|^2 |{\bf{b}}|^2}
	= 1,
	\label{eq:ub_cos_sin_rel}
\end{equation}
the cross helicity is expected to be related to the suppression of nonlinear mixing coming from ${\bf{u}} \times {\bf{b}}$.

\paragraph{(v) Boundedness}
	 The magnitude of cross helicity is bounded by that of MHD energy since
\begin{equation}
	\left( { {\bf{u}} \pm {\bf{b}} } \right)^2 \ge 0
	\label{eq:norm_positive}
\end{equation}
or equivalently
\begin{equation}
	\frac{|{\bf{u}} \cdot {\bf{b}}|}{({\bf{u}}^2 + {\bf{b}}^2)/2} \le 1.
	\label{eq:ch_boundedness}
\end{equation}
This inequality should be locally satisfied. So, the magnitude of local cross helicity (density) is always bounded by the local MHD energy (density). As will be shown in the later sections, this boundedness of cross helicity gives important constraints for the the magnitude of the cross-helicity effect, model constants for cross-helicity evolution equations, etc.

\begin{figure}[htpb]
\begin{center}
\includegraphics[width=.60\textwidth]{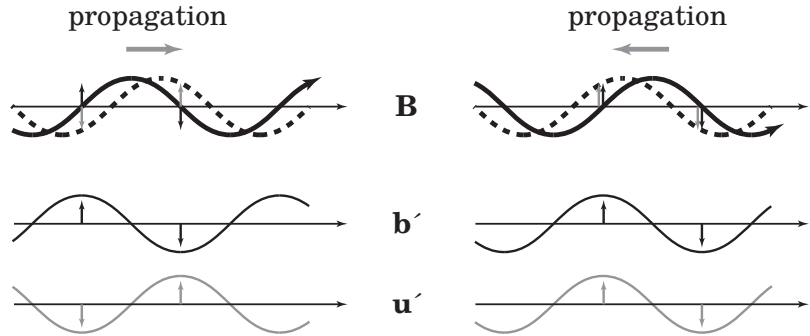}%
\caption{Alfv\'{e}n wave propagation and sign of helicity.}%
\label{fig:alfven_propagate}
\end{center}
\end{figure}

\paragraph{(vi) Alfv\'{e}n wave}
	Cross helicity is related to the asymmetry of Alfv\'{e}n waves. If the Alfv\'{e}n wave propagates in the direction parallel (or antiparallel) to the large-scale magnetic field, the velocity variation associated with the Alfv\'{e}n wave is antiparallel (or parallel) to the magnetic-field variation, contributing to a negative (or positive) turbulent cross helicity (Figure~\ref{fig:alfven_propagate}). If the Alfv\'{e}n waves equally propagate in the directions parallel and antiparallel to the large-scale magnetic field, the negative and positive contributions to cross helicity are canceled out, and the net cross helicity becomes zero. If we have asymmetry between these two directions, we have a finite cross helicity. In the case of solar wind, the source of oscillation is located on the surface of the Sun. As this result, the Alfv\'{e}n waves predominantly propagate outward direction from the Sun in the solar wind. This strong asymmetry of the Alfv\'{e}n wave propagation gives large positive and negative cross helicity in the solar-wind turbulence depending on the magnetic-field sectors. In a sector with magnetic field is outward (or inward) direction from (or toward) the Sun, we have negative (or positive) cross helicity. This is the reason why we observe exceptionally large magnitude of cross helicity in the solar-wind turbulence in general.

	This asymmetry of Alfv\'{e}n wave is related to the cross-helicity supply mechanism [Eq.~(\ref{eq:ch_gene_en_inhomo})]. This point will be referred to later in \S\ref{sec:ch_production} in relation to the cross-helicity generation mechanism in turbulence.

\section{Turbulent electromotive force\label{sec:turb_emf}}
\subsection{Mean and fluctuation}
	In order to investigate the flow--turbulence interaction, we have to simultaneously treat the mean fields and the fluctuation fields. In this section, we will show some basic results from the analytical statistical theory for the inhomogeneous magnetohydrodynamic (MHD) turbulence. The analytical expressions are often too complicated for the practical applications to the real-world turbulence in astro- and geophysical phenomena. Turbulence modelling on the basis of the statistical theory provides a powerful tool for analyzing the real-world turbulence. The methods of turbulence modelling will be also referred to in this section.

	For the sake of simplicity, in the following, we basically consider the incompressible flow. This treatment does not deny the importance of the compressibility in dynamo action at all. Rather, compressibility is often one of the most important ingredients of the dynamo processes. We will refer to the cross-helicity production mechanism in compressible MHD turbulence later. 

	With decomposition Eq.~(\ref{eq:Rey_decomp}), equations for the mean velocity and magnetic field are given as
\begin{equation}
	\frac{\partial {\bf{U}}}{\partial t}
	= {\bf{U}} \times {\bf{\Omega}}
	+ {\bf{J} \times {\bf{B}}}
	- \nabla \cdot \mbox{\boldmath$\cal R$}
	+ {\bf{F}}
	- \nabla \left( {
		P + \frac{1}{2} {\bf{U}}^2 
		+ \left\langle {\frac{1}{2} {\bf{b}}'{}^2} \right\rangle
	} \right),
	\label{eq:mean_U_eq_rot}
\end{equation}
\begin{equation}
	\frac{\partial {\bf{B}}}{\partial t}
	= \nabla \times \left( {
		{\bf{U}} \times {\bf{B}}
		+ {\bf{E}}_{\rm{M}}
	} \right)
	+ \eta \nabla^2 {\bf{B}},
	\label{eq:mean_B_eq}
\end{equation}
and the solenoidal conditions for the mean velocity and magnetic field:
\begin{equation}
	\nabla \cdot {\bf{U}} = \nabla \cdot {\bf{B}} = 0,
	\label{eq:mean_sol_cond}
\end{equation}
where $\mbox{\boldmath$\Omega$} (= \nabla \times {\bf{U}})$ is the mean vorticity and ${\bf{F}}$ is the mean part of the external force. Here the Reynolds stress $\mbox{\boldmath${\cal{R}}$}$ and the turbulent electromotive force ${\bf{E}}_{\rm{M}}$ are defined	by
\begin{equation}
	{\cal{R}}^{\alpha\beta} 
	\equiv \left\langle {
		u'{}^\alpha u'{}^\beta - b'{}^\alpha b'{}^\beta
	} \right\rangle,
	\label{eq:Re_strss_def}
\end{equation}
\begin{equation}
	{\bf{E}}_{\rm{M}}
	\equiv \left\langle {
		{\bf{u}}' \times {\bf{b}}'
	} \right\rangle.
	\label{eq:E_M_def}
\end{equation}
They are sole quantities representing the effects of fluctuation in the mean equations.

	On the other hand, equations for the velocity fluctuation ${\bf{u}}'$ and the magnetic-field counterpart ${\bf{b}}'$ are expressed in the rotational forms as
\begin{equation}
	\frac{\partial {\bf{u}}'}{\partial t}
	= {\bf{u}}' \times \mbox{\boldmath$\Omega$}
	+ {\bf{U}}\times \mbox{\boldmath$\omega$}'
	+ {\bf{j}}'\times {\bf{B}}
	+ {\bf{J}}\times {\bf{b}}'
	- \nabla p'
	- \nabla \cdot \mbox{\boldmath$\cal{R}$}
	+ {\bf{f}}'
	+ \nu \nabla^2 {\bf{u}}',
	\label{eq:fluct_u_rot_eq}
\end{equation}
\begin{equation}
	\frac{\partial {\bf{b}}'}{\partial t}
	= \nabla \times \left( {{\bf{u}}'\times {\bf{B}}} \right)
	+ \nabla \times \left( {{\bf{U}}\times {\bf{b}}'} \right)
	+ \nabla \times \left( {{\bf{u}}' \times {\bf{b}}'} \right)
	- \nabla \times {\bf{E}}_{\rm{M}}
	+ \eta \nabla^2 {\bf{b}}',
	\label{eq:fluct_b_rot_eq}
\end{equation}
with the solenoidal conditions for the fluctuation fields:
\begin{equation}
	\nabla \cdot {\bf{u}}' = \nabla \cdot {\bf{b}}' = 0
	\label{eq:fluct_sol_cond}
\end{equation}
(${\bf{f}}'$: fluctuation part of the external force).

\subsection{Reynolds stress and turbulent electromotive force}
	From the two-scale direct-interaction approximation (TSDIA), analytical theory of inhomogeneous MHD turbulence, the Reynolds stress and the turbulent electromotive force are expressed as
\begin{equation}
	{\cal{R}}^{\alpha\beta}
	= \frac{2}{3} K_{\rm{R}} \delta^{\alpha\beta}
	- \nu_{\rm{K}} {\cal{S}}^{\alpha\beta}
	+ \nu_{\rm{M}} {\cal{M}}^{\alpha\beta}
	+ \Omega^\alpha \Gamma^\beta
		+ \Omega^\beta \Gamma^\alpha
		- \frac{1}{3} \delta^{\alpha\beta} 
			\mbox{\boldmath$\Omega$} \cdot \mbox{\boldmath$\Gamma$},
	\label{eq:Re_strss_exp}
\end{equation}
\begin{equation}
	{\bf{E}}_{\rm{M}}
	= - \beta {\bf{J}} + \alpha {\bf{B}} + \gamma \mbox{\boldmath$\Omega$}
	\label{eq:E_M_exp}
\end{equation}
\citep{yos1990}. Here $K_{\rm{R}} (\equiv \langle {{\bf{u}}'{}^2 - {\bf{b}}'{}^2} \rangle/2)$ is the turbulent MHD residual energy, $\mbox{\boldmath$\Omega$} (= \nabla \times {\bf{U}})$ is the mean vorticity, and $\mbox{\boldmath$\Gamma$}$ depends on the gradient of the kinetic-helicity spectrum, $\nabla H_{uu}$ [For the definition of $H_{uu}$, see Eq.~(\ref{eq:def_spctral_fn_Huu_Hbb}) in Appendix]. In Eqs.~(\ref{eq:Re_strss_exp}) and (\ref{eq:E_M_exp}), $\nu_{\rm{K}}$, $\nu_{\rm{M}}$, $\beta$, $\alpha$, and $\gamma$ are the transport coefficients, which are connected to the statistical properties of turbulence. As will be shown later, $\nu_{\rm{K}}$ and $\nu_{\rm{M}}$ are directly related to $\beta$ and $\gamma$, respectively. In equation (\ref{eq:Re_strss_exp}), $\mbox{\boldmath${\cal{S}}$}$ and $\mbox{\boldmath${\cal{M}}$}$ are the strain rates of the mean velocity and magnetic fields, respectively. They are defined by
\begin{equation}
	{\cal{S}}^{\alpha\beta}
	= \frac{\partial U^\alpha}{\partial x^\beta} 
	+  \frac{\partial U^\beta}{\partial x^\alpha},
	\label{eq:mean_vel_strain}
\end{equation}
\begin{equation}
	{\cal{M}}^{\alpha\beta}
	= \frac{\partial B^\alpha}{\partial x^\beta} 
	+  \frac{\partial B^\beta}{\partial x^\alpha}.
	\label{eq:mean_mag_strain}
\end{equation}

	Outline of how to derive Eqs.~(\ref{eq:Re_strss_exp}) and (\ref{eq:E_M_exp}) from Eqs.~(\ref{eq:fluct_u_rot_eq}) and (\ref{eq:fluct_b_rot_eq}) with Eq.~(\ref{eq:fluct_sol_cond}) is given in Appendix. The main assumptions used in derivation is homogeneity and isotropy of the lowest-order fields (basic fields) of turbulence at very high Reynolds number. Effects of inhomogeneity of the mean quantities, the mean magnetic field, and the mean vortical motion (including the system rotation) are taken into account in a perturbational manner. Equations~(\ref{eq:Re_strss_exp}) and (\ref{eq:E_M_exp}) are obtained by the analysis up to the first-order calculations [$O(\delta)$ calculation, $\delta$: scale parameter in the expansion, see Eqs.~(\ref{eq:fast_slow_vars}), (\ref{eq:two-scale_deriv}), and (\ref{eq:scl-prmtr-expnd})]. The higher-order derivatives appear in the expressions of $\mbox{\boldmath${\cal{R}}$}$ and ${\bf{E}}_{\rm{M}}$ in the higher-order calculations. For example, a term representing the magnetic pumping appears in the second-order calculation. Also the term related to $\mbox{\boldmath$\Omega$} \times {\bf{J}}$ should appear in the higher-order [$O(\delta^2)$] analysis. Note that in Eq.~(\ref{eq:Re_strss_exp}) both $\mbox{\boldmath$\Omega$}$ and $\mbox{\boldmath$\Gamma$} (\propto \nabla H_{uu})$ contain a spatial derivative of the mean quantities. Thus the $\mbox{\boldmath$\Omega$}$-related terms originally come from the $O(\delta^2)$ calculation of the TSDIA. By making an analysis in a rotating frame, we selectively derive an expression for the helicity effect in the lower-order (first-order) calculations. Later in this paper these $\mbox{\boldmath$\Omega$}$-related term is often dropped from the $\mbox{\boldmath${\cal{R}}$}$ expression. 

		In Eq.~(\ref{eq:E_M_exp}), the turbulent electromotive force ${\bf{E}}_{\rm{M}}$ and the electric-current density ${\bf{J}}$ are polar vectors, which do not change their sign under the inversion of the coordinate system, whereas the magnetic field ${\bf{B}}$ and the vorticity $\mbox{\boldmath$\Omega$}$ are axial vectors, which do. Considering this symmetry, we see that the coefficients $\alpha$ and $\gamma$ are pseudoscalars which change their sign under the inversion, while $\beta$ is a usual (pure-)scalar.
	
	Substituting ${\bf{E}}_{\rm{M}}$ [Eq.~(\ref{eq:E_M_exp})] into the mean induction equation, we have
\begin{equation}
	\frac{\partial {\bf{B}}}{\partial t}
	= \nabla \times \left( {{\bf{U}} \times {\bf{B}}} \right)
	- \nabla \times \left[ { \left( {\eta + \beta} \right) \nabla \times {\bf{B}}} \right]
	+ \nabla \times \left( {
		\alpha {\bf{B}} + \gamma \mbox{\boldmath$\Omega$}
	} \right).
	\label{eq:mean_ind_eq_exp}
\end{equation}
The second term shows that the effective magnetic diffusivity is enhanced by turbulence as $\eta \rightarrow \eta + \beta$, with spatiotemporal variation of $\beta$. The third term represents the effects of pseudoscalars $\alpha$ and $\gamma$ (both of them, as well as $\beta$, show spatiotemporal variations), which may balance the $\beta$ effect or the turbulent magnetic diffusivity to suppress the enhanced transport of the mean magnetic field. Note that higher-order caluculations of the TSDIA analysis give rise to a contribution deviating from the isotropic expression of the transport coefficients.

	In the traditional and authentic mean-field dynamo theory, the mean-field dependence of the turbulent electromotive force ${\bf{E}}_{\rm{M}}$ has been calculated on the basis of fluctuation equations [Eqs.~(\ref{eq:fluct_u_rot_eq}) and (\ref{eq:fluct_b_rot_eq})]. At the same time, in some cases, ${\bf{E}}_{\rm{M}}$ is given without referring to the relation with the fluctuation equations, by using the Ansatz: the turbulent electromotive force should be expressed by a linear combination of the mean magnetic field and its derivatives as
\begin{equation}
	\left\langle {{\bf{u}}' \times {\bf{b}}'} \right\rangle^a
	= \alpha^{ab} B^b + \beta^{abc} \frac{\partial B^b}{\partial x^c} + \cdots.
	\label{eq:E_M_ansatz}
\end{equation}
This gives a clear insight on the expression of the turbulent electromotive force from the mathematical viewpoint. However, we should note that this ``generic'' form of ${\bf{E}}_{\rm{M}}$ is a direct consequence of putting ${\bf{U}} = \langle {{\bf{u}}} \rangle = 0$ in Eqs.~(\ref{eq:fluct_u_rot_eq}) and (\ref{eq:fluct_b_rot_eq}). If we retain the mean velocity in Eqs.~(\ref{eq:fluct_u_rot_eq}) and (\ref{eq:fluct_b_rot_eq}), we naturally have additional terms related to ${\bf{U}}$ in Eq.~(\ref{eq:E_M_ansatz}). Thus the ``generic'' form [Eq.~(\ref{eq:E_M_ansatz})] should be extended to the one with ${\bf{U}}$. In this context, we should note that there are quite a few papers in which the possibility of the other contributions to the turbulent electromotive force is discussed which do not depend on ${\bf{B}}$ \citep{rae1976,rae2000}. The reader is also referred to the arguments extended by \citet{rae2010}, where the ${\bf{U}}$ dependence of ${\bf{E}}_{\rm{M}}$ is considered in a general manner.

	In principle, expression for any correlations of the velocity and magnetic-field fluctuations can be derived from the evolution equations of the velocity and magnetic fluctuations [Eqs.~(\ref{eq:fluct_u_rot_eq}) and (\ref{eq:fluct_b_rot_eq})]. They are equivalently expressed as
\begin{eqnarray}
		\frac{\partial {\bf{u}}'}{\partial t}
		&=& - \left( { {\bf{U}} \cdot\nabla } \right){\bf{u}}'
		- \left( { {\bf{u}}' \cdot\nabla } \right){\bf{U}}
		+ \left( { {\bf{B}} \cdot\nabla } \right){\bf{b}}'
		+ \left( { {\bf{b}}' \cdot\nabla } \right){\bf{B}}
		\nonumber\\
		& & - \left( { {\bf{u}}' \cdot\nabla } \right){\bf{u}}'
		+ \left( { {\bf{b}}' \cdot\nabla } \right){\bf{b}}'
		- \nabla \cdot \mbox{\boldmath${\cal{R}}$}
		- \nabla p'_{\rm{M}}
		+ \nu \nabla^2 {\bf{u}}'
		+ {\bf{f}}',
	\label{eq:fluct_u_eq}
\end{eqnarray}
\begin{eqnarray}
		\frac{\partial {\bf{b}}'}{\partial t}
		&=& \left( {{\bf{B}} \cdot\nabla} \right){\bf{u}}'
		- \left( {{\bf{u}}' \cdot\nabla} \right){\bf{B}}
		- \left( {{\bf{U}} \cdot\nabla} \right){\bf{b}}'
		+ \left( {{\bf{b}}' \cdot\nabla} \right){\bf{U}}
		\nonumber\\
		& & - \left( {{\bf{u}}' \cdot\nabla} \right){\bf{b}}'
		+ \left( {{\bf{b}}' \cdot\nabla} \right){\bf{u}}'
		- \nabla\times{\bf{E}}_{\rm{M}}
		+ \eta \nabla^2 {\bf{b}}',
	\label{eq:fluct_b_eq}
\end{eqnarray}
where $p'_{\rm{M}}$ is the fluctuation part of the MHD pressure $p_{\rm{M}} = p + {\bf{b}}^2/2$.

	With the aid of an analytical theory of inhomogeneous turbulence, it was shown that the transport coefficients in Reynolds stress $\mbox{\boldmath$\cal{R}$}$ [Eq.~(\ref{eq:Re_strss_exp})] and the turbulent electromotive force ${\bf{E}}_{\rm{M}}$ [Eq.~(\ref{eq:E_M_exp})] are expressed as
\begin{equation}
	\alpha = \frac{1}{3}
	\int d{\bf{k}} \int_{-\infty}^\tau \!\!\!d\tau_1
	G(k,{\bf{x}};\tau,\tau_1,t) \left[ {
	-H_{uu}(k,{\bf{x}};\tau,\tau_1,t) + H_{bb}(k,{\bf{x}};\tau,\tau_1,t)
	} \right],
	\label{eq:alpha_spec_exp}
\end{equation}
\begin{equation}
	\beta = \frac{1}{3}
	\int d{\bf{k}} \int_{-\infty}^\tau \!\!\!d\tau_1
	G(k,{\bf{x}};\tau,\tau_1,t) \left[ {
	Q_{uu}(k,{\bf{x}};\tau,\tau_1,t) + Q_{bb}(k,{\bf{x}};\tau,\tau_1,t) 
	} \right],
	\label{eq:beta_spec_exp}
\end{equation}
\begin{equation}
	\gamma = \frac{1}{3}
	\int d{\bf{k}} \int_{-\infty}^\tau \!\!\!d\tau_1
	G(k,{\bf{x}};\tau,\tau_1,t) \left[ {
	Q_{ub}(k,{\bf{x}};\tau,\tau_1,t) + Q_{bu}(k,{\bf{x}};\tau,\tau_1,t)
	} \right],
	\label{eq:gamma_spec_exp}
\end{equation}
\begin{equation}
	\mbox{\boldmath$\Gamma$}
	= \frac{1}{15} \int k^{-2} d{\bf{k}} \int_{-\infty}^\tau \!\!\!d\tau_1
	G(k,{\bf{x}};\tau,\tau_1,t) \nabla H_{uu}(k,{\bf{x}};\tau,\tau_1,t)
	\label{eq:Gamma_spec_exp}
\end{equation}
with relations
\begin{equation}
	\nu_{\rm{K}} = \frac{7}{5} \beta,\;\;\; \nu_{\rm{M}} = \frac{7}{5} \gamma
	\label{eq:nu_beta_gamma_rel}
\end{equation}
\citep{yos1990}. Here $G$ is the response function of inhomogeneous turbulence, and $Q_{uu}$, $Q_{bb}$, $Q_{ub}$, $H_{uu}$, and $H_{bb}$ are the spectral functions of the turbulent kinetic energy, magnetic energy, cross helicity, kinetic helicity, and current helicity, respectively. For details of derivation of Eqs.~(\ref{eq:Re_strss_exp}) and (\ref{eq:E_M_exp}) with Eqs.~(\ref{eq:alpha_spec_exp})-(\ref{eq:nu_beta_gamma_rel}), the reader is referred to \citet{yos1990}.

	Equations (\ref{eq:alpha_spec_exp})-(\ref{eq:gamma_spec_exp}) indicate that the transport coefficients $\alpha$, $\beta$, and $\gamma$ can be modeled by the statistical quantities multiplied by the time scale of turbulence as 
\begin{subequations}
\begin{equation}
	\alpha  = C_\alpha \tau H\;\;\;
	\mbox{with}\;\;\; 
	H = \langle { - {\bf{u}}' \cdot \mbox{\boldmath$\omega$}' 
		+ {\bf{b}}' \cdot {\bf{j}}' } \rangle,
	\label{eq:alpha_model}
\end{equation}
\begin{equation}
	\beta = C_\beta \tau K\;\;\;
	\mbox{with}\;\;\;
	K = \langle { {\bf{u}}'{}^2 + {\bf{b}}'{}^2 } \rangle /2,
	\label{eq:beta_model}
\end{equation}
\begin{equation}
	\gamma = C_\gamma \tau W\;\;\;
	\mbox{with}\;\;\;
	W = \langle { {\bf{u}}' \cdot {\bf{b}}' } \rangle,
	\label{eq:gamma_model}
\end{equation}
\end{subequations}
where $\tau$ is the time scale of turbulence. Here, $C_\alpha$, $C_\beta$, and $C_\gamma$ are the model constant. They should be estimated from Eqs.~(35)-(37). There have been some attempts to evaluate them. According to these studies, they are estimated as
\begin{equation}
	C_\alpha = O(10^{-2}),\; C_\beta = O(10^{-1}),\; C_\gamma = O(10^{-1})
	\label{eq:model_consts}
\end{equation}
\citep{ham1992,yos1998}. Further work on estimating these constants using the high-resolution direct numerical simulations (DNS's) of the MHD turbulence is desired.

	As we see in Eqs.~(\ref{eq:beta_spec_exp}) and (\ref{eq:beta_model}), in the TSDIA analysis up to the $O(\delta)$ calculation ($\delta$: scale parameter in the expansion), the turbulent magnetic diffusivity $\beta$ depends both on the turbulent kinetic and magnetic energies. This is in disagreement with the results of the first-order calculation in the traditional mean-field theories such as the first-order smoothing approximation (FOSA), the $\tau$ approximation, etc. In the latter, the first-order calculation shows that $\beta$ depends on $\langle {{\bf{u}}'{}^2} \rangle$ but shows no dependence on $\langle {{\bf{b}}'{}^2} \rangle$ if the mean magnetic field is much smaller than the equipartition field \citep{rae2003,bra2005,rae2007}.

	As far as the TSDIA analysis is concerned, this rise of ``discrepancy'' is connected to the point how the solenoidal condition and magnetic pressure are treated in the TSDIA formalism.

	The Lorentz force in the momentum equation is rewritten as
\begin{equation}
	{\bf{j}} \times {\bf{b}}
	= ({\bf{b}} \cdot \nabla) {\bf{b}}
	- \nabla \left( {\frac{1}{2} {\bf{b}}^2} \right),
	\label{eq:Lorentz_force}
\end{equation}
and the second or magnetic-energy-related part is absorbed into the magnetohydrodynamic (MHD) pressure defined by
\begin{equation}
	p_{\rm{M}} = p + {\bf{b}}^2/2
	\label{eq:mhd_pressure_def}
\end{equation}
($p$: gas pressure). Applying the Reynolds decomposition [Eq.~(\ref{eq:Rey_decomp})], $p_{\rm{M}}$ is divided into the mean MHD pressure $P_{\rm{M}}$ and the fluctuation around it, $p'_{\rm{M}}$. They are expressed as
\begin{subequations}
\begin{equation}
	P_{\rm{M}} = P + \frac{1}{2} {\bf{B}}^2 
	+ \frac{1}{2} \langle {{\bf{b}}'{}^2} \rangle,
	\label{eq:mean_pm}
\end{equation}
\begin{equation}
	p'_{\rm{M}} = p' + {\bf{b}}' \cdot {\bf{B}}
	+\frac{1}{2} {\bf{b}}'^2 
	- \frac{1}{2} \langle {{\bf{b}}'{}^2} \rangle.
	\label{eq:fluct_pm}
\end{equation}
\end{subequations}
We see from Eq.~(\ref{eq:fluct_pm}) that the mean magnetic field ${\bf{B}}$ is included in the fluctuation MHD pressure $p'_{\rm{M}}$. 

	In the incompressible turbulence analysis, the fluctuation pressure ($p'_{\rm{M}}$ in the present case) is eliminated by using the solenoidal condition of the fluctuation velocity. At the same time, in the TSDIA analysis, the solenoidal condition (\ref{eq:FourierSlndl}) is satisfied by the solenoidal fluctuation (\ref{eq:phisDef}). This suggests that some higher-order calculation is needed for treating the solenoidal fluctuation field.

	If we recall the external parameter (${\bf{B}}$, $\mbox{\boldmath$\omega$}_{\rm{F}}$) expansion in the present TSDIA formalism [Eq.~(\ref{eq:ext-prmtr-expnd}) in Appendix], we see the magnetic-energy-related contribution emerges at higher-order calculations. Actually higher-order contributions in the TSDIA were examined without resorting to the Elsasser formulation. It was found that the magnetic fluctuation contribution is canceled by the higher-order contributions \citep{ham2008}. However, it is also probable that, if we proceed to further higher-order calculations, magnetic fluctuation dependence of the turbulent magnetic diffusivity would recover again. In relation to this point, we should note that in some literature the magnetic fluctuation contribution to the turbulent magnetic diffusivity has been reported \citep{rog2001,rog2004,kle2007}. In these papers, the turbulent transport coefficients are analyzed for the arbitrary ratio of the mean magnetic field to the equipartition field. If the mean magnetic field is not so small compared to the equipartition field, the turbulent magnetic diffusivity depends on $\langle {{\bf{b}}'{}^2} \rangle$ as well as on $\langle {{\bf{u}}'{}^2} \rangle$. However, if the mean magnetic magnetic field is much smaller than the equipartition, their results are in agreement with \citet{rae2003,bra2005,rae2007}.

	The dependence of the transport coefficients on the turbulent quantities itself can be derived easily without resorting to any elaborated closure theory for inhomogeneous turbulence. From Eqs.~(\ref{eq:fluct_u_eq}) and (\ref{eq:fluct_b_eq}), we write the equation of the turbulent electromotive force ${\bf{E}}_{\rm{M}} = \langle {{\bf{u}}' \times {\bf{b}}'} \rangle$. Multiplying $b'{}^b$ to the $a$ component of Eq.~(\ref{eq:fluct_u_eq}) and $u'{}^a$ to the $b$ component of Eq.~(\ref{eq:fluct_b_eq}), and adding them, we obtain
\begin{eqnarray}
	b'{}^b \frac{\partial u'{}^a}{\partial t} + u'{}^a \frac{\partial b'{}^b}{\partial t}
	&=& -U^c b'{}^b \frac{\partial u'{}^a}{\partial x^c}
	-U^c u'{}^a \frac{\partial b'{}^b}{\partial x^c}
	+ B^c b'{}^b \frac{\partial b'{}^a}{\partial x^c}
	+ B^c u'{}^a \frac{\partial u'{}^b}{\partial x^c}
	\nonumber\\
	& & + b'{}^b b'{}^c \frac{\partial B^a}{\partial x^c}
	- u'{}^a u'{}^c \frac{\partial B^b}{\partial x^c}
	+ u'{}^c b'{}^c \frac{\partial U^b}{\partial x^c}
	- b'{}^b u'{}^c \frac{\partial U^a}{\partial x^c}
	\nonumber\\
	& & - b'{}^b \frac{\partial p'_{\rm{M}}}{\partial x^a}
	+ b'{}^b \frac{\partial}{\partial x^c} \langle {u'{}^c u'{}^a - b'{}^c b'{}^a} \rangle
	+ u'{}^a \frac{\partial}{\partial x^c} \langle {u'{}^c b'{}^b - b'{}^c u'{}^a} \rangle
	\nonumber\\
	& & - b'{}^b u'{}^c \frac{\partial u'{}^a}{\partial x^c}
	+ b'{}^b b'{}^c \frac{\partial b'{}^a}{\partial x^c}
	- u'{}^c \frac{\partial}{\partial x^c} u'{}^a b'{}^b
	\nonumber\\
	& & + b'{}^c \frac{\partial}{\partial x^c} u'{}^a u'{}^b
	+ \eta u'{}^a \frac{\partial^2 b'{}^b}{\partial x^c \partial x^c}
	+ \nu b'{}^b \frac{\partial^2 u'{}^a}{\partial x^c \partial x^c}
	+ b'{}^b f'{}^a.
	\label{eq:emf_eq_1}
\end{eqnarray}
We multiply Eq.~(\ref{eq:emf_eq_1}) by the alternating tensor $\epsilon^{\alpha ab}$ and take the ensemble average $\langle \cdots \rangle$ of each term. We have
\begin{eqnarray}
	\left\langle{
	\frac{\partial}{\partial t} \left( {
		{\bf{u}}' \times {\bf{b}}'
	} \right)
} \right\rangle^\alpha
&=& \left\langle {
	\epsilon^{\alpha ab} \left( {
		b'{}^b \frac{\partial u'{}^a}{\partial t}
		+ u'{}^a \frac{\partial b'{}^b}{\partial t}
	} \right)
} \right\rangle
\nonumber\\
&=& - U^c \frac{\partial}{\partial x^c} \left\langle {
		\epsilon^{\alpha ab} u'{}^a b'{}^b
	} \right\rangle
\nonumber\\
& & + \frac{1}{3} \left\langle {
	b'{}^b \epsilon^{bca} 
		\frac{\partial b'{}^a}{\partial x^c}
	- u'{}^b \epsilon^{bca} 
		\frac{\partial u'{}^a}{\partial x^c}
} \right\rangle B^\alpha
- \frac{1}{3} \left\langle{ 
	u'{}^b u'{}^b + b'{}^b b'{}^b
} \right\rangle \epsilon^{\alpha ca} \frac{\partial B^a}{\partial x^c}
\nonumber\\
& & + \frac{2}{3} \left\langle {
	u'{}^b b'{}^b
} \right\rangle \epsilon^{\alpha ba} \frac{\partial U^a}{\partial x^b}
+ {\rm{R.T.}},
	\label{eq:emf_eq_2}
\end{eqnarray}
where ${\rm{R.T.}}$ stands for the higher-order terms. Here, use has been made of an approximation that the fluctuating field is statistically homogeneous and isotropic:
\begin{equation}
	\left\langle {f'_1{}^\alpha f'_2{}^\beta}\right\rangle
	+ \left\langle {f'_2{}^\alpha f'_1{}^\beta}\right\rangle
	= \frac{2}{3} \delta^{\alpha\beta} 
		\left\langle {f'_1{}^a f'_2{}^a}\right\rangle.
	\label{eq:isotropic_approx}
\end{equation}
The residual or higher-order terms $\rm{R.T.}$ include a term arising from the fluctuating MHD pressure $p'_{\rm{M}}$, $- \epsilon^{\alpha ab} \langle {b'{}^b (\partial p'_{\rm{M}} / \partial x^a)} \rangle$. As Eq.~(\ref{eq:fluct_pm}) shows, $p'_{\rm{M}}$ depends on the mean magnetic field ${\bf{B}}$. If we write the ${\bf{B}}$-related part of $p'_{\rm{M}}$ as $p'_{\rm{MB}} = {\bf{b}}' \cdot {\bf{B}}$, its contribution is written as
\begin{eqnarray}
	- \epsilon^{\alpha ab} \left\langle {
		b'{}^b \frac{\partial p'_{\rm{MB}}}{\partial x^a}
	} \right\rangle
	&=& - \epsilon^{\alpha ab} \left\langle {
		b'{}^b \frac{\partial}{\partial x^a} b'{}^c B^c
	} \right\rangle\nonumber\\
	&=& - \epsilon^{\alpha ab} \left\langle {
		b'{}^b b'{}^c
	} \right\rangle \frac{\partial B^c}{\partial x^a}
	- \epsilon^{\alpha ab} \left\langle {
		b'{}^b \frac{\partial b'{}^c}{\partial x^a}
	} \right\rangle B^c
	\nonumber\\
	&=& - \frac{1}{3} \left\langle {
		b'{}^d b'{}^d
	} \right\rangle \epsilon^{\alpha ab} \frac{\partial B^b}{\partial x^a}
	- \epsilon^{\alpha ab} \left\langle {
		b'{}^b \frac{\partial b'{}^c}{\partial x^a}
	} \right\rangle B^c.
	\label{eq:p_M_contribution}
\end{eqnarray}
Here use has been made of Eq.~(\ref{eq:isotropic_approx}). The first term in Eq.~(\ref{eq:p_M_contribution}) gives a contribution to the mean electric-current term while the second term gives a contribution to the magnetic pumping term.

	We should note that the fluctuation fields in general are neither statistically homogeneous nor isotropic. In this sense, we have to treat the fluctuation equations in more elaborative manners as the statistical closure theory for inhomogeneous turbulence which leads to Eqs.~(\ref{eq:alpha_spec_exp})-(\ref{eq:nu_beta_gamma_rel}).

	The treatment leading to Eq.~(\ref{eq:emf_eq_2}) given above is very primitive. It does not pay any special attention to the closure of the correlation moments. In this sense, this should be considered as an expedient to get a broad grasp of physics relevant to the electromotive force. The simple $\tau$ approach described by \citet{rae2007} is a much more elaborated method to understand the dependence of turbulent electromotive force on the mean magnetic field and velocity. The point here is that even the simplest possible approach like Eq.~(\ref{eq:emf_eq_1}) provides some insight into the mean velocity-related term if we retain the mean velocity in the fundamental equations.

\subsection{Physical interpretation of each effect\label{sec:phys_interprt}}
	In what follows, we shall examine the effects of each term that is directly linked to ${\bf{U}}$ and ${\bf{B}}$ in Eqs.~(\ref{eq:fluct_u_eq})-(\ref{eq:fluct_b_eq}). Such arguments should be employed with caution. Since each term reflects only a portion of the effects of ${\bf{U}}$, ${\bf{B}}$, etc., some effects may be canceled by other terms as will be suggested concerning to the magnetic fluctuation effect on the turbulent magnetic diffusivity (\S\ref{sec:betaK}). However, as long as we bear this point in mind, it serves a useful way for abstracting the physical origins of the field-destruction and -generation mechanisms due to turbulence.

\subsubsection{Electromotive force due to turbulent motion; $\beta$-related term\label{sec:betaK}}

\begin{figure}[htpb]
\begin{center}
\includegraphics[width=.35\textwidth]{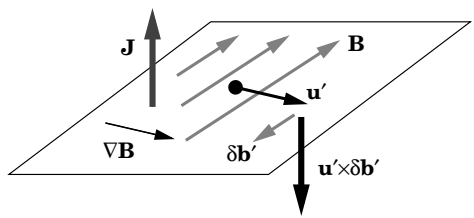}%
\caption{Turbulent kinetic energy effects.}%
\label{fig:kin_beta_effect}
\end{center}
\end{figure}

\paragraph{Velocity-fluctuation effect}
	We start with the field-destruction mechanism due to the turbulent motion. Let us consider a fluid element moving in a shearing mean magnetic field ${\bf{B}}$. From equation~(\ref{eq:fluct_b_eq}), the magnetic-field variation due to the mean magnetic shear, $\delta {\bf{b}}'$, is written as
\begin{equation}
	\delta {\bf{b}}'
	= - \tau_{\beta{\rm{K}}} ({\bf{u}}' \cdot \nabla){\bf{B}}
	\label{eq:fluct_b_dueto_GradB}
\end{equation}
($\tau_{\beta{\rm{K}}}$ is the time scale of the fluctuation). Equation~(\ref{eq:fluct_b_dueto_GradB}) shows that, if the fluid element fluctuates and moves in the mean magnetic field shear, the magnetic-field variation is induced in the direction of the mean magnetic field. The magnetic-field variation due to the fluctuating motion ${\bf{u}}'$, $\delta {{\bf{B}}}'$, is induced so that the induced variation may relax the original shear of the mean magnetic field. In case the element moves in the direction that ${\bf{B}}$ increases, $\delta {\bf{b}}$ is induced in the direction antiparallel to ${\bf{B}}$ (Figure~\ref{fig:kin_beta_effect}), resulting in the relaxation of the original gradient of ${\bf{B}}$. This  manifests itself the ``diamagnetic" nature of plasmas and is in common with the turbulent transport. For example, in the case of the turbulent diffusion of a passive scalar, the transport due to turbulence occurs in the direction opposite to the mean scalar gradient. The contribution to the turbulent electromotive force, ${\bf{u}}' \times \delta{\bf{b}}'$, resulting from the coupling of the mean magnetic shear and the turbulent motion, is in the direction antiparallel to the mean electric-current density ${\bf{J}}$ (Figure~\ref{fig:kin_beta_effect}). Note that the direction of ${\bf{u}}' \times \delta{\bf{b}}'$ is always antiparallel to ${\bf{J}}$ regardless of the fluctuation direction relative to the gradient ${\bf{B}}$. This point is consistent with the expression for the ${\bf{J}}$-proportional term in ${\bf{E}}_{\rm{M}}$ [Eq.~(\ref{eq:E_M_exp})]. Then the electromotive force due to the velocity fluctuation may be written as
\begin{equation}
	\langle {{\bf{u}}'\times{\bf{b}}'} \rangle_{\beta{\rm{K}}}
	= - \beta_{\rm{K}} {\bf{J}},
	\label{eq:E_M_beta_K}
\end{equation}
with the positive coefficient $\beta_{\rm{K}}$ whose magnitude is determined by the intensity of the velocity fluctuation.

\begin{figure}[htpb]
\begin{center}
\includegraphics[width=.35\textwidth]{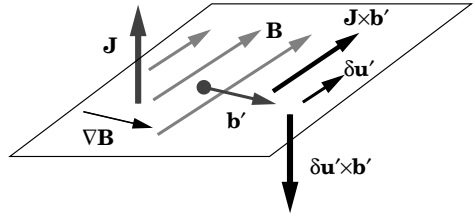}%
\caption{Turbulent magnetic energy effects. This figure should be viewed with caution. See also Figure~\ref{fig:mag_en_effect_2}.}%
\label{fig:mag_beta_effect}
\end{center}
\end{figure}

\paragraph{Magnetic-fluctuation effect}
	Next, we proceed to the effect of the magnetic-field fluctuation ${\bf{b}}'$ on the electromotive force. As is similar to the velocity-fluctuation case, we consider the shearing magnetic field or the mean electric current (Figure~\ref{fig:mag_beta_effect}). The same line of argument holds in the magnetic-field fluctuation case. However, physical origin of the magnetic-fluctuation effect may become much clearer if we consider the mean electric-current density ${\bf{J}}$  instead of ${\bf{B}}$ itself. In the presence of ${\bf{J}}$, the velocity variation due to the magnetic
fluctuation ${\bf{b}}'$, $\delta{\bf{u}}'$, is subject to the fluctuating Lorentz force as
\begin{equation}
	\delta {\bf{u}}'
	= \tau_{\beta{\rm{M}}}{\bf{J}}\times{\bf{b}}',
	\label{eq:fluct_u_dueto_J}
\end{equation}
where $\tau_{\beta{\rm{M}}}$ is the time scale of the motion. From equation (\ref{eq:fluct_u_dueto_J}), the contribution to the electromotive force is given as
\begin{equation}
	\delta {\bf{u}}'\times{\bf{b}}'
	= \tau_{\beta{\rm{M}}}({\bf{J}}\times{\bf{b}}') \times {\bf{b}}'
	= - \tau_{\beta{\rm{M}}} {\bf{b}}'^2 {\bf{J}}.
	\label{eq:delta_u_b_beta_M}
\end{equation}
Equation (\ref{eq:delta_u_b_beta_M}) shows that the contribution is in the direction antiparallel to ${\bf{J}}$  (Figure~\ref{fig:mag_beta_effect}). Note that since ${\bf{b}}'$ longitudinal to ${\bf{J}}$  makes no contribution to ${\bf{J}}\times{\bf{b}}'$, we consider only ${\bf{b}}'$ that is transverse to ${\bf{J}}$  in Figure~\ref{fig:mag_beta_effect}. Clearly, this argument has no dependence on the direction of ${\bf{b}}'$ to the magnetic-field gradient associated with ${\bf{J}}$ . Then, the electromotive force due to the magnetic fluctuation may be written as
\begin{equation}
	\langle {\bf{u}}'\times{\bf{b}}'\rangle_{\beta{\rm{M}}}
	= - \beta_{\rm{M}} {\bf{J}},
	\label{eq:E_M_beta_M}
\end{equation}
with $\beta_{\rm{M}}$ being the positive coefficient whose magnitude is determined by the magnitude of magnetic-field fluctuation.

\begin{figure}[htpb]
\begin{center}
\includegraphics[width=.35\textwidth]{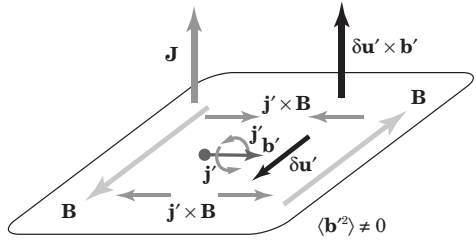}%
\caption{Higher-order turbulent magnetic energy effects.}%
\label{fig:mag_en_effect_2}
\end{center}
\end{figure}

\paragraph{Higher-order magnetic-fluctuation effect}
	As we have just derived above, the magnetic-field fluctuation contributes to the turbulent electromotive force antiparallel to the mean electric-current density [Eq.~(\ref{eq:E_M_beta_M})]. The higher-order contribution of the magnetic-field fluctuation may reduce this magnetic fluctuation contribution \citep{ham2008}. Let us consider the mean electric-current density ${\bf{J}}$, which corresponds to the sheared mean magnetic-field configuration as in Figure~\ref{fig:mag_en_effect_2}. We consider the magnetic-field fluctuation ${\bf{b}}'$, whose direction is the same as in Figure~\ref{fig:mag_beta_effect}. We have the electric-current fluctuation ${\bf{j}}'$ associated with the magnetic fluctuation ${\bf{b}}'$ as in Figure~\ref{fig:mag_en_effect_2}. The fluctuating Lorentz force ${\bf{j}}' \times {\bf{B}}$ due to the fluctuation electric current density ${\bf{j}}'$ is exerted in the diverging and converging directions on the near and far sides, respectively. Hence, the magnetic pressure on the near side becomes lower than the pressure on the far side. Because of this magnetic pressure gradient, velocity is induced in the direction from far to near sides as 
\begin{equation}
	\delta {\bf{u}}' = - \tau_{\beta{\rm{M2}}} \nabla p'_{\rm{M}}.
\end{equation}
	As this result, we have a contribution to the turbulent electromotive force parallel to the mean electric-current density:
\begin{equation}
	\left\langle {\delta {\bf{u}}' \times {\bf{b}}'} \right\rangle_{\beta{\rm{M2}}}
	= + \beta_{\beta\rm{M2}} {\bf{J}}.
	\label{eq:E_M_beta_M2}
\end{equation}

	For the same magnetic fluctuation, the contribution from Eq.~(\ref{eq:E_M_beta_M2}) is in the opposite direction to the one from Eq.~(\ref{eq:E_M_beta_M}). This implies that the magnetic fluctuation effect in the turbulent magnetic diffusivity may be suppressed due to the magnetic pressure, in particular in the case of incompressible MHD flow. This result is consistent with the traditional mean-field theories \citep{rae2003,bra2005,rae2007,ham2008}.

\subsubsection{Electromotive force due to turbulent helicity; $\alpha$-related term}
\begin{figure}[htpb]
\begin{center}
\includegraphics[width=.35\textwidth]{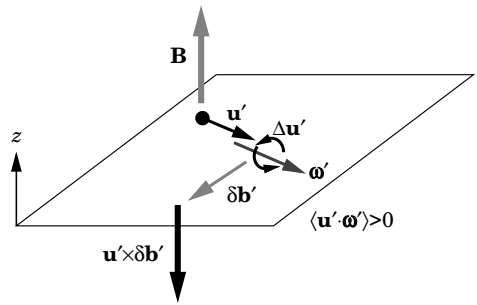}%
\caption{Turbulent kinetic helicity effects.}%
\label{fig:kin_alpha_effect}
\end{center}
\end{figure}

\paragraph{Kinetic-helicity effect\label{sec:kin_hel_effect}}
	We consider a fluid element placed in the uniform magnetic field ${\bf{B}}$ whose direction is coincident with the $z$ axis (Figure~\ref{fig:kin_alpha_effect}). The direct effects of the mean magnetic field ${\bf{B}}$ enter through the first term in equation (\ref{eq:fluct_b_rot_eq}) or $\nabla\times({\bf{u}}'\times{\bf{B}})$. It is useful to divide ${\bf{u}}'$ into two parts; the velocity fluctuation parallel to ${\bf{B}}$, ${\bf{u}}'_{||}$, and the one perpendicular to ${\bf{B}}$, ${\bf{u}}'_\bot$. Since ${\bf{u}}'_{||} \times {\bf{B}}=0$, we see from equation (\ref{eq:fluct_b_rot_eq}) that in the context of the direct effects of ${\bf{B}}$, only ${\bf{u}}'_\bot$ is relevant to the evolution of ${\bf{b}}'$. From equation (\ref{eq:fluct_b_eq}), the magnetic-field fluctuation due to the mean magnetic field is given as
\begin{equation}
	\delta {\bf{b}}'
	= \tau_{\alpha \rm{K}}({\bf{B}}\cdot\nabla){\bf{u}}'
	\label{eq:FluctbduetoB}
\end{equation}
with $\tau_{\alpha \rm{K}}$ being the time scale of the motion. Equation~(\ref{eq:FluctbduetoB}) can be rewritten as
\begin{equation}
	\delta {\bf{b}}'
	= \tau_{\alpha \rm{K}}\left| {\bf{B}} \right|
		\frac{\partial {\bf{u}}'}{\partial z}
	= \tau_{\alpha \rm{K}}\left| {\bf{B}} \right|
		\frac{\Delta {\bf{u}}'}{\Delta z},
	\label{eq:FluctbduetoBDif}
\end{equation}
where $\Delta {\bf{u}}'$ is the variation of ${\bf{u}}'$ with the displacement $\Delta z$ along ${\bf{B}}$. Equation~(\ref{eq:FluctbduetoBDif}) states that under the concept of magnetic-flux freezing, the magnetic-field line slightly bends as ${\bf{u}}'$ changes with $z$, which leads to the transverse magnetic field $\delta {\bf{b}}'$ proportional to $\Delta{\bf{u}}'/\Delta z$. The changes of ${\bf{u}}'$, $\Delta{\bf{u}}'$, in magnitude and in direction are determined by the topological properties of the turbulent field. The one in magnitude does not contribute to the turbulent electromotive force since, in this case, $\Delta{\bf{u}}'(\propto \delta{\bf{b}}')$  is parallel to the original ${\bf{u}}'$, leading to ${\bf{u}}'\times\delta{\bf{b}}'=0$. Then we shall consider the change in direction. In this context, we should recall that the turbulent kinetic helicity $\langle{\bf{u}}'\cdot\mbox{\boldmath$\omega'$}\rangle(=\langle{\bf{u}}'\cdot\nabla\times{\bf{u}}'\rangle)$ represents the helical property of the turbulent velocity field. In the presence of positive $\langle{\bf{u}}'\cdot\mbox{\boldmath$\omega'$}\rangle$, the fluctuation vorticity \mbox{\boldmath$\omega'$} is parallel to ${\bf{u}}'$ in a statistical sense. In other words, the velocity variation associated with \mbox{\boldmath$\omega'$}, $\Delta{\bf{u}}'$, tends to head for the right-skew direction to ${\bf{u}}'$ as is seen in Figure~\ref{fig:kin_alpha_effect}. It follows from equation (\ref{eq:FluctbduetoBDif}) that the magnetic field variation $\delta{\bf{b}}'$ is in the direction parallel to $\Delta{\bf{u}}'$. As a result, the contribution to the turbulent electromotive force, $\langle{\bf{u}}'\times\delta {\bf{b}}'\rangle$, becomes antiparallel to ${\bf{B}}$. On the other hand, the contribution is parallel to ${\bf{B}}$ in the case of negative $\langle{\bf{u}}'\cdot\mbox{\boldmath$\omega'$}\rangle$. Then the electromotive force due to the kinetic helicity may be expressed as
\begin{equation}
		\langle {\bf{u}}' \times {\bf{b}}' \rangle_{\alpha \rm{K}}
		= \alpha_{\rm{K}} {\bf{B}},
\end{equation}
where $\alpha_{\rm{K}}$ is the kinetic-helicity-related coefficient whose sign is equal to that of $-\langle{\bf{u}}'\cdot\mbox{\boldmath$\omega'$}\rangle$. We see from the above argument that the kinetic-helicity effect is originated from the emergence of the magnetic-fluctuation variation that is not aligned with the velocity fluctuation. Such a magnetic fluctuation is induced by the helical property of the turbulent velocity field through the intermediary of the magnetic-flux freezing.

\begin{figure}[htpb]
\begin{center}
\includegraphics[width=.35\textwidth]{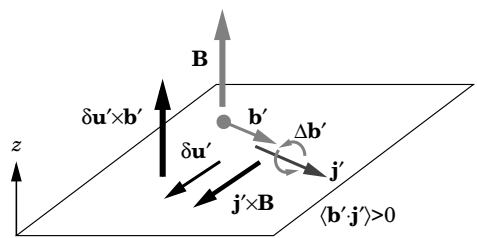}%
\caption{Turbulent current helicity effects.}%
\label{fig:mag_alpha_effect}
\end{center}
\end{figure}

\paragraph{Current-helicity effect}
	Next, we proceed to the electromotive force due to the turbulent current helicity $\langle{\bf{b}}'\cdot{\bf{j}}'\rangle(=\langle{\bf{b}}'\cdot\nabla\times{\bf{b}}'\rangle)$. For simplicity of discussion, we consider the magnetic-field fluctuation ${\bf{b}}'$ normal to the mean magnetic field ${\bf{B}}$ (Figure~\ref{fig:mag_alpha_effect}). Starting from equation (\ref{eq:fluct_u_eq}) with the argument similar to the kinetic-helicity case presented in \S\ref{sec:kin_hel_effect}, we can derive the velocity variation due to the mean magnetic field ${\bf{B}}$, $\delta{\bf{u}}'$. As to the current-helicity effect, however, the understanding would be more easily facilitated if we consider the effect of the fluctuating Lorentz force. From equation (\ref{eq:fluct_u_rot_eq}), the velocity variation due to ${\bf{B}}$, $\delta{\bf{u}}'$, is given by
\begin{equation}
	\delta {\bf{u}}' = \tau_{\alpha {\rm{M}}}{\bf{j}}'\times {\bf{B}},
\end{equation}
with $\tau_{\alpha {\rm{M}}}$ being the time scale of the motion. This variation results from the tension of the magnetic field, and is closely connected to the topological properties of the turbulent magnetic field. In contrast to the kinetic helicity $\langle{\bf{u}}'\cdot\mbox{\boldmath$\omega'$}\rangle$, the turbulent current helicity $\langle{\bf{b}}'\cdot{\bf{j}}'\rangle$ represents the helical property of the turbulent magnetic field. The positive $\langle{\bf{b}}'\cdot{\bf{j}}'\rangle$ means that the electric-current fluctuation $\bf j'$ is statistically parallel to ${\bf{b}}'$ as shown in Figure~\ref{fig:mag_alpha_effect}. As a result, the contribution to the electromotive force, $\langle\delta{\bf{u}}'\times{\bf{b}}'\rangle$, is parallel to ${\bf{B}}$ if $\langle{\bf{b}}'\cdot{\bf{j}}'\rangle>0$. On the other hand, the contribution is antiparallel to ${\bf{B}}$ in the case of negative $\langle{\bf{b}}'\cdot{\bf{j}}'\rangle$. Then the electromotive force due to the current helicity may be expressed as
\begin{equation}
	\langle{{\bf{u}}'\times{\bf{b}}'}\rangle_{\alpha {\rm{M}}}
	=\alpha_{\rm{M}} {\bf{B}},
\end{equation}
with $\alpha_{\rm{M}}$ being the current-helicity-related coefficient whose sign is equal to that of $\langle{\bf{b}}'\cdot{\bf{j}}'\rangle$.

\subsubsection{Electromotive force due to turbulent cross helicity; $\gamma$-related term\label{sec:gamma_term}}
\begin{figure}[htpb]
\begin{center}
\includegraphics[width=.35\textwidth]{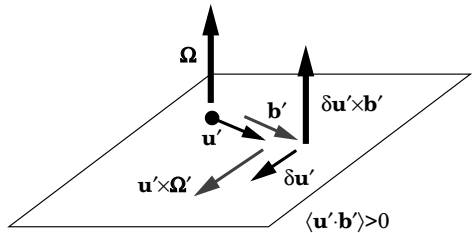}%
\caption{Turbulent cross-helicity effects.}%
\label{fig:gamma_effect}
\end{center}
\end{figure}

\paragraph{Cross-helicity effect}
	Thus far, we have treated the cases with the mean magnetic field ${\bf{B}}$; namely, the current density $\bf J$, as the curl of ${\bf{B}}$, and ${\bf{B}}$ itself. Here we shall treat a case with the mean velocity ${\bf{U}}$, and consider a fluid element fluctuating in the mean vorticity field ${\mbox{\boldmath$\Omega$}}(=\nabla\times{{\bf{U}}})$ (Figure~\ref{fig:gamma_effect}). If the element moves (${\bf{u}}'$) in the plane perpendicular to $\mbox{\boldmath$\Omega$}$, the force ${\bf{u}}'\times{\mbox{\boldmath$\Omega$}}$ acts on it because of the local angular-momentum conservation. Then the element is accelerated in the direction perpendicular both to ${\bf{u}}'$ and to $\mbox{\boldmath$\Omega$}$ as
\begin{equation}
	\delta{\bf{u}}' = \tau_\gamma {\bf{u}}'\times{\mbox{\boldmath$\Omega$}},
\end{equation}
with $\tau_\gamma$ being the time scale of the motion. In this context, we should recall that the turbulent cross helicity $\langle{\bf{u}}'\cdot{\bf{b}}'\rangle$ characterizes the cross correlation between ${\bf{u}}'$ and ${\bf{b}}'$. Non-vanishing turbulent cross helicity indicates that ${\bf{u}}'$ and ${\bf{b}}'$ are statistically aligned with each other. In the case of $\langle{\bf{u}}'\cdot{\bf{b}}'\rangle>0$, ${\bf{b}}'$ is parallel to ${\bf{u}}'$ in a statistical sense, while ${\bf{b}}'$ is antiparallel if $\langle{\bf{u}}'\cdot{\bf{b}}'\rangle<0$. As a result, the contribution to the electromotive force, $\langle\delta{\bf{u}}'\times{\bf{b}}'\rangle$, is parallel to $\mbox{\boldmath$\Omega$}$ in case $\langle{\bf{u}}'\cdot{\bf{b}}'\rangle>0$ and antiparallel in case $\langle{\bf{u}}'\cdot{\bf{b}}'\rangle<0$. Then the electromotive force due to the cross helicity may be expressed as
\begin{equation}
	\langle{{\bf{u}}'\times{\bf{b}}'}\rangle_\gamma 
	= \gamma{\mbox{\boldmath$\Omega$}},
\end{equation}
where $\gamma$ is the cross-helicity-related coefficient whose sign is equal to that of $\langle{\bf{u}}'\cdot{\bf{b}}'\rangle$. We see from the above consideration that the key ingredients for the cross-helicity effect are the local angular-momentum conservation in the mean vorticity field and the cross correlation between the turbulent velocity and magnetic fields.

\section{Cross-helicity effects\label{sec:ch_effect}}
	In the history of turbulent dynamo study, the cross-helicity-related term has been missing. Dropping the $\gamma$ or cross-helicity-related term in  Eq.~(\ref{eq:E_M_exp}), we have the usual $\alpha$ dynamo, where the main balancer against the $\beta$ or turbulent magnetic diffusivity effect is the $\alpha$ or helicity effect. We can consider the other limit: If we drop the $\alpha$ or helicity term in Eq.~(\ref{eq:E_M_exp}), the main balancer against $\beta$ is the $\gamma$ or cross-helicity effect. The latter situation may be called as the cross-helicity dynamo, in contrast to the former situation is called as the $\alpha$ dynamo:
\begin{equation}
	\overunderbraces{&\br{2}{\alpha\ \mbox{dynamo}}}%
	{{\bf{E}}_{\rm{M}} 
	 = &\alpha {\bf{B}} &- \beta {\bf{J}}&+&\gamma \mbox{\boldmath$\Omega$}}%
	 {& &\br{3}{\mbox{cross-helicity dynamo}}}.
 	\label{eq:E_M_full}
 \end{equation}
Without the $\gamma$ or cross-helicity effect, we have the usual $\alpha$ or helicity dynamo:
\begin{equation}
	{\bf{E}}_{\rm{M}} 
	= \alpha {\bf{B}} - \beta {\bf{J}}.
	\label{eq:E_M_alpha_dynamo}
\end{equation}
In this case, the main balancer against the turbulent magnetic diffusivity $\beta$ is the helicity or $\alpha$ effect. This main balance suggests the mean-field configuration with the alignment of the mean electric-current density ${\bf{J}}$ with the mean magnetic field ${\bf{B}}$. Consequently, the generated mean-field configuration gives a force-free state (${\bf{J}} \times {\bf{B}}$). This is one of the most prominent features of the $\alpha$ dynamo.

\begin{figure}[htpb]
\begin{center}
\includegraphics[width=.30\textwidth]{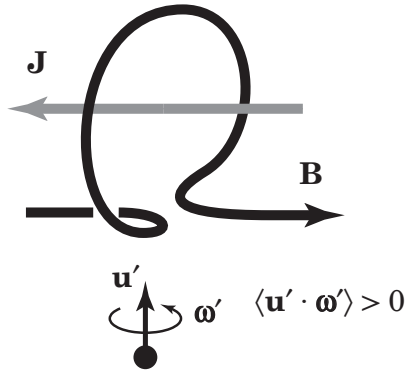}%
\caption{$\alpha$ dynamo.}%
\label{fig:alpha_dynamo}
\end{center}
\end{figure}

	A physical interpretation of the $\alpha$ or helicity dynamo is often presented with Figure~\ref{fig:alpha_dynamo}. If turbulence possesses a helical property, a configuration of the mean electric-current density ${\bf{J}}$ parallel or antiparallel to the mean magnetic field ${\bf{B}}$ can be generated. We should note that in the original idea of this figure, neither ${\bf{B}}$ nor ${\bf{J}}$ represent the mean fields. There supposed the instantaneous magnetic and current density fields, ${\bf{b}}$ and $\bf{J}$, in our notations. 

	At first glance, the physical picture of the helicity dynamo may be clear. In the description of the former dynamo, use has been often made of the concept of the bend-and-twist mechanism, where the magnetic-field loop originated from the bending of the magnetic-field line turns round under the helical nature of the turbulent field. The resultant effect is characterized by the electric-current configuration parallel or antiparallel to the original magnetic field \citep{kra1980,rob1993}. Concerning this description, we should remark upon the following points. In the picture, the diffusion or reconnection as well as the bend and twist of the magnetic field is indispensable for the dynamo process. The electric-current configuration aligned with the original magnetic field can not be attained to from an arbitrary magnetic diffusion. In other words, in order for the alignment configuration  to be realized, the twisting process due to the helicity should be delicately balanced with the diffusion process due to the turbulence. These points indicate that the bend-and-twist picture includes both the helicity effect (explicitly) and the turbulent-diffusion effect (implicitly) as the key ingredients. In handling the bend-and-twist picture, we should keep the above reservations in mind.

	If we substitute the turbulent electromotive force expression [Eq.~(\ref{eq:E_M_exp})] with the $\alpha$ effect dropped:
\begin{equation}
	{\bf{E}}_{\rm{M}} = - \beta {\bf{J}} + \gamma \mbox{\boldmath$\Omega$}
	\label{eq:Em_exp_beta_gamma}
\end{equation}
into the mean induction equation (\ref{eq:mean_B_eq}), we have
\begin{equation}
	\frac{\partial {\bf{B}}}{\partial t}
	= \nabla \times \left( {
		{\bf{U}} \times {\bf{B}} 
		- \beta {\bf{J}} 
		+ \gamma \mbox{\boldmath$\Omega$}
	} \right).
	\label{eq:mean_ind_eq_beta_gamma}
\end{equation}
Here we have neglected the molecular magnetic diffusivity $\eta$ since the turbulent magnetic diffusivity $\beta$ is much larger than $\eta$.

	For stationary state, Eq.~(\ref{eq:mean_ind_eq_beta_gamma}) has particular solutions
\begin{equation}
	{\bf{B}} = \frac{\gamma}{\beta} {\bf{U}} 
	= C_W \frac{W}{K} {\bf{U}},
	\label{eq:special_B_sol}
\end{equation}
\begin{equation}
	{\bf{J}} = \frac{\gamma}{\beta} \mbox{\boldmath$\Omega$} 
	= C_W \frac{W}{K} \mbox{\boldmath$\Omega$} 
	\label{eq:special_J_sol}
\end{equation}
with $C_W$ being a model constant\citep{yos1993b}.

	Equation~(\ref{eq:special_B_sol}) indicates that in the presence of cross helicity in turbulence, we have a magnetic field proportional to the mean velocity. The proportional coefficient $\gamma / \beta$ is expressed by the turbulent cross helicity scaled by the turbulent MHD energy. Since turbulent cross helicity is a pseudoscalar, it may have positive or negative value. If we have positive (or negative) cross helicity in turbulence, we have the mean magnetic field parallel (or antiparallel) to the mean velocity. 

\begin{figure}[htpb]
\begin{center}
\includegraphics[width=.35\textwidth]{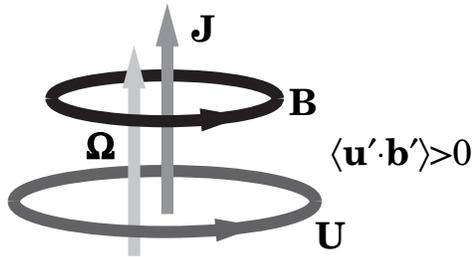}%
\caption{Cross-helicity dynamo.}%
\label{fig:cross-helicity_effect}
\end{center}
\end{figure}

	As Eq.~(\ref{eq:Em_exp_beta_gamma}) indicates, in the turbulent electromotive force, the turbulent magnetic diffusivity term $\beta {\bf{J}}$ is mainly balanced by the cross-helicity effect $\gamma \mbox{\boldmath$\Omega$}$. A schematic figure for the cross-helicity dynamo is given as Figure~\ref{fig:cross-helicity_effect}. In the presence of positive turbulent cross helicity $\langle {{\bf{u}}' \cdot {\bf{b}}'} \rangle > 0$, we have the mean electric-current ${\bf{J}}$ configuration parallel to the mean vorticity $\mbox{\boldmath$\Omega$}$. At the same time we have the mean magnetic-field ${\bf{B}}$ configuration parallel to the mean velocity ${\bf{U}}$ (${\bf{U}} \cdot {\bf{B}} > 0$). With normal cascade of the turbulent cross helicity, this sign is consistent with the positive turbulent cross helicity.

\begin{figure}[htpb]
\begin{center}
\includegraphics[width=.50\textwidth]{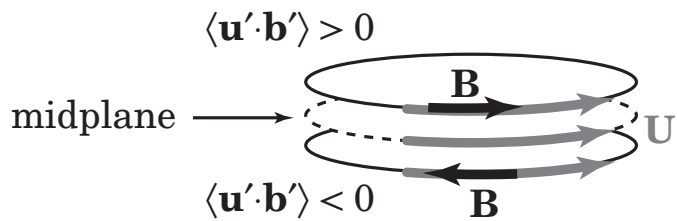}%
\caption{Antisymmetric magnetic configuration in the cross-helicity dynamo. Case with the positive turbulent cross helicity $W= \langle {{\bf{u}}' \cdot {\bf{b}}'} \rangle > 0$ in the upper half domain and the negative turbulent cross helicity in the lower half domain is presented here.}%
\label{fig:antisymmetry_c-h_dynamo}
\end{center}
\end{figure}

	Due to the pseudoscalar nature, the turbulent cross helicity is likely to be distributed antisymmetrically in space. For instance, we can consider a situation where the turbulent cross helicity is positive and negative in the respective regions upper and lower to the midplane (Figure~\ref{fig:antisymmetry_c-h_dynamo}). In this case, from Eq.~(\ref{eq:special_B_sol}), we have magnetic field parallel and antiparallel to the mean velocity in the upper and lower regions, respectively. With a mean velocity distribution symmetric with respect to the midplane, we have antisymmetric magnetic field. How antisymmetric distribution of the turbulent cross helicity is generated is discussed later in \S\ref{sec:ch_generation}, where we examine the production mechanisms of the turbulent cross helicity.

	Finally in this section, we should note the non-conservation of the cross helicity. In the picture of closure theory of turbulence at very high Reynolds number, the fully nonlinear mode coupling, typically represented by the response-function equation [Eq.~(\ref{eq:GreenFnDef})], is the subject of main interests. It appears that the molecular viscosity and diffusivity play only a subsidiary role. However, it is not the case. The turbulent cross helicity, as well as the turbulent energy, cascades from larger to smaller scales, and is dissipated at the smallest scales of turbulence. The dissipation due to the molecular viscosity and diffusivity and the cascade associated with such dissipation are essential ingredients of this turbulence cascade. In this sense, the non-conservation of the cross helicity in the real-world turbulence is a matter of course.

	The cascade property of a quantity $G$ is usually represented by the transfer rate of the quantity, $\Pi_G$, which is equivalent to the dissipation rate of $G$, $\varepsilon_G$, in the cascade picture of Richardson or Kolmogorov:
\begin{equation}
	\Pi_G = \varepsilon_G
\end{equation}
with $G=(K,W)$.

	The viscosity and diffusivity effects appear only in the definitions of the dissipation rates [Eqs.~(\ref{eq:eps_K_def}) and (\ref{eq:eps_W_def})]  and in some viscosity transport terms. Instead of considering the evolution of the dissipation rates, we consider the transfer rates and construct the equations for them. No viscosity effects explicitly appear in the model on the basis of the assumption that the turbulent Reynolds numbers are high enough everywhere. In hydrodynamic turbulence, however, in the immediate vicinity of the wall, the viscosity or diffusion effect should be taken into account irrespective of how high the bulk Reynolds number may be. This issue is often called the ``low-Reynolds number correction'' in turbulence modeling \citep{dur2011,han2011}. In wall boundary-layer turbulence, we have several important and established laws that the mean velocity and the turbulent correlations should obey. They include the logarithmic wall law of the mean velocity, the asymptotic behaviour of each components of the Reynolds stresses as the distance from the wall approaching to zero. In contrast, in the astrophysical applications we often have no wall boundaries. Then we have no definite laws of the mean magnetic field or the Reynolds (and turbulent Maxwell) stresses.

	Equations~(\ref{eq:special_B_sol}) and (\ref{eq:special_J_sol}) [and also Eqs.~(\ref{eq:delta_U_sol}) and (\ref{eq:delta_U_sol_phys_unit}) in \S~7.3] are obtained with the assumption that the turbulent transports are much larger than the molecular counterparts ($\beta \gg \eta$). If we retain the molecular magnetic diffusivity- or $\eta$-related term in Eq.~(\ref{eq:mean_ind_eq_beta_gamma}), the counterparts of Eqs.~(\ref{eq:special_B_sol}) and (\ref{eq:special_J_sol}) are written as
\begin{equation}
	{\bf{B}} = \frac{\gamma}{\beta + \eta} {\bf{U}}
	= \left( {1- \frac{1}{Rm^{(\rm{T})}}} \right) \frac{\gamma}{\beta} {\bf{U}}
	= \left( {1- \frac{1}{Rm^{(\rm{T})}}} \right) C_W \frac{W}{K} {\bf{U}},
	\label{low_Rm_B_sol}
\end{equation}
\begin{equation}
	{\bf{J}} = \frac{\gamma}{\beta + \eta} \mbox{\boldmath$\Omega$}
	= \left( {1- \frac{1}{Rm^{(\rm{T})}}} \right) \frac{\gamma}{\beta} 
		\mbox{\boldmath$\Omega$}
	= \left( {1- \frac{1}{Rm^{(\rm{T})}}} \right) C_W \frac{W}{K} 
		\mbox{\boldmath$\Omega$},
	\label{eq:low_Rm_J_sol}
\end{equation}
respectively. Here $Rm^{(\rm{T})}$ is the turbulent magnetic Reynolds number defined by $Rm^{(\rm{T})} = \beta / \eta$.

	In the astrophysical applications, where $Rm^{(\rm{T})}$ is usually huge, corrections due to the molecular magnetic diffusivity are expected to be negligibly small. However, for numerical simulations with intermediate $Rm^{(\rm{T})}$, such corrections may lead to a substantial difference.

\section{Cross-helicity generation mechanisms\label{sec:ch_generation}}
\subsection{Turbulence modelling and statistical quantities}
	The most straightforward approach to investigate turbulent flows is to directly solve the system of fundamental equations. However, for most geo/astrophysical flows of interests, with huge Reynolds number ($Re$) and magnetic Reynolds number ($Rm$), it is impossible to perform direct numerical simulations (DNS's) in the foreseeable future. In this situation, turbulence models provide a very useful and strong tool for investigating turbulent flows at high $Re$ and $Rm$. 
	
	In turbulence modelling, the statistical properties of unresolved motions have to be modelled by using some quantities that represent such properties. The simplest model is the mixing-length theory of the eddy viscosity $\nu_{\rm{T}}$, where the turbulent or eddy viscosity is expressed in terms of the typical velocity and length scales of turbulence as
\begin{equation}
	\nu_{\rm{T}} \sim u \ell.
	\label{eq:nu_T_mixing_length}
\end{equation}
If typical velocity scale $u$ is estimated by using the mean velocity shear $|dU/dx|$ as
\begin{equation}
	u \sim \ell \left| {\frac{dU}{dx}} \right|
	\label{eq:mixing_length_def}
\end{equation}
with a length scale of turbulence $\ell$ called the mixing length, the eddy viscosity $\nu_{\rm{T}}$ is expressed as
\begin{equation}
	\nu_{\rm{T}} \sim u \ell \sim \ell^2 \left| {\frac{dU}{dx}} \right|.
	\label{eq:mixing_length_nu_T}
\end{equation}
At this moment, the problem is reduced to the point how to estimate the mixing length.

\begin{figure}[htpb]
\begin{center}
\includegraphics[width=.50\textwidth]{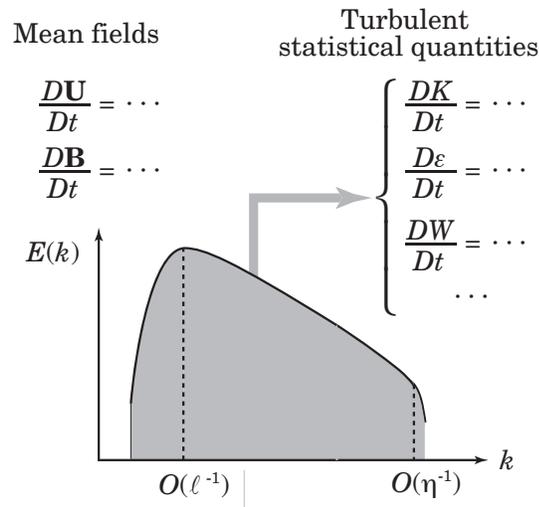}%
\caption{Turbulence modeling. Mean fields and turbulent statistical quantities.}%
\label{fig:modeling-rans}
\end{center}
\end{figure}

	A more elaborated modeling approach is to consider appropriate statistical quantities that represent the statistical properties of the unresolved motions and to construct the transport equations of these statistical quantities. Since the statistical quantities evolve depending on the mean fields or resolved motions, equations of the statistical quantities should be solved with the mean or resolved field equations. Since both mean and turbulence fields are solved simultaneously, the whole system of equations is self-consistently treated in this approach. This is the main reason why this type of modeling approach can provide a very strong tool for investigating turbulent flows. Here the problems are reduced to the points: which statistical quantities we choose, and how to construct proper transport equations for the statistical quantities. The schematic methodology of the basic notion of turbulence model is depicted in Figure~\ref{fig:modeling-rans}.
	
	Statistical analytical theory of inhomogeneous turbulence can provide a firm basis for the turbulence modeling \citep{yos1984}. For details on how to construct the MHD turbulence model on the basis of inhomogeneous turbulence closure theory, the reader is referred to \citet{yok2006,yok2011a,yok2007,yok2008}.

\subsection{Transport equation of turbulent cross helicity (incompressible case)}
	From the viewpoint of constructing transport equations, the turbulent statistical quantities that are related to the conservative property of the fundamental equations are very important. Transport equations for such turbulent statistical quantities can be written in a simple and clear form.
	 
	The magnetohydrodynamic (MHD) energy $\int_V ({\bf{u}}^2 + {\bf{b}}^2)/2 dV$ and $\int_V {\bf{u}} \cdot {\bf{b}} dV$ are inviscid invariants of the MHD equation. The local densities of these quantities, the turbulent MHD energy (density) $K = \langle {{\bf{u}}'{}^2 + {\bf{b}}'{}^2} \rangle / 2$ and the turbulent cross helicity (density) $W = \langle {{\bf{u}}' \cdot {\bf{b}}' } \rangle$ obey a simple evolution equation: 
\begin{equation}
	\left( {
	\frac{\partial}{\partial t}
	+ {\bf{U}} \cdot \nabla
	} \right) G
	= P_G - \varepsilon_G + T_G
	\label{eq:K_W_eq}
\end{equation}
with $G=(K, W)$. Here, $P_G$, $\varepsilon_G$, and $T_G$ are the production, dissipation, and transport rates defined as
\begin{subequations}\label{eq:K_P_eps_T}
\begin{equation}
	P_K 
	= - {\cal{R}}^{ab} \frac{\partial U^a}{\partial x^b}
	- {\bf{E}}_{\rm{M}} \cdot {\bf{J}},
	\label{eq:P_K_def}
\end{equation}
\begin{equation}
	\varepsilon_K = \nu \left\langle {
	\frac{\partial u'{}^a}{\partial x^b} \frac{\partial u'{}^a}{\partial x^b}
	} \right\rangle
	+ \eta \left\langle {
	\frac{\partial b'{}^a}{\partial x^b} \frac{\partial b'{}^a}{\partial x^b}
	} \right\rangle
	(\equiv \varepsilon),
	\label{eq:eps_K_def}
\end{equation}
\begin{equation}
	T_K = {\bf{B}} \cdot \nabla W 
	+ \left\langle {{\bf{f}}' \cdot {\bf{u}}'} \right\rangle
	+ \nabla \cdot {\bf{T}}'_K,
	\label{eq:T_K_def}
\end{equation}
\end{subequations}
\begin{subequations}\label{eq:W_P_eps_T}
\begin{equation}
	P_W 
	= - {\cal{R}}^{ab} \frac{\partial B^a}{\partial x^b}
	- {\bf{E}}_{\rm{M}} \cdot \mbox{\boldmath$\Omega$},
	\label{eq:P_W_def}
\end{equation}
\begin{equation}
	\varepsilon_W = (\nu + \eta) \left\langle {
	\frac{\partial u'{}^a}{\partial x^b} \frac{\partial b'{}^a}{\partial x^b}
	} \right\rangle,
	\label{eq:eps_W_def}
\end{equation}
\begin{equation}
	T_W = {\bf{B}} \cdot \nabla K 
	+ \left\langle {{\bf{f}}' \cdot {\bf{b}}'} \right\rangle
	+ \nabla \cdot {\bf{T}}'_W
	\label{eq:T_W_def}
\end{equation}
\end{subequations}
[${\bf{f}}'$: fluctuation external force]. Here, ${\bf{T}}'_K$ and ${\bf{T}}'_W$ are the transport rates of the turbulent MHD energy $K$ and the turbulent cross helicity $W$, respectively. They are explicitly written as
\begin{equation}
	{\bf{T}}'_K =  \left\langle {
	- \left( {
	\frac{{\bf{u}}'{}^2 + {\bf{b}}'{}^2}{2} + p'_{\rm{M}}
	} \right) {\bf{u}}'
	+ \left( {{\bf{u}}' \cdot {\bf{b}}'} \right) {\bf{b}}'
	} \right\rangle,
\end{equation}
\begin{equation}
	{\bf{T}}'_W = \left\langle {
	- \left( {{\bf{u}}' \cdot {\bf{b}}'} \right) {\bf{u}}'
	+ \left( {
	\frac{{\bf{u}}'{}^2 + {\bf{b}}'{}^2}{2} - p'_{\rm{M}}
	} \right) {\bf{b}}'
	} \right\rangle.
\end{equation}

	In order to perform a realistic calculation of the turbulent transport with the aid of MHD turbulence model, it is of fundamental importance to properly estimate how much turbulent quantities such as the turbulent MHD energy, the turbulent cross helicity, etc.\ are generated and dissipated. In the context of the cross-helicity dynamo, the production rate of the turbulent cross helicity, which is directly related to the mean-field configurations, is most important. We will focus on the cross-helicity generation mechanisms in the following subsection (\S~\ref{sec:ch_production}).

	It is also very important to properly estimate how and how much the cross helicity is dissipated. In practical calculations, the dissipation rate of the turbulent cross helicity is often estimated by using the algebraic model as
\begin{equation}
	\varepsilon_W = C_W \frac{W}{\tau},
	\label{eq:eps_W_alg_model}
\end{equation}
where $\tau$ is the characteristic time of turbulence and $C_W$ is the model constant. This algebraic model is the simplest possible expression for the cross-helicity dissipation rate. More elaborated model for $\varepsilon_W$ has been also proposed. For the detailed discussions on the cross-helicity evolution, including the theoretical derivation of the $\varepsilon_W$ equation, the reader is referred to \citet{yok2011a}.

\subsection{Cross-helicity production mechanisms\label{sec:ch_production}}
	Production rates represent turbulence generation arising from the coupling between the fluctuation and mean-field inhomogeneity. If the Reynolds stress, the correlations between the fluctuation velocities and the fluctuation magnetic fields, is coupled with the mean-velocity shear, the turbulent energy can be generated through $- {\cal{R}}^{ab} \partial U^a / \partial x^b$ [the first term in Eq.~(\ref{eq:P_K_def})]. If we adopt the eddy-viscosity representation, the simplest possible model expression for the Reynolds stress is
\begin{equation}
	{\cal{R}}^{\alpha\beta} 
	= \frac{2}{3} K_{\rm{R}} \delta^{\alpha\beta}
	- \nu_{\rm{K}} {\cal{S}}^{\alpha\beta},
	\label{eq:Rey_strss_eddy_visc}
\end{equation}
with the residual energy $K_{\rm{R}} \equiv \langle {{\bf{u}}'{}^2 - {\bf{b}}'{}^2} \rangle /2$, the production rate related to $\mbox{\boldmath$\cal{R}$}$ is written as
\begin{equation}
	- {\cal{R}}^{ab} \frac{\partial U^a}{\partial x^b}
	= + \frac{1}{2} \nu_{\rm{K}} \left( {{\cal{S}}^{ab}} \right)^2.
\end{equation}
Since the eddy viscosity is positive ($\nu_{\rm{K}} > 0$), the mean velocity strain always enhances the turbulent energy.

	If we only consider the turbulent magnetic diffusivity expression for the turbulent electromotive force as
\begin{equation}
	{\bf{E}}_{\rm{M}} = - \beta {\bf{J}},
	\label{eq:Em_beta_exp}
\end{equation}
the production rate related to ${\bf{E}}_{\rm{M}}$ reads
\begin{equation}
	- {\bf{E}}_{\rm{M}} \cdot {\bf{J}}
	= + \beta {\bf{J}}^2.
\end{equation}
Since the turbulent magnetic diffusivity is positive ($\beta > 0$), the mean electric current always enhances the turbulent energy. This is the enhancement of the Joule heating due to turbulence.

	A similar argument can be applied to the production rate of turbulent cross helicity, $P_W$. However, due to the non positive-definite nature of the cross helicity, the results are different. The Reynolds stress coupled with the mean magnetic-field shear $- {\cal{R}}^{ab} (\partial B^a / \partial x^b)$ [the first term in Eq.~(\ref{eq:P_W_def})] gives the cross-helicity production. If we adopt the eddy-viscosity representation [Eq.~(\ref{eq:Rey_strss_eddy_visc})], we have
\begin{equation}
	- {\cal{R}}^{ab} \frac{\partial B^a}{\partial x^b}
	= + \frac{1}{2} \nu_{\rm{K}} {\cal{S}}^{ab} {\cal{M}}^{ab}.
	\label{eq:P_W_Rey_est}
\end{equation}
Positive or negative turbulent cross helicity is generated depending on the configuration of the mean velocity and magnetic-field strains. This is also the case for the cross-helicity generation related to the turbulent electromotive force [the second term in Eq.~(\ref{eq:P_W_def})]. The turbulent magnetic diffusivity representation of ${\bf{E}}_{\rm{M}}$ [Eq.~(\ref{eq:Em_beta_exp})] leads to the cross-helicity generation
\begin{equation}
	- {\bf{E}}_{\rm{M}} \cdot \mbox{\boldmath$\Omega$}
	= + \beta {\bf{J}} \cdot \mbox{\boldmath$\Omega$}.
	\label{eq:P_W_Em_est}
\end{equation}
This indicates that a positive or negative turbulent cross helicity is generated depending on the configuration of the mean electric current and vorticity. If the mean electric current and vorticity are aligned in a parallel (or antiparallel) manner, positive (or negative) turbulent cross helicity is generated.
\begin{subequations}
\begin{eqnarray}
	P_W > 0 \hspace{20pt} \mbox{for} \hspace{10pt} 
	{\bf{J}} \cdot \mbox{\boldmath$\Omega$} > 0,\\
	P_W < 0 \hspace{20pt} \mbox{for} \hspace{10pt} 
	{\bf{J}} \cdot \mbox{\boldmath$\Omega$} < 0.
\end{eqnarray}
\end{subequations}
The results expressed by Eqs.~(\ref{eq:P_W_Rey_est}) and (\ref{eq:P_W_Em_est}) show that the cross-correlation between the velocity and magnetic field in turbulence entirely depends on the mean-field configurations. If we have a particular configuration between the mean velocity and magnetic field, the turbulent cross helicity has a particular preference for its sign.

	Another important cross-helicity generation mechanism arising from the coupling between the mean field and turbulent correlation is the inhomogeneity of the turbulent energy along the mean magnetic field [the first term in Eq.~(\ref{eq:T_W_def})]. This mechanism is related to the cross-helicity generation expressed by Eq.~(\ref{eq:ch_gene_en_inhomo}), and shows a property entirely different from the production rates of the turbulent cross helicity, $P_W$ [Eq.~(\ref{eq:P_W_def})], (and that of the turbulent MHD energy, $P_K$ [Eq.~(\ref{eq:P_K_def})]).

	From the mean velocity and magnetic field equations, the evolution equation for the mean-flow MHD energy, $({\bf{U}}^2 + {\bf{B}}^2) /2$ is written as
\begin{equation}
	\left( {\frac{\partial}{\partial t} + {\bf{U}} \cdot \nabla} \right)
	\frac{1}{2} \left( {
		{\bf{U}}^2 + {\bf{B}}^2 
	} \right)
	= P_{K{\rm{M}}} - \varepsilon_{K{\rm{M}}} + T_{K{\rm{M}}}.
\end{equation}
Here, $P_{K\rm{M}}$, $\varepsilon_{K\rm{M}}$, and $T_{K\rm{M}}$ are the production, dissipation, and transport rates of the mean MHD energy. They are defined as
\begin{subequations}\label{eq:P_eps_T_GM}
\begin{equation}
		P_{K\rm{M}}
		= + {\cal{R}}^{ab} \frac{\partial U^b}{\partial x^a}
		+ {\bf{E}}_{\rm{M}} \cdot {\bf{J}}
		= - P_K,	
	\label{eq:product_KM}
\end{equation}
\begin{equation}
		\varepsilon_{K\rm{M}}
		= \nu \left( {\frac{\partial U^a}{\partial x^b}} \right)^2
		+ \lambda \left( {\frac{\partial B^a}{\partial x^b}} \right)^2,
	\label{eq:dissip_KM}
\end{equation}
\begin{equation}
		T_{K\rm{M}} = T_{K\rm{MT}} + T_{K\rm{MB}},
	\label{eq:trans_KM}
\end{equation}
\end{subequations}
where $T_{K\rm{MT}}$ is the transport rate of the mean MHD energy arising from the fluctuation correlations: 
\begin{equation}
		T_{K\rm{MT}}
		= \nabla \cdot \left( {
		- {\bf{U}} : \mbox{\boldmath${\cal{R}}$}
		+ {\bf{E}}_{\rm{M}} \times {\bf{B}}
		} \right)
	\label{eq:T_KMT_def}
\end{equation}
[$({\bf{U}}:\mbox{\boldmath${\cal{R}}$})^\alpha= U^b {\cal{R}}^{b\alpha}$].
The other term $T_{K\rm{MB}}$ arises from the mean magnetic field and velocity field:
\begin{equation}
		T_{K\rm{MB}}
		= {\bf{B}} \cdot \left[ {
			\nabla \left( {
			{\bf{U}} \cdot {\bf{B}}
			} \right)
		} \right]
		+ {\bf{U}} \cdot \left( {{\bf{F}} - \nabla P_{\rm{M}}} \right)
	\label{eq:T_KMB_def}
\end{equation}
(${\bf{F}}$: mean external force).

	On the other hand, the equation of the mean-field cross helicity, ${\bf{U}} \cdot {\bf{B}}$, are written as
\begin{equation}
	\left( {\frac{\partial}{\partial t} + {\bf{U}} \cdot \nabla} \right)
	\left( { {\bf{U}} \cdot {\bf{B}} } \right)
	= P_{W{\rm{M}}} - \varepsilon_{W{\rm{M}}} + T_{W{\rm{M}}}.
\end{equation}
Here, $P_{W\rm{M}}$, $\varepsilon_{W\rm{M}}$, and $T_{W\rm{M}}$ are the production, dissipation, and transport rates of the mean cross helicity. They are defined by
\begin{subequations}\label{eq:P_eps_T_WM}
\begin{equation}
		P_{W\rm{M}}
		= + {\cal{R}}^{ab} \frac{\partial B^b}{\partial x^a}
		+ {\bf{E}}_{\rm{M}} \cdot \mbox{\boldmath$\Omega$}
		= - P_W,	
	\label{eq:product_WM}
\end{equation}
\begin{equation}
		\varepsilon_{W\rm{M}}
		= (\nu + \lambda) \frac{\partial U^a}{\partial x^b}
			\frac{\partial B^a}{\partial x^b},
	\label{eq:dissip_WM}
\end{equation}
\begin{equation}
		T_{W\rm{M}} = T_{W\rm{MT}} + T_{W\rm{MB}},
	\label{eq:trans_WM}
\end{equation}
\end{subequations}
where $T_{W\rm{MT}}$ is the transport rate of the mean cross helicity arising from the fluctuation correlations:
\begin{equation}
		T_{W\rm{MT}}
		= \nabla \cdot \left( {
		- {\bf{B}} : \mbox{\boldmath${\cal{R}}$}
		+ {\bf{E}}_{\rm{M}} \times {\bf{U}}
		} \right)
	\label{eq:T_WMT_def}
\end{equation} 
[$({\bf{B}}:\mbox{\boldmath${\cal{R}}$})^\alpha= B^b {\cal{R}}^{b\alpha}$]. The other term arises from the mean magnetic field:
\begin{equation}
		T_{W\rm{MB}}
		= {\bf{B}} \cdot \left[ {
			\nabla \left( {
			\frac{{\bf{U}}^2 + {\bf{B}}^2}{2}
			} \right)
			+ {\bf{F} - \nabla P_{\rm{M}}}
		} \right].
	\label{eq:T_WMB_def}
\end{equation}

	The generation mechanisms of the cross helicity can be divided into two categories: those related to the production rate $P_W$ [Eq.~(\ref{eq:P_W_def})]; and those related to the transport rate $T_W$ [Eq.~(\ref{eq:T_W_def})].
 
	As Eqs.~(\ref{eq:product_KM}) and (\ref{eq:product_WM}) show, the production rates of the mean-flow MHD energy and the mean-flow cross helicity, $P_{K{\rm{M}}}$ and $P_{W{\rm{M}}}$, are exactly the same expressions as the turbulent counterparts but with the opposite signs ($P_{K{\rm{M}}} = - P_{K}$ and $P_{W{\rm{M}}} = - P_{W}$). This shows that the production rates of the turbulent MHD energy and cross helicity correspond to the drain or sink of the mean-flow counterparts. This reflects the cascade nature of the MHD energy and the cross helicity.
	
	On the other hand, the generation by $T_W$ is related to the asymmetric distribution of the energy in the volume. This mechanism is not related to the cascade process, and does not necessarily need mean-fields, either. Unlike the production rates, $P_K$ and $P_W$, we have no drain-like term for the ${\bf{B}} \cdot \nabla K$ in the mean cross-helicity equation. This reflects the fact that the cross-helicity generation due to the inhomogeneity along the mean magnetic field, ${\bf{B}} \cdot \nabla K$, is not related to the cascade nature of turbulence, but is related to the cross-helicity injection through the boundary [Eq.~(\ref{eq:ch_gene_en_inhomo})]. This feature gives a special position for this mechanism in cross helicity generation in real-world turbulence.

	We should note that cross-helicity generation due to ${\bf{B}} \cdot \nabla K$ is related to the Alfv\'{e}n-wave interpretation of the turbulent cross helicity. If we assume that turbulence is a collective motion associated with the Alfv\'{e}n waves, a region with large (or small) turbulent energy corresponds to one with the large (or small) number of Alfv\'{e}n-wave packets. In this picture, the energy inhomogeneity along the mean magnetic field is interpreted as the spatial inhomogeneous distribution of the number of Alfv\'{e}n-wave packets. We can expect that the number of Alfv\'{e}n waves propagating along the mean magnetic field from region with larger turbulent energy to one with smaller turbulent energy is larger than the one propagating in the other direction: from smaller to larger turbulent energy regions. Due to this asymmetry with respect to the directions along the mean magnetic field, we have a finite cross helicity in turbulence.
\begin{figure}[htpb]
\begin{center}
\includegraphics[width=.45\textwidth]{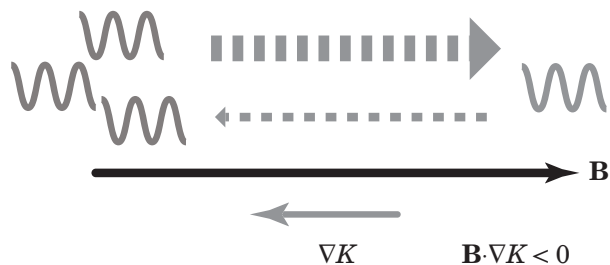}%
\caption{Turbulence inhomogeneity along the mean magnetic field.}%
\label{fig:inhomogeneity_along_B}
\end{center}
\end{figure}

	If we have no mechanism of turbulent cross-helicity generation independent of the mean magnetic field, the cross-helicity generation is just related to the dynamo instability itself. On the basis of this thought, \citet{rue2011} examined importance of the non-conservation of the cross helicity in turbulent media. In some situation, however, the external force term $\langle {{\bf{f}}' \cdot {\bf{b}}'} \rangle$ may play an essential role in the turbulent cross-helicity generation. If we consider a convective flow with buoyancy force ${\bf{f}} = - \alpha_0 \theta {\bf{g}}$ ($\alpha_0$: thermal expansion coefficient, $\theta$: temperature, ${\bf{g}}$: gravitational acceleration), this expression
\begin{equation}
	\left\langle {{\bf{f}}' \cdot {\bf{b}}'} \right\rangle
	= - \alpha_0 \left\langle {\theta' {\bf{b}}'} \right\rangle \cdot {\bf{g}}
	\label{eq:W_gen_buoy}
\end{equation}
gives rise to an important turbulent cross-helicity generation independent of the mean magnetic field.

	This is a direct consequence of the non-positive definiteness of the cross helicity, one of the prominent characteristics of the cross helicity, common to other pseudoscalars such as the kinetic, magnetic, and current helicities. Even if the cross helicity is zero when it is averaged through the total volume, positive and negative cross helicities can be distributed spatially or temporally. This property gives the generation mechanism of the cross helicity very distinctive in comparison with that of the energy. The generation mechanism arising from the transport terms [Eq.~(\ref{eq:T_W_def})] is related to such a spatial distribution of the cross helicity.

\begin{figure}[htpb]
\begin{center}
\includegraphics[width=.50\textwidth]{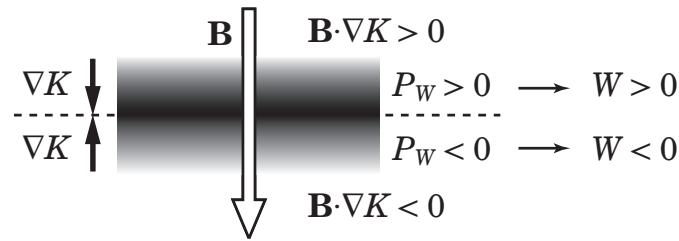}%
\caption{Cross-helicity generation due to the inhomogeneity along the magnetic field.}%
\label{fig:gradK_along_B}
\end{center}
\end{figure}

	Coupled with the mean magnetic-field shear in the momentum equation and with the mean vorticity in the magnetic-field equation, a finite cross helicity existing in a local region may play an important role in the modification of transports there. If we take a volume average, however, the average of the cross helicity is identically zero. This shows that the averaging thorough the total volume is not appropriate for capturing the cross helicity existing locally in space. The cross-helicity distribution reflects the breakage of symmetry due to the inhomogeneity of turbulence and directions of the mean-field quantities. We should properly define the averaging procedure depending on the asymmetry of the mean-field configurations. If the generation mechanism is due to ${\bf{B}} \cdot \nabla K$ [the first term in Eq.~(\ref{eq:T_W_def})], the domain for average should reflect the asymmetry of the turbulence inhomogeneity and the magnetic field direction. For example, let us consider the cross-helicity generation in an accretion disk shown later in Figure~\ref{fig:gradK_along_B}. In this case, the midplane is the plane of symmetry for the turbulence inhomogeneity. It follows that the positive and negative cross helicities are distributed in the northern and southern hemispheres, respectively. This shows that the average should be taken separately in the northern and southern hemisphere. Otherwise, this feature of the cross-helicity distribution can not be captured at all.

\subsection{Transport equation of turbulent cross helicity (compressible case)\label{sec:W_eq_comp}}
	In the compressible magnetohydrodynamic case, it is useful to write the density dependence explicitly. In this subsection we express the magnetic field in the original physical unit. The turbulent cross helicity is defined by
\begin{equation}
	W_{\ast} \equiv \left\langle { {\bf{u}}' \cdot {\bf{b}}'_{\ast} } \right\rangle
	\label{eq:cross_helicity_def}
\end{equation}
Here, subscript $\ast$ denotes that the magnetic field is measured in the original physical unit (not in Alfv\'{e}n-speed units).

	The transport equation of $W_{\ast}$ is given as
\begin{subequations}\label{eq:comp_W_eq_full}
\begin{eqnarray}
	\frac{DW_{\ast}}{Dt} 
	&\equiv& \left( {
		\frac{\partial}{\partial t} + {\bf{U}} \cdot \nabla
	} \right) W_{\ast}\nonumber\\
	&=& - \frac{1}{2} \left\langle {
	u'{}^a u'{}^b
	- \frac{1}{\mu_0 \overline{\rho}} 
		b'_{\ast}{}^a b'_{\ast}{}^b
	} \right\rangle \left( {
		\frac{\partial B_{\ast}^b}{\partial x^a}
			+ \frac{\partial B_{\ast}^a}{\partial x^b}
	} \right)\label{eq:comp_W_R}\\
	&-& \left\langle {{\bf{u}}' \times {\bf{b}}'_{\ast}} \right\rangle \cdot 
			\mbox{\boldmath$\Omega$}\label{eq:comp_W_emf}\\
	&-& (\gamma_0 - 1) \frac{1}{\overline{\rho}} 
		\left\langle {\rho' {\bf{b}}'_{\ast}} \right\rangle \cdot \nabla Q\\
	&-& (\gamma_0 - 1) \frac{1}{\overline{\rho}} 
		\left\langle{ q' {\bf{b}}'_{\ast}} \right\rangle 
		\cdot \nabla \overline{\rho}\label{eq:grad_rho}\\
	&-& \frac{1}{\overline{\rho}} \left\langle {
	\rho' {\bf{b}}'_{\ast}
} \right\rangle \cdot \frac{D{\bf{U}}}{Dt}\\
	&-& W_{\ast} \nabla \cdot {\bf{U}}\label{eq:comp_W_divU}\\
	&+& {\bf{B}}_{\ast} \cdot \nabla \left\langle {
		\frac{1}{2} {\bf{u}}'{}^2
	} \right\rangle\label{eq:comp_W_inhomo}\\
	&+& \left\langle {{\bf{f}}' \cdot {\bf{b}}'} \right\rangle
	\label{eq:comp_W_force}\\
	&-& \varepsilon_{W_{\ast}}
	+ T_{W_{\ast}}
	+ {\rm{R.T.}},
\end{eqnarray}
\end{subequations}
where ${\rm{R.T.}}$ denotes the residual terms arising from the higher order terms.
Here $q (= C_V \theta)$ is the internal energy ($C_V$: specific heat at constant volume, $\theta$: the temperature), $\gamma_0 (= C_P / C_V)$ the ratio of specific heats ($C_P$: specific heat at constant pressure), and ${\bf{f}}'$ the fluctuation of the external force per unit mass. The internal energy is divided into the mean $Q$ and fluctuation around it, $q'$ ($q = Q + q'$). The plasma pressure is assumed to satisfy the ideal gas relation $p = R \rho \theta = (\gamma_0 - 1) \rho q$ ($R$: gas constant).

	In Eq.~(\ref{eq:comp_W_eq_full}), $\varepsilon_{W_{\ast}}$ and $T_{W_{\ast}}$ are the dissipation and transport rates, respectively, whose detailed expressions are suppressed here. Among the other terms, (\ref{eq:comp_W_R}), (\ref{eq:comp_W_emf}), and (\ref{eq:comp_W_inhomo}) are incompressible terms. They have counterparts in Eq.~(\ref{eq:K_W_eq}) with Eq.~(\ref{eq:W_P_eps_T}).

	Equation~(\ref{eq:comp_W_eq_full}) indicates that, in the compressible case, even if we dropped the density-fluctuation effect ($\rho' = 0$), we still have some production mechanisms of $W_{\ast}$ that is not directly connected to the mean magnetic field. Terms labeled (\ref{eq:grad_rho}) and (\ref{eq:comp_W_divU}) are such terms. The importance of (\ref{eq:grad_rho}) is discussed in the context of the local magneto-convection in the Sun. On the other hand, (\ref{eq:comp_W_divU}) indicates that the magnitude of $W$ increases irrespective of the sign of $W$, if the mean flow is converging ($\nabla \cdot {\bf{U}} < 0$).
	
	It is in general very difficult to simultaneously measure three components of the fluctuating velocity and magnetic field by remote observations. However, there are some attempts to estimate the turbulent cross helicity in terms of mean-field quantities which are easier to measure \citep{kle2003,rue2011}. It would be useful to compare Eq.~(\ref{eq:comp_W_eq_full}) with the previous estimate of the cross helicity. For the inhomogeneous and density stratified turbulence, Kleeorin et al.\ (2003) derived an expression for the turbulent cross helicity in their Eq.~(11) as
\begin{equation}
	\left\langle {{\bf{u}}' \cdot {\bf{b}}'} \right\rangle
	= \frac{3}{2} \beta \Lambda_u^{-1} B^r
	+ \phi_{\rm{ch}}(B) \left( {{\bf{B}} \cdot \nabla} \right) {\bf{B}}^2,
	\label{eq:ch_estimate_kle}
\end{equation}
where $\Lambda_u^{-1} = \nabla \langle {{\bf{u}}'{}^2} \rangle / \langle {{\bf{u}}'{}^2} \rangle$ is the reciprocal of the turbulence inhomogeneity scale, $B^r$ is the radial mean velocity, and $\phi_{\rm{ch}}(B)$ is a quenching function expressed in terms of  the toroidal field $B$. Another expression was proposed by \citet{rue2011} in their Eq.~(15) as
\begin{equation}
	\left\langle {{\bf{u}}' \cdot {\bf{b}}'} \right\rangle
	= \beta {\bf{G}} \cdot {\bf{B}}
		+ \left( {
		\frac{\beta}{2} + \frac{2\hat{\eta}}{3}
	} \right) {\bf{B}} \cdot \nabla \ln \langle {{\bf{u}}'{}^2} \rangle,
	\label{eq:ch_estimate_rue}
\end{equation}
where ${\bf{G}} = \nabla \ln \overline{\rho}
= \nabla \overline{\rho} / \overline{\rho}$ is the reciprocal of density scale height ($\overline{\rho}$: mean density).

	The first term of Eq.~(\ref{eq:ch_estimate_kle}) and the second term of Eq.~(\ref{eq:ch_estimate_rue}) is quite similar: the turbulent energy inhomogeneity along the mean magnetic field. This is an important factor generating the turbulent cross helicity. This contribution is expressed as Eq.~(\ref{eq:comp_W_inhomo}). As was referred to previously, this mechanism arises not from the production rate due to cascade but from the transport rate. Equation ~(\ref{eq:comp_W_inhomo}) is rewritten as
\begin{equation}
	{\bf{B}}_\ast \cdot \nabla \left\langle {\frac{1}{2} {\bf{u}}'{}^2} \right\rangle
	= \frac{1}{2} {\bf{B}}_\ast \left\langle {{\bf{u}}'{}^2} \right\rangle 
		\cdot \frac{1}{\left\langle {{\bf{u}}'{}^2} \right\rangle}  
		\nabla \left\langle {{\bf{u}}'{}^2} \right\rangle
	\sim \frac{1}{\tau} \frac{\beta}{2} {\bf{B}}_\ast
		\cdot \nabla\ln \left\langle {{\bf{u}}'{}^2} \right\rangle,
	\label{eq:en_inhomo_B_cmpr}
\end{equation}
where use has been made of $\beta \sim \langle {{\bf{u}}'{}^2} \rangle \tau$. Equation~(\ref{eq:en_inhomo_B_cmpr}) corresponds to the second term of Eq.~(\ref{eq:ch_estimate_rue}).

	The effect of the mean density stratification appears in Eq.~(\ref{eq:grad_rho}). The fluctuation of the internal energy is expressed as
\begin{equation}
	q' = \frac{1}{\gamma_0 - 1} \frac{p'_\ast}{\overline{\rho}}
	\label{eq:fluct_q-p_rel}
\end{equation}
for the ideal gas. With Eq.~(\ref{eq:fluct_q-p_rel}), the correlation of the internal-energy and magnetic-field fluctuation is written as
\begin{equation}
	\left\langle {q' {\bf{b}}'_\ast} \right\rangle
	= \frac{1}{\gamma_0 - 1}\frac{1}{\overline{\rho}}
		\left\langle {p'_\ast {\bf{b}}'_\ast} \right\rangle
\end{equation}
Then, Eq.~(\ref{eq:grad_rho}) yields to
\begin{equation}
	- (\gamma_0 - 1) \left\langle {q' {\bf{b}}'_\ast} \right\rangle
		\frac{1}{\overline{\rho}}\nabla \overline{\rho}
	= \frac{1}{\overline{\rho}} \left\langle {p' {\bf{b}}'_\ast} \right\rangle
	\frac{1}{\overline{\rho}}\nabla \overline{\rho}
	\sim \left\langle {{\bf{u}}'{}^2 {\bf{b}}'_\ast} \right\rangle 
	\frac{1}{\overline{\rho}}\nabla \overline{\rho}
\end{equation}
where use has been made of $p' \sim \overline{\rho} {\bf{u}}'{}^2$. If we estimate the triple correlation as $\langle {u'{}^2 {\bf{b}}'_\ast} \rangle \sim - \beta {\bf{B}}_\ast /\tau$, we finally obtain
\begin{equation}
	- (\gamma_0 - 1) \left\langle {q' {\bf{b}}'_\ast} \right\rangle
	\frac{1}{\overline{\rho}}\nabla \overline{\rho}
	\sim \frac{1}{\tau} \beta {\bf{B}} \nabla \ln \overline{\rho}.
\end{equation}
This corresponds to the first term in Eq.~(\ref{eq:ch_estimate_rue}).

The Lorentz force is divided into the Maxwell-tensor and the magnetic-pressure parts. The cross-helicity generation due to the turbulent Maxwell stress is expressed by the second part of Eq.~(\ref{eq:comp_W_R}):
\begin{equation}
	P_{W_\ast}^{(\rm{mxw})} 
	= + \frac{1}{2} \frac{1}{\mu_0 \overline{\rho}}
	\left\langle {b'_\ast{}^a b'_\ast{}^b} \right\rangle \left( {
		\frac{\partial B_\ast^b}{\partial x^a}
		+ \frac{\partial B_\ast^a}{\partial x^b}
	} \right).
\end{equation}
If we simply model the turbulent Maxwell stress as $\left\langle {b'_\ast{}^a b'_\ast{}^b} \right\rangle \sim B_\ast^a B_\ast^b$, the production rate due to the turbulent Maxwell stress yields to
\begin{equation}
	P_{W_\ast}^{(\rm{mxw})} 
	\sim \frac{1}{\mu_0 \overline{\rho}} 
		B_\ast^a \frac{\partial}{\partial x^a} {\bf{B}}_\ast^2.
\end{equation}
This is similar to the second term of Eq.~(\ref{eq:ch_estimate_kle}). But of course, in order to take the quenching effect into account, we should include the reduction mechanisms of the turbulent cross helicity such as the dissipation rate $\varepsilon_W$ [in Eq.~(\ref{eq:grad_rho})] and the $\alpha$ effect [in Eq.~(\ref{eq:comp_W_emf}), also see Eq.~(\ref{eq:solar_W_red_alpha})].
	
	The turbulent cross helicity generation in the compressible MHD turbulence will be further reported in the future work \citep{yok2012b}.

\section{Illustrative examples of cross-helicity effects\label{sec:ch_examples}}
	In this section, we present several examples of the application of the cross-helicity effects to astrophysical and fusion plasma phenomena.

\subsection{Galactic magnetic field\label{sec:galaxy_dynamo}}
\begin{figure}[htpb]
\begin{center}
\includegraphics[width=.35\textwidth]{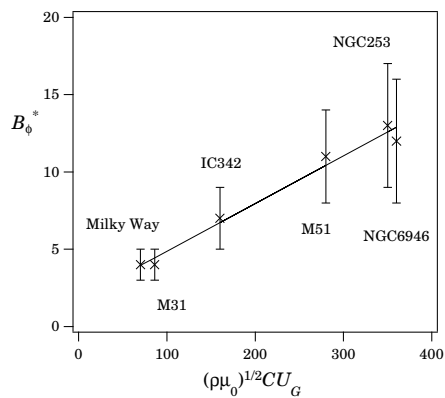}%
\caption{Magnetic field strength of several galaxies against their rotation speed.}%
\label{fig:gmf_strength}
\end{center}
\end{figure}

	As we have seen, if the main balancer against the turbulent magnetic diffusivity $\beta$ is the cross-helicity effect $\gamma$, we have the mean-field configuration with the mean vorticity $\mbox{\boldmath$\Omega$}$ aligned with the mean electric current density ${\bf{J}}$. We substitute model expressions for $\beta$ and $\gamma$ into the special solution for the stationary magnetic field [Eq.~(\ref{eq:special_B_sol})]. The magnetic field measured in the Alfv\'{e}n speed unit, ${\bf{b}}$, is related to the original magnetic field (measured in the physical unit), ${\bf{b}}_\ast$, as in Eq.~(\ref{eq:alfven_unit}). Thus the mean magnetic field ${\bf{B}}_\ast$ is expressed as	
\begin{equation}
	{\bf{B}}_\ast = (\mu_0 \rho)^{1/2} {\bf{B}} 
	= (\mu_0 \rho)^{1/2} C_W \frac{W}{K} {\bf{U}}.
	\label{eq:galactic_mag}
\end{equation}
Rotation curves of several galaxies show a very flat profile in outer regions. The rotation speed of galaxies can be represented by this constant part of the rotation profile. In Figure~\ref{fig:gmf_strength}, we plot the observed magnetic-field strength of several galaxies against the rotational speed. This figure shows that the magnetic-field strength of galaxies is approximately proportional to the rotational speed of them. The inclination angle of the plot gives the value of the cross helicity scaled by the turbulent MHD energy, $|W|/K$. We see from Figure~\ref{fig:gmf_strength} that $|W|/K$ is estimated as
\begin{equation}
	\frac{|W|}{K} = 0.03.
	\label{eq:scaled_W_galax}
\end{equation}
This estimate indicates that the scaled cross helicity of $|W|/K = O(10^{-2})$ is large enough for explaining the galactic field strength.

	Detailed analysis of the Faraday rotation measure (RM) of several galaxies has revealed the basic properties of the galactic magnetic fields. They may be summarized as \citep{sof1986}
\begin{enumerate}
\item Strength of the mean magnetic field is much smaller than the total magnetic field estimated by using the Zeeman effect;
\item Direction of the global magnetic field is approximately along the spiral arms rather than along the global velocity;
\item Most ubiquitous configuration of the global magnetic field is the ``bisymmetric spiral (BSS)''. The directions of global magnetic field is in outward direction for one spiral arm and inward for the next arm.
\end{enumerate}

	In the framework of the cross-helicity dynamo, these features can be explained as follows. If the cross helicity is distributed antisymmetric with respect to the midplane of galaxy with symmetric distribution of global velocity, from Eq.~(\ref{eq:galactic_mag}) we have a global magnetic field whose strength is the same but the direction is opposite in the upper and lower half domains of the galactic disk. In such a case, the rotation measure observed from a remote place may be canceled out. We need some additional breakage of symmetry with respect to the midplane of the galactic disk. It is known that, due to a sort of corrugation, the density of galactic gas is distributed asymmetrically with respect to the midplane \citep{wea1974}. The reference density is the same between the upper and lower half domains, but the actual global distribution of gas density corrugates and deviates from the reference value approximately up to $\pm 10 \%$. This asymmetry gives the residual contribution for the Farady rotation measure (RM). Taking this into account, three features of galactic magnetic field listed above can be elucidated to some extent in the framework of the cross-helicty dynamo with a simple expression for the mean magnetic field [Eq.~(\ref{eq:galactic_mag})] \citep{yok1996a}.

\subsection{Accretion disks\label{sec:accretion_dynamo}}
	A gas surrounding a compact massive object accretes to the central object rotating around it. This is called accretion disk, and is ubiquitously observed in several astrophysical bodies such as protostar, binary stars, active galactic nuclei, black holes, etc. Bipolar jets, ejected from the central region of the accretion disk to both directions perpendicular to the disk, are often observed. These jets are called astrophysical jets. One of the prominent features of astrophysical jets is their high collimation. The expansion rate estimated by the spatial dimensions for the vertical to parallel directions is very small: $O(10^{-6})$. One possible explanation of this extremely high collimation is confinement of plasma gas by the magnetic fields associated with the accretion disk and jets.

	From the viewpoint of cross-helicity effects, we should note that the accretion disk geometry is favorable for the cross-helicity dynamo to work since the mean-field configurations are favorable for the production of turbulent cross helicity there. We consider a situation where a global magnetic field is threading the gas disk whose turbulence is strongest at the midplane. (Note that this particular direction of gradient or inhomogeneity is not essential for the following argument.) Due to this inhomogeneity along the mean magnetic field, we have different signs of cross-helicity generation between the upper and lower half domains. If the threading magnetic field is in the downward direction (from upper to lower domains), the production of turbulent cross helicity is positive and negative in the upper- and lower-half domains, respectively. According to Eq,~(\ref{eq:special_B_sol}), the global magnetic field is parallel (or anti-parallel) to the velocity in the upper (or lower) half domain. We have an antisymmetric magnetic field for a global rotational motion symmetric with respect to the midplane.
\begin{figure}[htpb]
\begin{center}
\includegraphics[width=.35\textwidth]{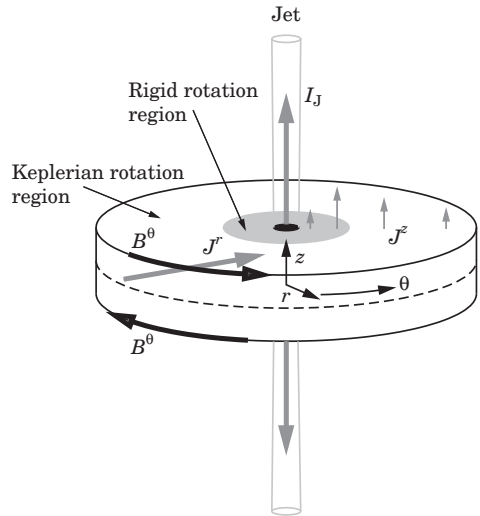}%
\caption{Mean field configuration of an accretion disk.}%
\label{fig:accretion}
\end{center}
\end{figure}

	The global magnetic-field configuration is dipolar-like, and we have a global electric-field current density in a radially inward direction. As this result, a global electric current is ejected from the center region of the accretion disk in the direction perpendicular to the disk or bipolar direction. The existence of global electric current ${\bf{J}}$ in the bipolar direction suggests that there is a self-sustaining mechanism for the turbulent cross helicity. The global bipolar electric current coupled with the global vorticity $\mbox{\boldmath$\Omega$}$ contributes to the turbulent cross helicity generation as
\begin{equation}
	\begin{array}{llll}
		\beta {\bf{J}} \cdot \mbox{\boldmath$\Omega$} > 0 \hspace{10pt} 
		&\rightarrow\hspace{10pt} 
		& P_W > 0\hspace{10pt}
		&{\mbox{for the upper half domain}},\\
		\beta {\bf{J}} \cdot \mbox{\boldmath$\Omega$} < 0 \hspace{10pt}
		&\rightarrow\hspace{10pt} 
		& P_W < 0\hspace{10pt}
		& {\mbox{for the lower half domain}}.
	\end{array}
	\label{Pw_accretion}
\end{equation}
These signs are equal to the original ones, leading to the self-sustained cross-helicity distribution for an accretion disk.

\subsection{Solar dynamos\label{sec:solar_dynamo}}
\subsubsection{$\alpha$ effect as a perturbation\label{sec:alpha_perturb}}
	One important point to note is that the magnetic field generated by the $\alpha$ or helicity effect may reduce the turbulent cross helicity originally presented\citep{yos2000b}. As we show in the following, this property is expected to play a very important role in the periodic behaviour of the solar magnetic field.

	To see this point, we consider a combination of the cross-helicity and $\alpha$ effects. We assume that the dominant dynamo effect is due to the cross-helicity effect (reference state), and the $\alpha$ effect serves itself as a perturbation or modulation to the reference state. In this sense, the perturbations ${\bf{B}}_1$ and ${\bf{J}}_1$ are smaller than the ${\bf{B}}_0$ and ${\bf{J}}_0$ fields.  We write the mean magnetic field and electric-current density as
\begin{equation}
	{\bf{B}} = {\bf{B}}_0 + {\bf{B}}_1,\;\;
	{\bf{J}} = {\bf{J}}_0 + {\bf{J}}_1.
	\label{eq:alpha_as _perturb}
\end{equation}
where ${\bf{B}}_0$ and ${\bf{J}}_0$ are zeroth-order in $\alpha$, and ${\bf{B}}_1$ and ${\bf{J}}_1$ first-order. Substituting Eq.~(\ref{eq:alpha_as _perturb}) into the mean induction equation [Eq.~(\ref{eq:mean_ind_eq_exp})], we obtain equations for the reference [$O(\alpha^0)$] and modulation [$O(\alpha^1)$] fields as
\begin{equation}
	\frac{\partial {\bf{B}}_0}{\partial t}
	= \nabla \times \left( {
		{\bf{U}} \times {\bf{B}}_0 
		- \beta {\bf{J}}_0 
		+ \gamma \mbox{\boldmath$\Omega$}
	} \right),
	\label{eq:solar_dyn_B0_eq}
\end{equation}
\begin{equation}
	\frac{\partial {\bf{B}}_1}{\partial t}
	= \nabla \times \left( {
		{\bf{U}} \times {\bf{B}}_1 
		+ \alpha {\bf{B}}_0
		- \beta {\bf{J}}_1 
	} \right).
	\label{eq:solar_dyn_B1_eq}
\end{equation}

	As we saw in \S~\ref{sec:ch_effect}, the reference-field equation [Eq.~(\ref{eq:solar_dyn_B0_eq})] has a special solution for the stationary state as
\begin{equation}
	{\bf{B}}_0 = \frac{\gamma}{\beta} {\bf{U}}.
	\label{eq:solar_dyn_B0_sol}
\end{equation}
Substituting Eq.~(\ref{eq:solar_dyn_B0_sol}) into Eq.~(\ref{eq:solar_dyn_B1_eq}), we have the modulation-field equation as
\begin{equation}
	\frac{\partial {\bf{B}}_1}{\partial t}
	= \nabla \times \left( {
		{\bf{U}} \times {\bf{B}}_1 
		+ \frac{\alpha\gamma}{\beta} {\bf{U}}
		- \beta {\bf{J}}_1 
	} \right).
	\label{eq:solar_dyn_B1_eq_mod}
\end{equation}
We approximate the mean velocity ${\bf{U}}$ in the polar spherical coordinate $(r, \theta, \phi)$ by the toroidal velocity as ${\bf{U}} = (U^r, U^\theta, U^\phi) \simeq (0, 0, U^\phi)$. In this section (\S\ref{sec:alpha_perturb}) and also in the following section (\S\ref{sec:several_models}), axisymmetry of the mean velocity and magnetic field, ${\bf{U}}$ and ${\bf{B}}$, is assumed. In the low latitude region, the radial component of the mean magnetic field is small ($B_1^r \simeq 0$), and the latitudinal gradient of the toroidal mean velocity is also small ($\partial U^\phi / \partial \theta \simeq 0$). Using these approximations, we estimate 
\begin{equation}
	\nabla \times \left( {{\bf{U}} \times {\bf{B}}} \right)
	\simeq \left( {0, 0, B_1^r \frac{\partial U^\phi}{\partial r}
		+ B_1^\theta \frac{1}{r} \frac{\partial U^\phi}{\partial \theta}} \right)
	\simeq \left( {0, 0, 0} \right).
\end{equation}
Under these considerations, we see that
\begin{equation}
	{\bf{J}}_1 = \frac{\alpha}{\beta} {\bf{B}}_0 
	= \frac{\alpha\gamma}{\beta^2} {\bf{U}}
\end{equation}
is an approximate solution for the stationary state of Eq.~(\ref{eq:solar_dyn_B1_eq_mod}). This corresponds to the poloidal field ${\bf{B}}_1$  generation from the toroidal field ${\bf{B}}_0$ through the $\alpha$ effect.

	Here one remark should be put on the role of differential rotation in the cross-helicity dynamo. A prominent feature of the cross-helicity dynamo lies in the point that it produces a toroidal magnetic field from a poloidal one without resorting to the differential rotation. This does not deny the importance of the differential rotation, which is essential to sustain turbulence. Without turbulence, turbulent cross helicity also vanishes.

	Since the modulated field ${\bf{B}}_1$ associated with the mean electric-current density ${\bf{J}}_1 (= \nabla \times {\bf{B}}_1)$ is the poloidal one, ${\bf{B}}_1$ is aligned with the local rotation vector $\mbox{\boldmath$\Omega$}$. Since $\alpha$ and $\gamma$ are pseudoscalars, both of them change their signs between the northern and southern hemispheres. Consequently, the directions of ${\bf{J}}$ are the same for both hemispheres, leading to a dipole-like magnetic-field configuration.
	
	We consider the evolution equation of the turbulent cross helicity. There is a contribution to the cross-helicity production arising from the poloidal magnetic field ${\bf{B}}_1$ induced by the $\alpha$ effect, $P_{W1}$, as
\begin{equation}
	\frac{\partial W}{\partial t} = \cdots  \underbrace{- \alpha {\bf{B}}_1 \cdot \mbox{\boldmath$\Omega$}}_{P_{W1}} + \cdots.
	\label{eq:solar_W_red_alpha}
\end{equation}

\begin{figure}
\begin{center}
\includegraphics[width=.60\textwidth]{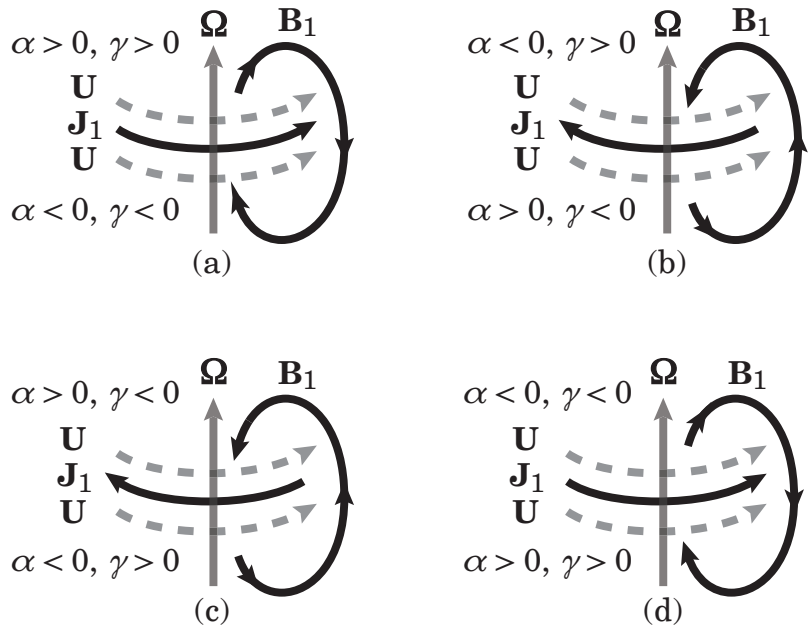}%
\caption{Combination of the cross-helicity and $\alpha$ effects. Depending on the signs of the turbulent cross helicity ($\gamma$) and turbulent residual helicity ($\alpha$), the mean magnetic-field configuration changes. }%
\label{fig:perturb_alpha}
\end{center}
\end{figure}

	First we consider a situation where the turbulent cross helicity is positive ($\gamma >  0$) in the northern hemisphere. If the turbulent residual helicity is also positive ($\alpha > 0$) there, the mean electric-current density ${\bf{J}}_1$ induced by the $\alpha$ effect is parallel to the mean vorticity as
\begin{equation}
	{\bf{J}}_1 = \frac{\alpha\gamma}{\beta^2} {\bf{U}}\;\;\;
	\mbox{with}\;\;\; \frac{\alpha\gamma}{\beta^2} > 0.
\end{equation}
In this case, the mean magnetic field ${\bf{B}}_1$ induced by the $\alpha$ effect is parallel to the mean vorticity $\mbox{\boldmath$\Omega$}$ as in Figure~\ref{fig:perturb_alpha}(a). Thus we have a negative turbulent cross-helicity generation due to the $\alpha$ effect as
\begin{equation}
	P_{W1} = - \alpha {\bf{B}}_1 \cdot \mbox{\boldmath$\Omega$} < 0\;\;\;
	\mbox{for}\;\;\; \alpha > 0, \gamma > 0.
	\label{eq:alpha_p_gamma_p}
\end{equation}
If the turbulent residual helicity is negative ($\alpha <0$) there, the mean electric-current density ${\bf{J}}_1$ induced by the $\alpha$ effect is antiparallel to the mean velocity as
\begin{equation}
	{\bf{J}}_1 = \frac{\alpha\gamma}{\beta^2} {\bf{U}}\;\;\;
	\mbox{with}\;\;\; \frac{\alpha\gamma}{\beta^2} < 0.
\end{equation}
In this case, the mean magnetic field ${\bf{B}}_1$ induced by the $\alpha$ effect is antiparallel to the mean vorticity $\mbox{\boldmath$\Omega$}$ as in Figure~\ref{fig:perturb_alpha}(b). Thus again, a negative turbulent cross helicity is generated by the $\alpha$ effect as
\begin{equation}
	P_{W1} = - \alpha {\bf{B}}_1 \cdot \mbox{\boldmath$\Omega$} < 0\;\;\;
	\mbox{for}\;\;\; \alpha < 0, \gamma > 0.
	\label{eq:alpha_n_gamma_p}
\end{equation}

	For situations where the turbulent cross helicity is negative ($\gamma <0$) in the northern hemisphere, a similar argument can be applied. We have a positive turbulent cross-helicity generation due to the $\alpha$ effect as
\begin{equation}
	P_{W1} = - \alpha {\bf{B}}_1 \cdot \mbox{\boldmath$\Omega$} > 0\;\;\;
	\mbox{for}\;\;\; \alpha_<^> 0, \gamma < 0
	\label{eq:alpha_pn_gamma_n}
\end{equation}
[Figures~\ref{fig:perturb_alpha}(c) and (d)].

	We see from the above arguments that for both situations with the positive and negative turbulent cross helicity, the $\alpha$ effect works as the reduction of the original turbulent cross helicity. This suggests that the cross-helicity dynamo coupled with the $\alpha$ effect gives the possibility of the periodic magnetic-field reversal through the oscillatory behaviour of the turbulent cross helicity.

\subsubsection{Several levels of models\label{sec:several_models}}
	On the basis of the mean induction equation~(\ref{eq:mean_ind_eq_exp}), we write the equations for the toroidal magnetic field $B$ and for toroidal component of the vector potential $A$ representing the mean poloidal magnetic field. One of the most simplified expressions is given as
\begin{equation}
	\frac{\partial}{\partial t}
	\begin{pmatrix}
		B\\ A
	\end{pmatrix}
	= \begin{pmatrix}
		\beta \nabla^2 & \hat{G}\\ \hat{\alpha} & \beta \nabla^2
	\end{pmatrix}
	\begin{pmatrix}
		B\\ A
	\end{pmatrix},
	\label{eq:generic_parker_eq}
\end{equation}
where $\hat{G}$ denotes the mean velocity shear and $\hat{\alpha}$ denotes the $\alpha$ effect. The mean velocity shear ($\hat{G}$) coupled with the mean poloidal magnetic field ($A$) induces the toroidal magnetic field ($B$). At the same time, the helical properties of turbulence, represented by $\alpha$ effect ($\hat{\alpha}$), coupled with the toroidal magnetic field ($B$) give rise to the poloidal field ($A$). 

	Keeping the arguments developed in \S\ref{sec:alpha_perturb} in mind, we should take into account the following two points:
\begin{enumerate}
	\item Transport equations for the transport coefficients;
	\item Inclusion of the cross-helicity effect.
\end{enumerate}

	As we see from Eqs.~(\ref{eq:alpha_spec_exp})-(\ref{eq:gamma_spec_exp}), the transport coefficients should be determined by the statistical properties of turbulence. For instance, $\alpha$ is determined by the turbulent residual helicity, the difference between the turbulent kinetic and current helicities defined by $\langle{ - {\bf{u}}' \cdot \mbox{\boldmath$\omega$}' + {\bf{b}}' \cdot {\bf{j}}'} \rangle$. However, since neither kinetic helicity $\int_V {{\bf{u}} \cdot \mbox{\boldmath$\omega$}} dV$ nor the current helicity $\int_V {{\bf{b}} \cdot {\bf{j}}} dV$ is inviscid invariant of the MHD equations, it is difficult to derive a model equation for $\alpha$ on a theoretically firm basis. On the other hand, since the magnetic helicity $\int_V {{\bf{a}} \cdot {\bf{b}}} dV$ is an inviscid invariant of the MHD equation, the transport equation for the turbulent magnetic helicity $\langle {{\bf{a}}' \cdot {\bf{b}}'} \rangle$ can be written in a simple form on a firm theoretical basis. In this line of thought, models for the $\alpha$ or magnetic-helicity evolution have been proposed \citep{kle1999,kle2000,kle2003}. Also a recent sophisticated mean-field models for the magnetic-helicity feedback reproduces the helicity pattern in close agreement with the observations \citep{pip2011a}. 

	From the viewpoint of the cross-helicity dynamo, inclusion of the cross helicity effect with its transport equation may be further important steps. As we have seen in \S\ref{sec:alpha_perturb}, the coupling of the poloidal magnetic field generated by the $\alpha$ effect with the turbulent cross-helicity generation is expected to play an essential role in the periodic reversal of the solar magnetic field. The essential ingredients of the field reversal process are the evolution equations of the toroidal and poloidal magnetic fields, $B_{\rm{T}}$ and $B_{\rm{P}}$:
\begin{equation}
	\frac{\partial {\bf{B}}}{\partial t}
	= \nabla \times \left( {
		{\bf{U}} \times {\bf{B}}
	} \right) 
	+ \nabla \times {\bf{E}}_{\rm{M}}
	+ \eta \nabla^2 {\bf{B}},
\end{equation}
and the evolution equations of the turbulent cross helicity $W$:
\begin{equation}
	\frac{\partial W}{\partial t}
	+ \left( {{\bf{U}} \cdot \nabla} \right) W
	= - {\cal{R}}^{ab} \frac{\partial B^a}{\partial x^b}
	- {\bf{E}}_{\rm{M}} \cdot \mbox{\boldmath$\Omega$} 
	+ {\bf{B}} \cdot \nabla K
	+ \nabla \cdot {\bf{T}}'{}_W,
\end{equation}
with the Reynolds stress $\mbox{\boldmath${\cal{R}}$}$ [Eq.~(\ref{eq:Re_strss_exp})] and the turbulent electromotive force ${\bf{E}}_{\rm{M}}$ [Eq.~(\ref{eq:E_M_exp})].

	The transport coefficients appearing in $\mbox{\boldmath${\cal{R}}$}$, $\nu_{\rm{K}}$ and $\nu_{\rm{M}}$, and in ${\bf{E}}_{\rm{M}}$, $\alpha$, $\beta$, and $\gamma$, are not adjustable constants. They should represent statistical properties of turbulence as Eqs.~(\ref{eq:alpha_spec_exp})-(\ref{eq:gamma_spec_exp}) show. The simplest possible expressions for them are the mixing-length type ones. A further elaborated approach is to construct evolution equations of the transport coefficients themselves or equations of statistical quantities determining the transport coefficients.

\paragraph{(i) Toy model}
	We can construct a minimal model for the periodic behaviour of solar magnetic field. This model is constituted of the equations that express (i) the toroidal-field generation due to the cross-helicity effect [Eq.~(\ref{eq:solar_dyn_B0_sol})], (ii) the poloidal-field generation due to the $\alpha$ effect [Eq.~(\ref{eq:solar_dyn_B1_eq})]; and (iii) the cross-helicity reduction due to the poloidal field [Eq.~(\ref{eq:solar_W_red_alpha})]. The model is expressed as
\begin{subequations}
\begin{equation}
	B_{\rm{T}} = \gamma^\ast U_{\rm{T}},
\end{equation}
\begin{equation}
	\frac{dB_{\rm{P}}}{dt} = \alpha^\ast B_{\rm{T}},
\end{equation}
\begin{equation}
	\frac{d \gamma^\ast}{dt} = \delta^\ast B_{\rm{P}},
\end{equation}
\end{subequations}
where $\alpha^\ast$ and $\delta^\ast$ are defined as
\begin{equation}
	\alpha^\ast = \frac{1}{\tau_{\rm{C}}} \frac{\alpha L_{\rm{C}}}{\beta},\;\;
	\delta^\ast = \tau_{\rm{C}} \frac{\alpha}{\beta} \omega_{\rm{F}}
\end{equation}
[$\tau_{\rm{C}}$: characteristic time scale of turbulence often modeled as $\tau_{\rm{C}} = K/\varepsilon$, $L_{\rm{C}}$: characteristic length scale of turbulence, $\omega_{\rm{F}}$: angular velocity of the Sun]. If we eliminate $B_{\rm{T}}$ and $B_{\rm{P}}$ from these equations, the equation for the cross helicity can be written as  
\begin{equation}
	\frac{d^2 \gamma^\ast}{dt^2}
	= - \omega_r^2 \gamma^\ast
	\label{eq:toy_gamma_eq}
\end{equation}
with 
\begin{equation}
	\omega_r = \sqrt{\alpha^\ast \delta^\ast U_\phi}.
\end{equation}
Equation ~(\ref{eq:toy_gamma_eq}) shows a simple sinusoidally oscillation of the cross helicity with the reversal frequency of $\omega_r$.

\begin{figure}
\begin{center}
\includegraphics[width=.65\textwidth]{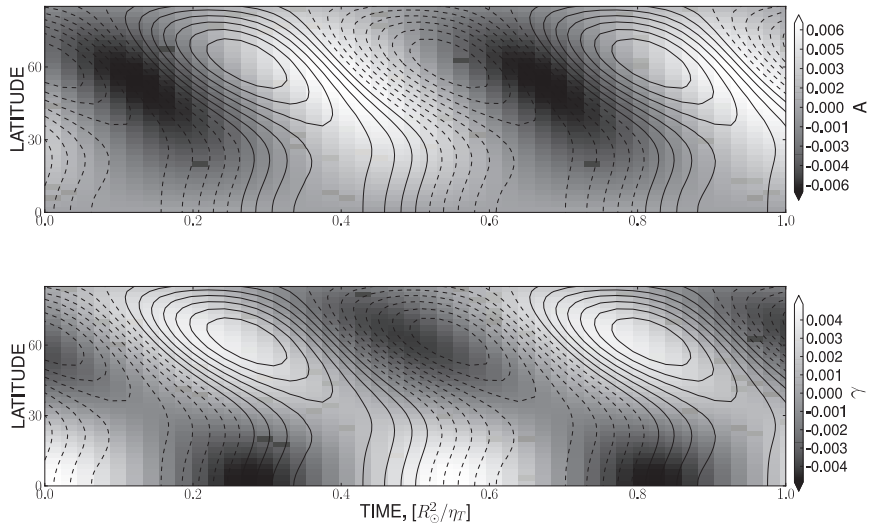}%
\caption{Temporal evolution of the magnetic fields and cross helicity. The spatiotemporal distributions of the poloidal magnetic field (upper) and the turbulent cross helicity (lower) are plotted in gray scale. The toroidal field is expressed as contours. Courtesy of Valery Pipin.}%
\label{fig:butterfly_ch}
\end{center}
\end{figure}

\paragraph{(ii) Models with cross-helicity evolution equation}
	Some other attempts have been started for treating more elaborated model equations. \citet{kuz2007} and \citet{pip2011b} solved a model transport equation of the turbulent cross helicity as well as the equations for the toroidal and poloidal magnetic fields. Here, as an example, we introduce a recent result by \citet{pip2011c}. If we construct model equations in a spherical coordinate system with only the latitudinal dependence retained, the equations for the toroidal field $B$, the toroidal component of the vector potential representing the poloidal field $A$, and the turbulent cross helicity $\gamma$, are given as
\begin{subequations}
\begin{equation}
	\frac{\partial B}{\partial t}
	= \frac{\partial}{\partial \theta} \frac{1}{\sin \theta} 
		\frac{\partial (B \sin\theta)}{\partial \theta}
	- 2 \gamma C_\gamma {\cal{D}} (x \sin\theta + 1) f(\theta),
\end{equation}
\begin{equation}
	\frac{\partial A}{\partial t}
	= \cos \theta B
	+ \frac{1}{\sin\theta} \frac{\partial}{\partial \theta}
		\frac{1}{\sin\theta} \frac{\partial A \sin\theta}{\partial \theta},
\end{equation}
\begin{equation}
	\frac{\partial \gamma}{\partial t}
	= - \frac{\xi}{\sin\theta} \frac{\partial A \sin\theta}{\partial\theta}
	+ \frac{1}{\sin\theta} \frac{\partial}{\partial \theta} 
		\sin\theta \frac{\partial \gamma}{\partial \theta},
\end{equation}
\end{subequations}
where ${\cal{D}}$ is the dynamo number, $f(\theta) = \partial \Omega / \partial x$ the radial derivative of the shear, $\xi$ the stratification parameter (varies from 5 at the bottom to 30), $C_\gamma$ the model constant related to the cross helicity generation. Here we omit the detailed description of the model equation. Using this system of equations, we can reproduce a periodic behaviour of the toroidal and poloidal magnetic field with the oscillation behaviour of the turbulent cross helicity without resorting to the mean-velocity shear or so-called $\Omega$ effect term (Figure~\ref{fig:butterfly_ch}).

	In the same line of thought, another simple equation can be also proposed. We write a system of equations in a local Cartesian coordinate sytem $(x,y,z)$ ($x$: colatitude, $y$: azimuthal, $z$: radial directions). For the sake of simplicity, we drop the azimuthal and radial dependence of the field quantities ($\partial/\partial y = \partial/\partial z = 0$). We assume that the mean velocity has only azimuthal component ${\bf{U}} = (U^x, U^y, U^z) = (0, U(x),0)$ and its latitudinal profile is prefixed (kinematic treatment). As for the turbulent transport coefficients, $\alpha$ and $\beta$ are treated as parameter (no spatial dependence), but the evolution equation for the cross-helicity-related coefficient $\gamma$ is solved. Under these assumptions and approximations, equations for the toroidal magnetic field $B^y (\equiv B)$, toroidal component of the vector potential $A^y (\equiv A)$, and $\gamma$ are written as
\begin{subequations}
\begin{equation}
	\frac{\partial B}{\partial t}
	= \beta \frac{\partial^2 B}{\partial x^2}
	+ \left( {\nabla \times \gamma \mbox{\boldmath$\Omega$}} \right)^y
	= \beta \frac{\partial^2 B}{\partial x^2}
	- \frac{\partial^2 U}{\partial x^2} \gamma
	- \frac{\partial U}{\partial x} \frac{\partial \gamma}{\partial x},
	\label{eq:simple_B_eq}
\end{equation}
\begin{equation}
	\frac{\partial A}{\partial t}
	= \beta \frac{\partial^2 A}{\partial x^2}
	+ \alpha B,
	\label{eq:simple_A_eq}
\end{equation}
\begin{equation}
	\frac{\partial \gamma}{\partial t}
	= \beta \frac{\partial^2 \gamma}{\partial x^2}
	- \alpha \tau {\bf{B}} \cdot \mbox{\boldmath$\Omega$}
	+ \beta \tau {\bf{J}} \cdot \mbox{\boldmath$\Omega$}
	= \beta \frac{\partial^2 \gamma}{\partial x^2}
	- \alpha \tau \frac{\partial U}{\partial x} \frac{\partial A}{\partial x}
	+ \beta \tau \frac{\partial U}{\partial x} \frac{\partial B}{\partial x},
	\label{eq:simple_gamma_eq}
\end{equation}
\end{subequations}
where $\tau$ is the timescale of turbulence. Again, apart from the diffusion term related to $\beta$ (the first term), we retain only the cross-helicity or $\gamma$ effect (the second and third terms) and dropped the $\alpha$ and $\Omega$ effects in the toroidal magnetic-field equation [Eq.~(\ref{eq:simple_B_eq})]. In the poloidal magnetic-field or vector-potential equation [Eq.~(\ref{eq:simple_A_eq})], we only retain the $\alpha$ effect (the second term) in addition to the diffusion or $\beta$-related term (the first term). As for the equation of the cross-helicity-related coefficient [Eq.~(\ref{eq:simple_gamma_eq})], we retain the $\alpha$-related reduction term (the second term) and the cross-helicity generation term (third term) in addition to the diffusion term (the first term). This linear system of equations is solved as an eigenvalue problem. The result will be reported in the forthcoming paper \citep{sch2012}.

\section{Flow generation\label{sec:flow_generation}}
\subsection{Cross-helicity effects in the momentum equation\label{sec:flow_dynamo}}
	The cross-helicity effects appear in the Reynolds stress $\mbox{\boldmath$\cal{R}$}$ [Eq.~(\ref{eq:Re_strss_exp})] in the mean momentum equation [Eq.~(\ref{eq:mean_U_eq_rot})], as well as in the turbulent electromotive force ${\bf{E}}_{\rm{M}}$ [Eq.~(\ref{eq:E_M_exp})] in the mean induction equation [Eq.~(\ref{eq:mean_B_eq})]. Substituting Eq.~(\ref{eq:Re_strss_exp}) into Eq.~(\ref{eq:mean_U_eq_rot}), we have
\begin{equation}
	\frac{\partial {\bf{U}}}{\partial t}
	= {\bf{U}} \times {\bf{\Omega}}
	+ {\bf{J} \times {\bf{B}}}
	+ \nu_{\rm{K}} \nabla^2 \left( {{\bf{U}} - \frac{\gamma}{\beta} {\bf{B}}} \right)
	+ {\bf{F}}
	- \nabla \left( {
		P + \frac{1}{2} {\bf{U}}^2 
		+ \left\langle {\frac{1}{2} {\bf{b}}'{}^2} \right\rangle
		+ \frac{2}{3} K_{\rm{R}}
	} \right).
	\label{eq:mean_U_eq_exp}
\end{equation}
Note that the $\gamma$-related term in Eq.~(\ref{eq:mean_U_eq_exp}), $- \nu_K \nabla^2 (\gamma {\bf{B}} / \beta)$, comes from the third term in the Reynolds stress expression [Eq.~(\ref{eq:Re_strss_exp})], $\nu_{\rm{M}} \mbox{\boldmath${\cal{M}}$}$. This suggests that the coupling of the turbulent cross helicity and the magnetic-field strain may effectively suppress the eddy-viscosity effect $\nu_{\rm{K}}$.

	The mean Ohm's law is written as
\begin{equation}
	{\bf{J}}
	= \sigma \left( {
	{\bf{E}}
	+ {\bf{U}} \times {\bf{B}} 
	+ {\bf{E}}_{\rm{M}}
	} \right).
	\label{eq:mean_Ohms_law}
\end{equation}
If we substitute Eq.~(\ref{eq:E_M_exp}) into Eq.~(\ref{eq:mean_Ohms_law}), and solve it with respect to ${\bf{J}}$, we have
\begin{equation}
	{\bf{J}}
	= \frac{1}{\beta} \left( {
	{\bf{U}} \times {\bf{B}} 
	+ \alpha {\bf{B}}
	+ \gamma \mbox{\boldmath$\Omega$} 
	- \frac{\partial {\bf{A}}}{\partial t}
	- \nabla \varphi
	} \right)
	\label{eq:mean_J_exp}
\end{equation}
(${\bf{A}}$: vector potential, $\varphi$: electrostatic potential). Note that $\eta {\bf{J}}$ was dropped as compared with $\beta {\bf{J}}$ since $\eta \ll \beta$. However we should keep in mind the discussions extended in the final part of \S~\ref{sec:ch_effect}.
It follows from Eq.~(\ref{eq:mean_J_exp}) that the mean-field Lorentz force ${\bf{J}} \times {\bf{B}}$ is expressed as
\begin{equation}
	{\bf{J}} \times {\bf{B}}
	= \frac{1}{\beta} 
	\left( {{\bf{U}} \times {\bf{B}}} \right) \times {\bf{B}}
	+ \frac{\gamma}{\beta}
	\mbox{\boldmath$\Omega$} \times {\bf{B}}
	- \frac{1}{\beta} \left( {
		\frac{\partial {\bf{A}}}{\partial t} + \nabla \varphi 
	} \right) \times {\bf{B}}.
	\label{eq:mean_Lorentz_force_exp}
\end{equation}
Note that the $\alpha$-related term has no contribution to Eq.~(\ref{eq:mean_Lorentz_force_exp}) since the $\alpha$ effect gives ${\bf{J}}$ parallel to ${\bf{B}}$. Substituting Eq.~(\ref{eq:mean_Lorentz_force_exp}) into Eq.~(\ref{eq:mean_U_eq_exp}), and taking the curl operation, we obtain the mean vorticity equation as
\begin{eqnarray}
	\frac{\partial {\bf{\Omega}}}{\partial t}
	&=& \nabla \times \left[ {
	\left( {
	{\bf{U}} - \frac{\gamma}{\beta} {\bf{B}}
	} \right)
	\times {\bf{\Omega}}
	+ \nu_{\rm{K}} \nabla^2 \left( {
	{\bf{U}} - \frac{\gamma}{\beta} {\bf{B}}
	} \right)
	} \right]\nonumber\\
	&+& \nabla \times \left[ {
	{\bf{F}}
	+ \frac{1}{\beta} 
	\left( {{\bf{U}} \times {\bf{B}}} \right) \times {\bf{B}}
	- \frac{1}{\beta} \left( { 
		\frac{\partial {\bf{A}}}{\partial t} + \nabla \varphi 
	} \right) \times {\bf{B}}
	} \right].
	\label{eq:mean_Omega_eq_exp}
\end{eqnarray}
This equation is fully utilized in the following examples.

\subsection{Plasma rotation in internal-transport-barrier mode in tokamaks}
	We consider the reversed magnetic shear confinement or reversed shear (RS) mode in tokamaks, where an internal transport barrier (ITB) is formed in the core region of plasmas. In the RS mode, a global plasma rotation in the poloidal direction is observed associated with the ITB formation. Here we address the generation of the poloidal rotation in the RS mode from the viewpoint of the turbulent dynamo \citep{yos1999,yok2008}. For a more general treatment of this phenomena the reader is referred to \citet{dia2010} and references therein.
	
\begin{figure}[htpb]
\begin{center}
\includegraphics[width=.70\textwidth]{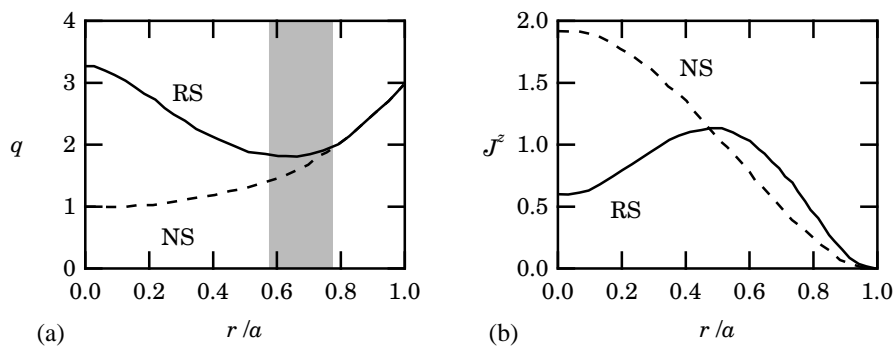}%
\caption{(a) Safety factor and (b) plasma current in the RS and NS (normal shear) modes.}%
\label{fig:rs_q_jz}
\end{center}
\end{figure}

	In the RS mode in tokamaks, the safety factor $q$ defined by the ratio of toroidal and poloidal twist numbers as
\begin{equation}
	q = \frac{r}{R} \frac{B^\phi}{B^\theta}
	\label{eq:safety_factor}
\end{equation}
($R$: major radius, $B^\phi$: toroidal magnetic field, $B^\theta$: poloidal magnetic field) shows a radial profile whose minimum is located in the core region of plasma [Figure~\ref{fig:rs_q_jz}(a)]. Such a radial profile of $q$ corresponds to the hollow radial profile of the plasma current in the RS mode [Figure~\ref{fig:rs_q_jz}(b)]. Namely, the plasma current $J^\phi (= J^z\; \mbox{in cylindrical approximation})$ shows a local minimum in the center of plasma ($r/a = 0$, $a$: minor radius).

	We approximate a torus by a cylinder (cylinder approximation) with a cylindrical  coordinate system ($r, \theta, z$) ($\theta$: poloidal direction, $z$: toroidal direction). We assume that the physical quantities depend only on the radius $r$ ($\partial / \partial \theta = \partial / \partial z = 0$). In this situation, we see from Eq.~(\ref{eq:mean_Omega_eq_exp}) that the toroidal or $z$ component of the mean vorticity, $\Omega^z$, obey
\begin{equation}
	\frac{\partial \Omega^z}{\partial t}
	= \nu_{\rm{K}} \nabla^2 \Omega^z - \nu_{\rm{M}} \nabla^2 J^z.
	\label{eq:RS_Omega_eq}
\end{equation}
Here we assumed that the spatial variation of $\nu_{\rm{M}} / \nu_{\rm{K}} = \gamma / \beta$ can be neglected. Equation~(\ref{eq:RS_Omega_eq}) shows that in the absence of turbulent cross helicity ($\nu_{\rm{M}} = \gamma = 0$) $\Omega^z$ is subject to only the decaying process due to the turbulent viscosity $\nu_{\rm{K}}$. In contrast, in the presence of the turbulent cross helicity, mean vorticity can be generated.

	The turbulent cross helicity $W$ is generated by the production term $P_W$ [Eq.~(\ref{eq:P_W_def})]. At the early stage of plasma rotation, where $|\mbox{\boldmath$\Omega$}|$ is small, the production rate can be expressed as
\begin{equation}
	P_W \simeq \beta {\bf{J}} \cdot \mbox{\boldmath$\Omega$}.
	\label{eq:RS_P_W}
\end{equation}
In this case, the cross-helicity evolution is subject to
\begin{equation}
	\frac{\partial W}{\partial t}
	= C_\beta \frac{K^2}{\varepsilon} J^z \Omega^z + \cdots.
	\label{eq:RS_W_eq}
\end{equation}

\begin{figure}[htpb]
\begin{center}
\includegraphics[width=.35\textwidth]{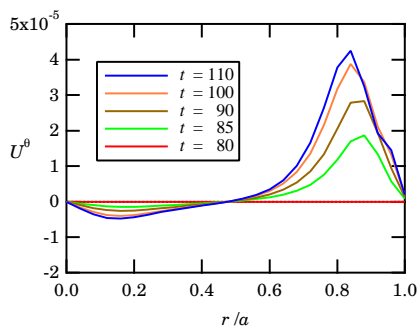}%
\caption{Poloidal flow induced by the cross helicity.}%
\label{fig:rs_pol_flow}
\end{center}
\end{figure}

	From Eqs.~(\ref{eq:RS_Omega_eq}) and (\ref{eq:RS_W_eq}), we obtain
\begin{equation}
	\frac{\partial^2 \Omega^z}{\partial t^2}
	- \left( {
		- \frac{5}{7} C_\beta C_\gamma 
			\frac{K^3}{\varepsilon^2} J^z \nabla^2 J^z
	} \right) \Omega^z
	= \cdots.
	\label{eq:RS_Omega_eq_fin}
\end{equation}
This suggests that the mean vorticity $\Omega^z$ increases with the growth rate $\chi_\Omega^2$ if
\begin{equation}
	\chi_\Omega^2
	= - \frac{5}{7} C_\beta C_\gamma  \frac{K^3}{\varepsilon^2} J^z \nabla^2 J^z
	> 0.
	\label{eq:RS_chi_def}
\end{equation}
From the radial distribution of the toroidal mean electric-current density $J^z$ shown in Figure~\ref{fig:rs_q_jz}(b), the radial distribution of $J^z \nabla^2 J^z$ can be calculated. In the region near $r/a \simeq 0.6$, $J^z \nabla^2 J^z < 0$, leading to a positive $\chi_\Omega^2$ in this region. This suggests that in the core region the mean toroidal vorticity will increases. For given magnetic-field profiles corresponding  to the radial profiles of the safety factor $q$ [Figure~\ref{fig:rs_q_jz}(a)] and the hollow mean electric current [Figure~\ref{fig:rs_q_jz}(b)], the mean-vorticity equation is numerically solved simultaneously with the evolution equations of the turbulent MHD energy $K$, its dissipation rate $\varepsilon$, and the turbulent cross helicity $W$. The poloidal velocity profiles are plotted in Figure~\ref{fig:rs_pol_flow}. We see in the core region that the poloidal velocity is generated and increases as time goes by. This numerical result confirms that the presence of turbulent cross helicity coupled with a mean magnetic-field reversed shear configuration causes a poloidal rotation in the core region.

\subsection{Torsional oscillation inside the Sun}
	Thanks to the remarkable developments in helioseismology research, the detailed configurations of plasma motions inside the Sun have been revealed with amazing accuracy in the past two decades. One of the most interesting features obtained by helioseismology is the torsional oscillation in the solar convective zone. The azimuthal or rotational motion inside the Sun shows oscillatory properties. This periodic motion shows a similarity in pattern with the solar magnetic activity. A typical period of oscillatory motion is a few years and the magnitude of the oscillating azimuthal velocity is 10 ${\rm{m}}\ {\rm{s}}^{-1}$. 
	
	In the context of the torsional oscillation of the Sun, the importance of the feedback effect due to the mean-field Lorentz force ${\bf{J}} \times {\bf{B}}$ (${\bf{B}} = \langle{\bf{b}}\rangle$, ${\bf{J}} = \nabla \times {\bf{B}}$) has been pointed out \citep{mal1975}. By investigating the turbulent transport through the Reynolds (and turbulent Maxwell) stress, \citet{rue1990} proposed the so-called $\Lambda$-effect quenching mechanism, where the balance between the turbulent transport and the mean-field effect is supposed to occur.
	
	Here we address this torsional oscillation phenomenon from the viewpoint of {\it flow dynamo}: flow generation due to the cross-helicity effect \citep{ito2005}.
	
	In order to extract the cross-helicity effects in the momentum equation, we divide the mean velocity and vorticity as
\begin{equation}
	{\bf{U}} = {\bf{U}}_0 + \delta {\bf{U}},\;\;
	\mbox{\boldmath$\Omega$} 
	= \mbox{\boldmath$\Omega$}_0 + \delta \mbox{\boldmath$\Omega$},
	\label{eq:U_ref_modulation}
\end{equation}
where ${\bf{U}}_0$ and $\mbox{\boldmath$\Omega$}_0$ are the reference fields without the cross-helicity effect and $\delta{\bf{U}}$ and $\delta\mbox{\boldmath$\Omega$}$ are the modulation fields due to the cross-helicity effect. Substituting Eq.~(\ref{eq:U_ref_modulation}) into the mean vorticity equation in a rotating frame, we obtain the equations for the reference and modulation mean vorticities as
\begin{equation}
	\frac{\partial {\bf{\Omega}}_0}{\partial t}
	= \nabla \times \left[ {
		{\bf{U}}_0 \times 2\mbox{\boldmath$\omega$}_{\rm{F}}
		+ \nu_{\rm{K}} \nabla^2 {\bf{U}_0}
		+ {\bf{F}}
		- \frac{1}{\beta} \left( {
			\frac{\partial {\bf{A}}}{\partial t} + \nabla \varphi
		} \right) \times {\bf{B}}
	} \right],
	\label{eq:Omega_0_eq}
\end{equation}
\begin{equation}
	\frac{\partial \delta{\bf{\Omega}}}{\partial t}
	= \nabla \times \left[ {
		\left( {\delta{\bf{U}} - \frac{\gamma}{\beta} {\bf{B}}} \right)
		\times 2 \mbox{\boldmath$\omega$}_{\rm{F}}
		+ \nu_{\rm{K}} \nabla^2 \left( {
		\delta{\bf{U}} - \frac{\gamma}{\beta} {\bf{B}}
		} \right)
	} \right].
	\label{eq:delta_Omega_eq}
\end{equation}
Here we dropped the mean vorticity $\mbox{\boldmath$\Omega$}$ as it is small compared with the system rotation $\mbox{\boldmath$\omega$}_{\rm{F}}$.

	If the time scale of the turbulent viscosity is rapid enough compared the temporal evolution of the periodic motion as
\begin{equation}
	\left| {\frac{\partial \delta \mbox{\boldmath$\Omega$}}{\partial t}} \right|
	\ll \left| {\nu_{\rm{K}} \nabla^2 \delta \mbox{\boldmath$\Omega$}} \right|,
	\label{eq:rapid_diffusion}
\end{equation}
we see from Eq.~(\ref{eq:delta_Omega_eq}) that the modulation velocity
\begin{equation}
	\delta {\bf{U}} = \frac{\gamma}{\beta} {\bf{B}}
	\label{eq:delta_U_sol}
\end{equation}
is a particular solution of Eq.~(\ref{eq:delta_Omega_eq}) in the stationary state in that Eq.~(\ref{eq:rapid_diffusion}) is satisfied.

	This solution indicates that the mean velocity is modulated by the mean magnetic field in the presence of the turbulent cross helicity. If we rewrite Eq.~(\ref{eq:delta_U_sol}) in physical units as
\begin{equation}
	\delta {\bf{U}} = \frac{\gamma}{\beta} {\bf{B}}
	= \frac{\gamma}{\beta} \frac{{\bf{B}}_\ast}{\sqrt{\mu \rho}},
	\label{eq:delta_U_sol_phys_unit}
\end{equation}
we see the following features of this solution:
\begin{enumerate}
\item The pattern of the periodic change of differential rotation follows the pattern of the solar magnetic cycle;
\item The flow oscillates in time;
\item The direction and magnitude of the flow change according to the changes of $W/K$ and ${\bf{B}}$;
\item $\delta{\bf{U}}$ is larger near the surface where the density $\rho$ is smaller.
\end{enumerate}

	As for the solar parameters, we adopt the number density of hydrogen as $O(10^{28}) {\rm{m}}^{-3}$ and the magnitude of magnetic field as $|{\bf{B}}| = 1 {\rm{T}}$ at the location of the relative solar radius of $r/R_{\odot} = 0.8-0.9$. If we assume the turbulent cross helicity scaled by the turbulent MHD energy to be of the order
\begin{equation}
	\left| {{W}/{K}} \right| = O(10^{-1.5}),
\end{equation}	
we obtain
\begin{equation}
	|\delta {\bf{U}}| \sim 10\ {\rm{m\ s}}^{-1},
	\label{eq:delta_U_estimate}
\end{equation}
which agrees with the result obtained by helioseismology. If the value of $|W/K| = O(10^{-4})$ or smaller, the estimate Eq.~(\ref{eq:delta_U_sol_phys_unit}) gives too small $|\delta {\bf{U}}|$. In such a case, the cross-helicity effect is not relevant for the torsional oscillation.

	In addition, Eq.~(\ref{eq:delta_U_sol}) cannot be applied to the case in which the phase difference between the magnetic and flow pattern is large. Here we should note that the expression for the modulation velocity [Eq.~(\ref{eq:delta_U_sol})] is time independent in the meaning of Eq.~(\ref{eq:rapid_diffusion}). In order to treat the phase difference between the magnetic and flow patterns, we have to consider the higher-order part of the modulation velocity that responds to the temporal variation of the turbulent cross helicity and the mean magnetic field. A report of such investigation is in preparation \citep{yok2012a}.
	
	The present mechanism using the cross-helicity effect is similar to the previous work in that it consider both the feedback due to the mean-field Lorentz force ${\bf{J}} \times {\bf{B}}$ and the turbulent transport through the Reynolds (and turbulent Maxwell) stress. The balance between the structure destruction and generation is considered to play an essential role in the torsional oscillation.
	
	The main difference between the present and previous work lies in the point: the former considers the mean-velocity effect through the turbulent cross helicity whereas the latter does not. In the present scenario, the effect of mean vortical motions on the magnetic-field generation is considered through the cross-helicity effect term in Eq.~(\ref{eq:E_M_exp}), $\gamma \mbox{\boldmath$\Omega$}$, and the mean vorticity effect on the momentum equation, coming from the $\mbox{\boldmath$\Gamma$} \mbox{\boldmath$\Omega$}$ terms in Eq.~(\ref{eq:Re_strss_exp}), is negligible as compared with the cross-helicity and mean magnetic strain term, $\nu_{\rm{M}} \mbox{\boldmath${\cal{M}}$}$. This may be expressed as
\begin{subequations}\label{eq:tors_osc_ch}
\begin{eqnarray}
	{\cal{R}}^{\alpha\beta}&:=& 
	- \nu_{\rm{K}} {\cal{S}}^{\alpha\beta}
	+ \nu_{\rm{M}} {\cal{M}}^{\alpha\beta},
	\label{eq:tors_osc_ch_re}\\
	{\bf{E}}_{\rm{M}}&:=&
	- \beta {\bf{J}}
	+ \gamma \mbox{\boldmath$\Omega$}
	\label{eq:tors_osc_ch_emf}
\end{eqnarray}
\end{subequations}
(``$:=$'' denotes ``is schematically expressed by'').

	On the other hand, in the previous work using the $\Lambda$-effect quenching, the cross-helicity effect is neglected. It may be schematically expressed as
\begin{subequations}\label{eq:tors_osc_lambda}
\begin{eqnarray}
	{\cal{R}}^{\alpha\beta}&:=& 
	- \nu_{\rm{K}} {\cal{S}}^{\alpha\beta}
	+ [\mbox{\boldmath$\Gamma$} \mbox{\boldmath$\Omega$}]^{\alpha\beta},
	\label{eq:tors_osc_lambda_re}\\
	{\bf{E}}_{\rm{M}}&:=&
	- \beta {\bf{J}}
	+ \alpha {\bf{B}}
	\label{eq:tors_osc_lambda_emf}
\end{eqnarray}
\end{subequations}
[The symbol ``$\mbox{\boldmath$\Lambda$}$'' in the $\Lambda$ effect is replaced by ``$\mbox{\boldmath$\Gamma$}$'' following our notation in Eq.~(\ref{eq:Re_strss_exp})]. This difference is reflected by the point that the mean velocity ${\bf{U}} = \langle {\bf{u}} \rangle$ is entirely neglected in the basic equations for the velocity and magnetic field in \citet{rue1990}. 

	In the TSDIA formalism, the $\mbox{\boldmath$\Gamma$} \mbox{\boldmath$\Omega$}$ [the last three terms in Eq.~(\ref{eq:Re_strss_exp})] arises from the higher-order [$O(\delta^2)$] calculation. This is the reason why we drop it at the first stage of research. If we have no cross helicity at all, the third term or $\nu_{\rm{M}} {\cal{M}}^{\alpha\beta}$ vanishes, so we have to retain the $\mbox{\boldmath$\Gamma$} \mbox{\boldmath$\Omega$}$ as a first candidate for balancing the turbulent viscosity effect $\nu_{\rm{K}}$. Actually this is the case in hydrodynamic turbulence \citep{yok1993}.

	The key question is which is the dominant effect in the torsional oscillation: cross helicity or helicity? As mentioned above, if we have the turbulent cross helicity normalized by the turbulent MHD energy of $|W|/K = O(10^{-1}) - O(10^{-2})$, the cross helicity effect may be relevant. Numerical experiments with realistic parameters using a more generalized form:
\begin{subequations}\label{eq:tors_osc_full}
\begin{eqnarray}
	{\cal{R}}^{\alpha\beta}&:=& 
	- \nu_{\rm{K}} {\cal{S}}^{\alpha\beta}
	+ \nu_{\rm{M}} {\cal{M}}^{\alpha\beta}
	+ [\mbox{\boldmath$\Gamma$} \mbox{\boldmath$\Omega$}]^{\alpha\beta},
	\label{eq:tors_osc_full_re}\\
	{\bf{E}}_{\rm{M}}&:=&
	- \beta {\bf{J}}
	+ \gamma \mbox{\boldmath$\Omega$}
	+ \alpha {\bf{B}}
	\label{eq:tors_osc_full_emf}
\end{eqnarray}
\end{subequations}
would be an interesting subject.

\subsection{Flow--turbulence interaction in magnetic reconnection}
	In order to get efficient magnetic reconnection, we need enhanced magnetic diffusivity. We also need some mechanism that will bridge the scale gap between the diffusion region of the magnetic field and the typical scale of the system where magnetic reconnection occurs. Turbulence is considered to be one of the candidates that contributes to the fast and localized reconnection process \citep{mat1985,mat1986,laz1999}. From the viewpoint of the cross-helicity effects, turbulent magnetic reconnection is a very interesting phenomenon where both of the effects in magnetic-field and flow generations play an essential role.

	If we substitute the electromotive force expression ${\bf{E}}_{\rm{M}}$ [Eq.~(\ref{eq:E_M_exp})] with the $\alpha$-related term dropped into the mean magnetic-field induction equation (\ref{eq:mean_B_eq}), we get Eq.~(\ref{eq:mean_ind_eq_beta_gamma}).
	
	To extract the magnetic field intrinsic to the cross-helicity effect, we divide the mean magnetic field and electric-current density as
\begin{equation}
	{\bf{B}} = {\bf{B}}_0 + \delta {\bf{B}},\;\;\;
	{\bf{J}} = {\bf{J}}_0 + \delta {\bf{J}},
	\label{eq:B_ref_modulation}
\end{equation}
where ${\bf{B}}_0$ and ${\bf{J}}_0$ are the reference fields without the cross-helicity effect, and $\delta{\bf{B}}$ and $\delta{\bf{J}}$ are the modulation fields due to the cross-helicity effect. Substituting Eq.~(\ref{eq:B_ref_modulation}) into Eq.~(\ref{eq:mean_ind_eq_beta_gamma}), we obtain
\begin{equation}
	\frac{\partial {\bf{B}}_0}{\partial t}
	= \nabla \times \left( {{\bf{U}} \times {\bf{B}}_0} \right)
	- \nabla \times \left( {\beta \nabla \times {\bf{B}}_0} \right)
\end{equation}
for the reference field, and
\begin{equation}
	\frac{\partial \delta{\bf{B}}}{\partial t}
	= \nabla \times \left( {{\bf{U}} \times \delta{\bf{B}}} \right)
	- \nabla \times \left( {
		\beta \nabla \times \delta{\bf{B}} 
		- \frac{\gamma}{\beta} \nabla \times {\bf{U}}
	} \right)
	\label{eq:B_modulation_eq}
\end{equation}
for the modulation field. Equation~(\ref{eq:B_modulation_eq}) has a particular solution
\begin{equation}
	\delta{\bf{B}} = \frac{\gamma}{\beta} {\bf{U}}
	\label{eq:B_modulation_sol}
\end{equation}
for a stationary state. Here we should note Eq.~(\ref{eq:B_modulation_sol}) is not so bad approximation even for the variable $\beta$ and $\gamma$ case. Reviewing the relation:
\begin{equation}
	\nabla \times \left( {\frac{\gamma}{\beta} {\bf{U}}} \right)
	= \frac{\gamma}{\beta} \nabla \times {\bf{U}}
	+ \nabla \left( {\frac{\gamma}{\beta}} \right) \times {\bf{U}},
	\label{eq:grad_gamma_beta}
\end{equation}
we see Eq.~(\ref{eq:B_modulation_sol}) holds as long as the magnitude of $\nabla (\gamma / \beta) \times {\bf{U}}$ is not so critical. Equation~(\ref{eq:B_modulation_sol}) shows that in the presence of the turbulent cross helicity, we have a modulation of the mean magnetic field aligned with the mean velocity and  the proportional coefficient is given by the scaled turbulent cross helicity. If the sign of the turbulent cross helicity is positive (or negative), we have a modulation magnetic field parallel (or antiparallel) to the mean velocity.

	In a similar manner, we consider the momentum equation. If we substitute Eq.~(\ref{eq:U_ref_modulation}) into the mean vorticity equation (\ref{eq:mean_Omega_eq_exp}), we obtain
\begin{equation}
	\frac{\partial {\bf{\Omega}}_0}{\partial t}
	= \nabla \times \left[ {
		{\bf{U}}_0 \times \mbox{\boldmath$\Omega$}_0
		+ \nu_{\rm{K}} \nabla^2 {\bf{U}_0}
		+ {\bf{F}}
		- \frac{1}{\beta} \left( {
			\frac{\partial {\bf{A}}}{\partial t} + \nabla \varphi
		} \right) \times {\bf{B}}
	} \right]
	\label{eq:Omega_ref_eq}
\end{equation}
for the reference mean vorticity $\mbox{\boldmath$\Omega$}_0$, and
\begin{equation}
	\frac{\partial \delta{\bf{\Omega}}}{\partial t}
	= \nabla \times \left[ {
		\left( {\delta{\bf{U}} - \frac{\gamma}{\beta} {\bf{B}}} \right)
		\times \mbox{\boldmath$\Omega$}_0
		+ \nu_{\rm{K}} \nabla^2 \left( {
		\delta{\bf{U}} - \frac{\gamma}{\beta} {\bf{B}}
		} \right)
	} \right]
	\label{eq:Omega_modulation_eq}
\end{equation}
for the modulation vorticity $\delta \mbox{\boldmath$\Omega$}$. Equation~(\ref{eq:Omega_modulation_eq}) has a particular solution
\begin{equation}
	\delta{\bf{U}} = \frac{\gamma}{\beta} {\bf{B}}
	\label{eq:U_modulation_sol}
\end{equation}
for a stationary state. This shows, in the presence of the turbulent cross helicity, we have a modulation of the mean velocity aligned with the mean magnetic field, and the proportional coefficient is given by the scaled turbulent cross helicity. If the sign of the turbulent cross helicity is positive (or negative), we have a modulation velocity parallel (or antiparallel) to the mean magnetic field.

\begin{figure}[htpb]
\begin{center}
\begin{minipage}{95mm}
\subfigure[Typical mean-field configurations.]{
\resizebox*{40mm}{!}{\includegraphics{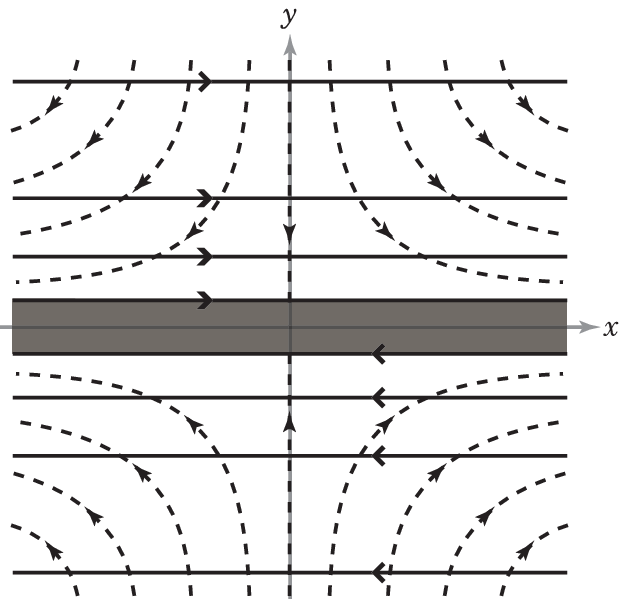}}}%
\hspace{15mm}
\subfigure[Spatial distribution of the turbulent cross helicity.]{
\resizebox*{40mm}{!}{\includegraphics{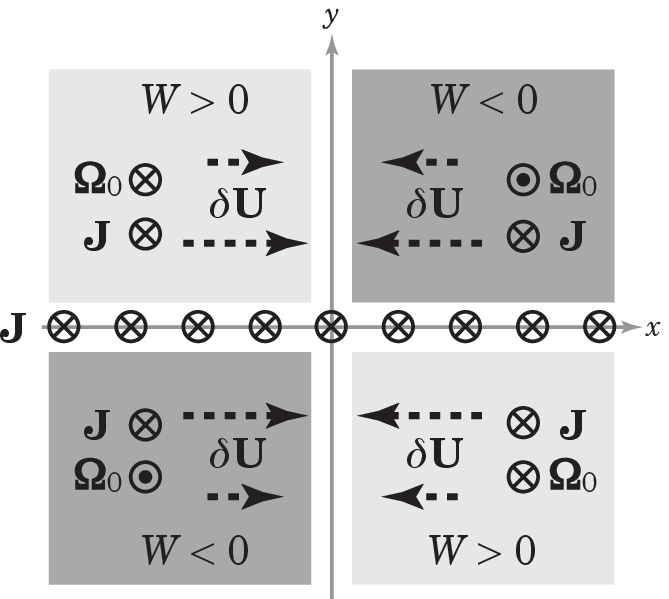}}}%
\caption{Mean and turbulent fields in magnetic reconnection.}%
\label{fig:mr_mean_turb}
\end{minipage}
\end{center}
\end{figure}

	Evolution of the turbulent cross helicity is subject to Eq.~(\ref{eq:K_W_eq}) with Eq.~(\ref{eq:W_P_eps_T}). As the production rate [Eq.~(\ref{eq:P_W_def})] shows, the spatial distribution of the turbulent cross helicity is determined by the mean-field configurations such as the combination of the mean velocity and magnetic strains, $\mbox{\boldmath$\cal{S}$}:\mbox{\boldmath$\cal{M}$}$ [Eq.~(\ref{eq:P_W_Rey_est})] and the combination of the mean vorticity and electric-current density, ${\bf{J}} \cdot \mbox{\boldmath$\Omega$}$ [Eq.~(\ref{eq:P_W_Em_est})] [Figure~\ref{fig:mr_mean_turb}(a)]. Considering such mean-field configurations around the magnetic reconnection, we see that the spatial distribution of the turbulent cross helicity is the quadrupole-like configuration [Figure~\ref{fig:mr_mean_turb}(b)].
	
	If we combine the modulation fields [Eqs.~(\ref{eq:U_modulation_sol}) and (\ref{eq:B_modulation_sol})] with the quadrupole-like spatial distribution of the turbulent cross helicity, we have a converging-type flow and a X-point-like magnetic-field configuration, which is favorable for fast reconnection. The basic role of the cross-helicity effect is balancing and suppressing the effect of turbulent magnetic diffusivity. However, the turbulent cross helicity is spatially distributed with positive and negative values and vanishing at the symmetry surfaces. This pseudoscalar property makes the reconnection region very narrow and thin, which contributes to the fast reconnection.

\begin{figure}[htpb]
\begin{center}
\includegraphics[width=.45\textwidth]{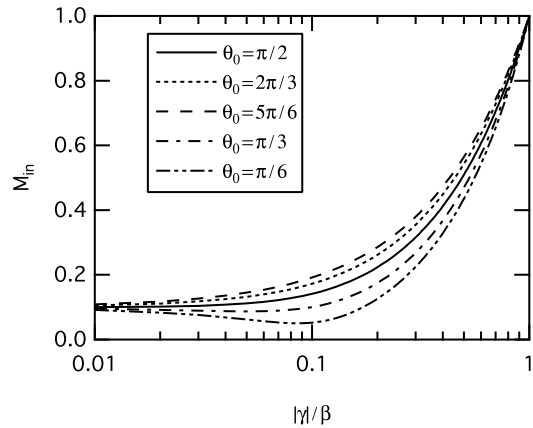}%
\caption{Magnetic reconnection rate $M_{\rm{in}}$ against the scaled turbulent cross helicity $\gamma / \beta$. $\theta_0$ is the angle between the reference inflow velocity and magnetic field.}%
\label{fig:mr_reconnection_rate}
\end{center}
\end{figure}

	Using Eqs.~(\ref{eq:U_modulation_sol}) and (\ref{eq:B_modulation_sol}), we can estimate the magnetic reconnection rate $M_{\rm{in}}$ as a function of the scaled turbulent cross helicity. In Figure~{\ref{fig:mr_reconnection_rate}}, we show how the magnetic reconnection rate $M_{\rm{in}}$ is enhanced by the turbulent cross helicity. For detailed arguments, the reader is referred to \citet{yok2011c}.

\section{Numerical tests\label{sec:numerical_tests}}
	In order to fully solve the mean-field dynamo equations under the combination of the helicity and cross-helicity effects without resorting to any approximate or perturbation methods, we have to utilize numerical simulation. Actually, what has been lacking in the study of cross-helicity dynamo is numerical tests of the basic notions. In this section, we present some results obtained by numerical simulations.
	
	\citet{bra1998} numerically solved the induction equation for the mean magnetic field with the turbulent electromotive force consisting of both the helicity and cross-helicity effects. They succeeded in explaining the rapid growth rate of the large-scale magnetic field in young galaxies, which the conventional helicity dynamo had failed to elucidate. However, the turbulence properties such as the profiles and magnitude of the turbulent diffusivity, helicity, etc.\ are presumed and fixed in their simulation. In this sense, the relationship between the helicity and cross-helicity effects still remains indeterminate. Recently, \citet{sur2009} examined the cross-helicity effect in the Archontis flow (a generalization of the Arnold--Beltrami--Childress flow). By performing direct numerical simulations in a situation with no helicity or $\alpha$ effect, they showed that a certain magnetic field can be generated genuinely by the cross-helicity effect. Performing a large-eddy simulation (LES) of magnetohydrodynamic (MHD) turbulence, \citet{ham2010} examined the turbulent electromotive force in a turbulent channel flow. With the aid of a subgrid-scale (SGS) model for LES, it was confirmed that the cross-helicity effect coupled with the large-scale vorticity plays a central role in producing the turbulent electromotive force and that the magnetic field is induced by the cross helicity dynamo in this case.  
	
	In the following, we present some other simulations which have been very recently performed.

\begin{figure}[htpb]
\begin{center}
\includegraphics[width=.50\textwidth]{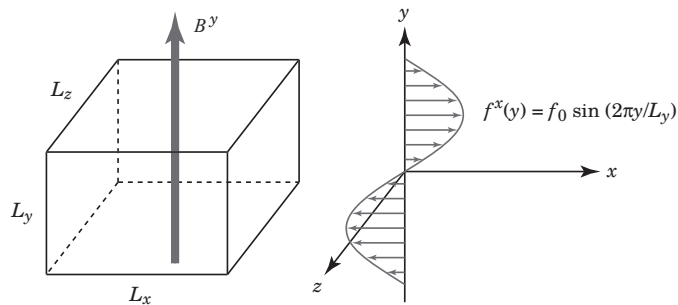}%
\caption{Kolmogorov flow with imposed uniform magnetic field (Left) and external forcing (Right).}%
\label{fig:kolmogorov_flow_setup}
\end{center}
\end{figure}

\paragraph{Kolmogorov flow with imposed magnetic field}
	For understanding basic properties of magnetohydrodynamic (MHD) flow, it is useful to consider a simple inhomogeneous flow configuration. Kolmogorov flow is a three-dimensional periodic flow with external forcing
\begin{equation}
	{\bf{f}} = \left( {f^x, f^y, f^z} \right)
	= \left( {f_0 \sin \frac{2\pi y}{L_y}, 0, 0} \right)
\end{equation}
($L_x$, $L_y$, $L_z$: box dimension). Due to the forcing, this flow is inhomogeneous in the $y$ direction, but homogeneous in $x$ and $z$ directions. Kolmogorov flow is known to be suitable for investigating three-dimensional inhomogeneous turbulent flow. Here, in order to examine basic MHD properties, we further impose a uniform magnetic field in the inhomogeneous or $y$ direction:
\begin{equation}
	{\bf{B}} = \left( {0, B_0, 0} \right)
\end{equation}
(Figure~\ref{fig:kolmogorov_flow_setup}). With this numerical setup, we perform a direct numerical simulation (DNS) to examine the turbulent electromotive force and its model \citep{yok2011b}. As for the statistics, we adopt the averaging over the homogeneous ($x$ and $z$) directions and ensemble  average over 70 independent realizations in time.

\begin{figure}[htpb]
\begin{center}
\includegraphics[width=.35\textwidth]{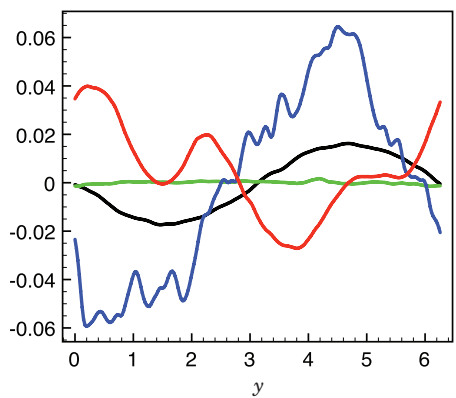}%
\caption{Turbulent electromotive force and its model in Kolmogorov flow. Spatial distribution of $\langle {{\bf{u}}' \times {\bf{b}}'} \rangle^z$ (black) is compared with each term in the turbulent electromotive force model, $\alpha B^z$ (green); $\beta J^z$ (blue); $\gamma \Omega^z$ (red).}%
\label{fig:kolmogorov_flow_emf}
\end{center}
\end{figure}

	A comparison of the turbulent electromotive force ${\bf{E}}_{\rm{M}} = \langle {{\bf{u}}' \times {\bf{b}}'} \rangle$ with each term of the model expression Eq.~(\ref{eq:E_M_exp}) is shown in Figure~\ref{fig:kolmogorov_flow_emf}. We see the $\alpha$-related term, $\alpha {\bf{B}}$, is negligibly small in the whole region of the flow. The main balance is held between the turbulent magnetic diffusivity or $\beta$-related term, $\beta {\bf{J}}$, and the turbulent cross-helicity or $\gamma$-related term, $\gamma \mbox{\boldmath$\Omega$}$, in this flow. This is because in this flow we have certain mechanisms for generating the turbulent cross helicity whereas there is no generation mechanisms for the turbulent residual helicity.

\begin{figure}[htpb]
\begin{center}
\includegraphics[width=.50\textwidth]{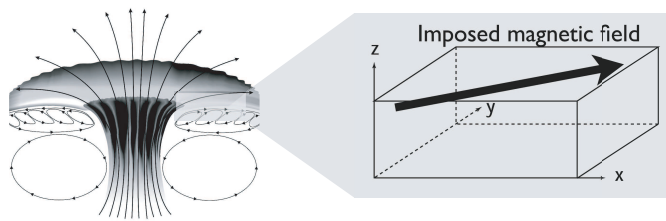}%
\caption{Numerical simulation mimicking the flow around a sunspot.}%
\label{fig:sunspot_flow_setup}
\end{center}
\end{figure}

\paragraph{Flow around the sunspot}
	How and how much turbulent cross helicity exists in real geo/astrophysical situations is a very important issue. With the aid of a realistic numerical simulations of the flow around the sunspot \citep{jac2008}, the spatial distribution of the turbulent cross helicity and its generation mechanisms are investigated \citep{yok2012b}. We consider a rectangular box mimicking a local flow region around a sunspot (Figure~\ref{fig:sunspot_flow_setup}). The depth of the box corresponds to the depth of the local convection zone (region of $0$ to $5\ {\rm{Mm}}$ from the solar surface).  A large-scale magnetic field inclined by 85$^\circ$ toward the surface (almost horizontal) is imposed. We perform numerical simulations with different magnetic-field strengths (600, 1200, and 1500 G). We assume the system to be periodic in the horizontal ($x$-$y$) directions. All the statistics is made using the horizontal plane average.

\begin{figure}[htpb]
\begin{center}
\includegraphics[width=.35\textwidth]{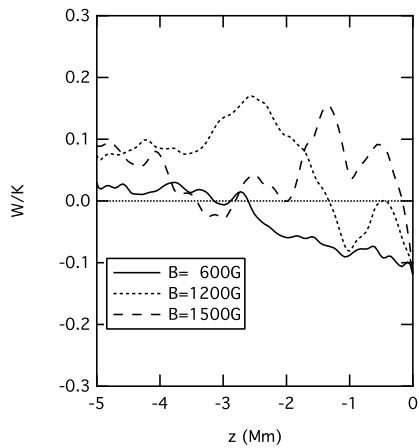}%
\caption{Spatial distribution of the scaled cross helicity.}%
\label{fig:sunspot_flow_W_over_K}
\end{center}
\end{figure}

	The spatial distribution of the turbulent cross helicity scaled by the turbulent MHD energy is shown in Figure~\ref{fig:sunspot_flow_W_over_K}. The statistics fluctuate from one realization to another, but we see the basic tendency. First, the magnitude of the scaled cross helicity is
\begin{equation}
	\frac{\left\langle {{\bf{u}}' \cdot {\bf{b}}'} \right\rangle}
		{\left\langle {{\bf{u}}'{}^2 + {\bf{b}}'{}^2} \right\rangle /2}  
	= \frac{\left\langle {
		{\bf{u}}' \cdot {\bf{b}}_{\rm{c}}' /\sqrt{4\pi \overline{\rho}}
	} \right\rangle}
	{\left\langle {
		{\bf{u}}'{}^2 + \frac{1}{4\pi \overline{\rho}} {\bf{b}}_{\rm{c}}'{}^2
	} \right\rangle/2}
	= O(10^{-1.5})-O(10^{-1})
\end{equation}
(subscript ${\rm{c}}$ denotes that magnetic field is measured in the physical cgs unit). This magnitude seems to be large enough for the cross-helicity effect to work. Secondly, the turbulent cross helicity is negative near the surface and positive in the deeper region.

\begin{figure}[htpb]
\begin{center}
\begin{minipage}{150mm}
\subfigure[B = 600 G.]{
\resizebox*{5cm}{!}{\includegraphics{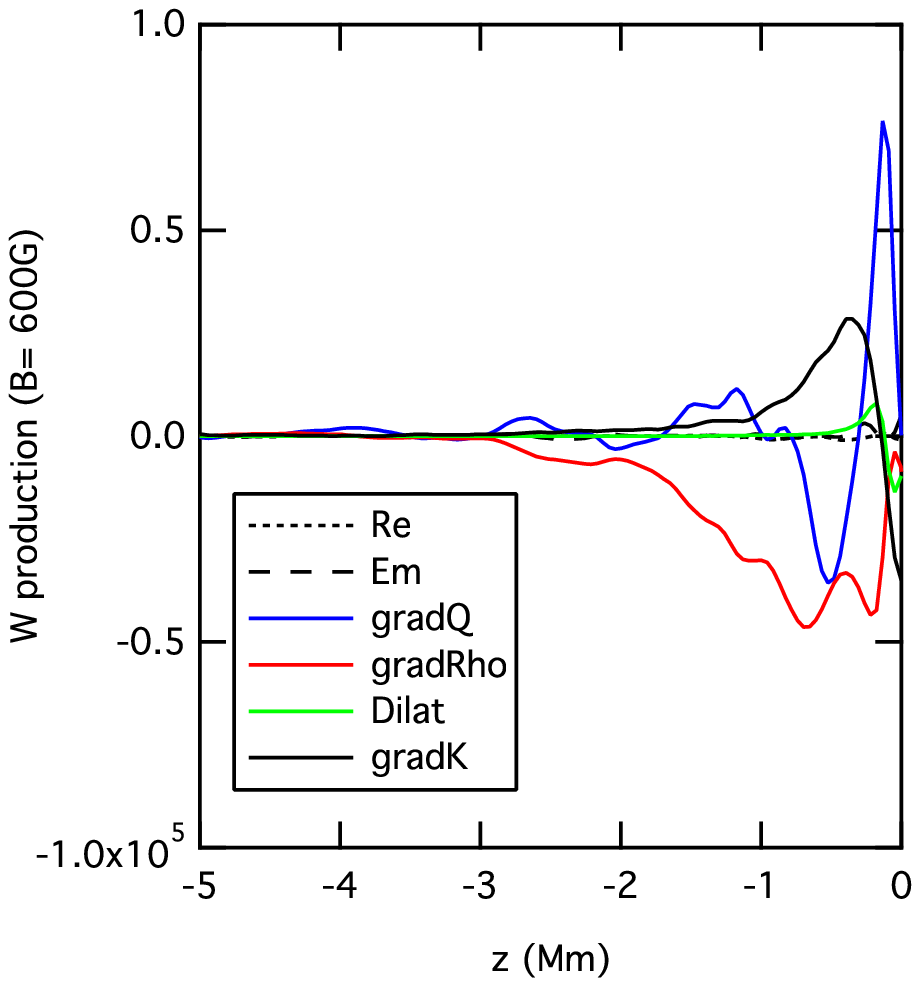}}}%
\subfigure[B = 1200 G.]{
\resizebox*{5cm}{!}{\includegraphics{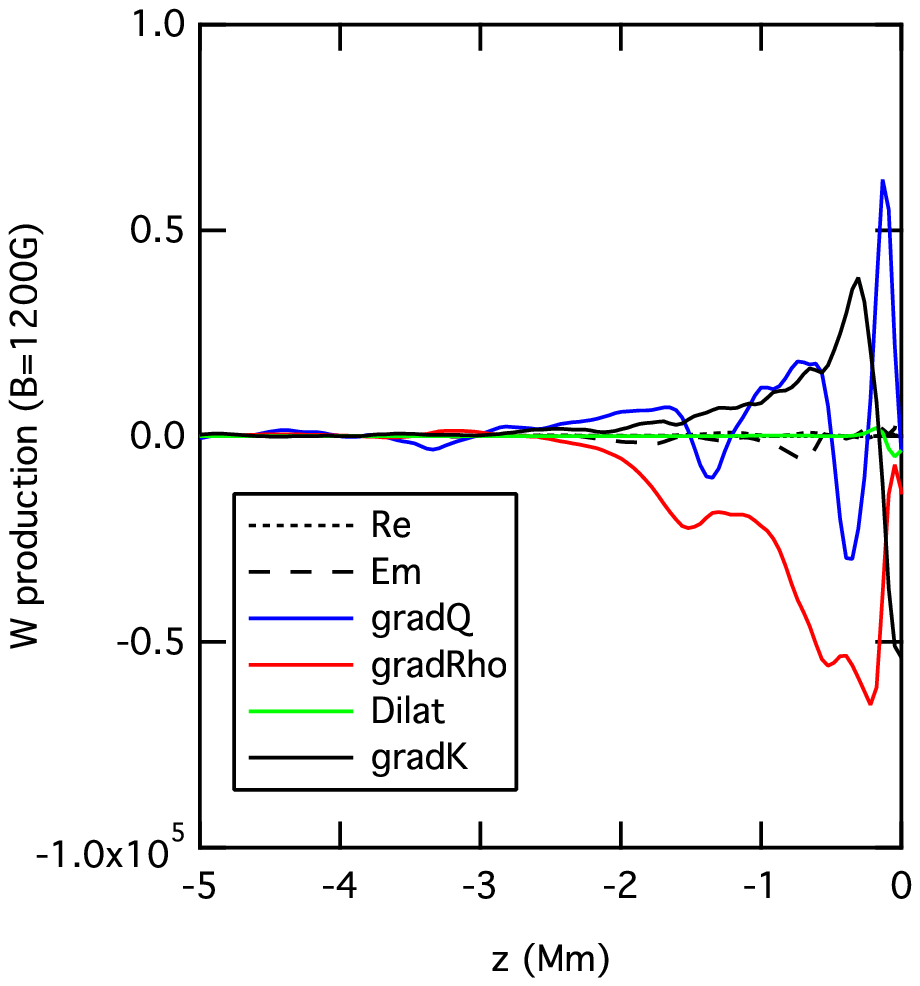}}}%
\subfigure[B = 1500 G.]{
\resizebox*{5cm}{!}{\includegraphics{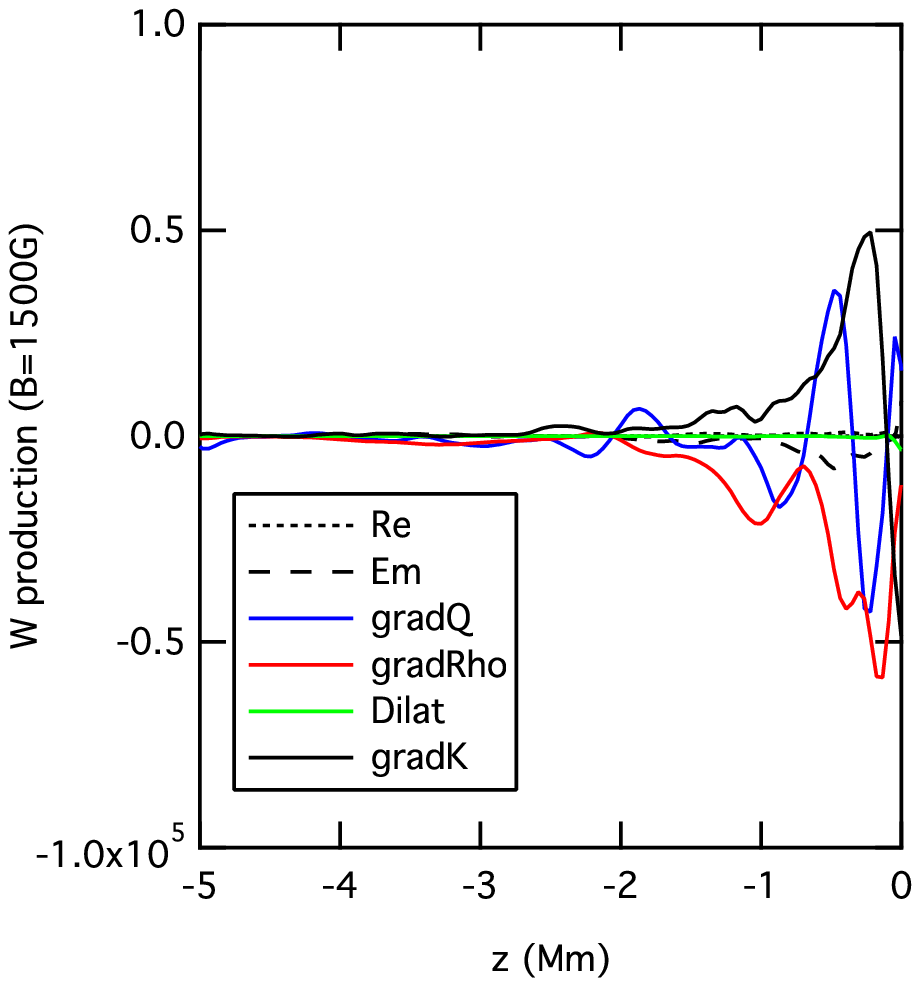}}}%
\caption{Spatial distribution of several production terms.}%
\label{fig:sunspot_flow_Pw}
\end{minipage}
\end{center}
\end{figure}

	As we showed in Eq.~(\ref{eq:comp_W_eq_full}) in \S\ref{sec:W_eq_comp}, in the compressible case, we have several cross-helicity generation mechanisms. Spatial distributions of several production terms are plotted in Figure~\ref{fig:sunspot_flow_Pw} for three cases with different magnetic-field strengths (600, 1200, and 1500 G). In all magnetic-field strength cases, the turbulent cross-helicity generation mechanism related to the mean density stratification [Eq.~(\ref{eq:grad_rho})]:
\begin{equation}
	P_{W\nabla\rho}
	= - (\gamma_0 - 1)\frac{1}{\overline{{\rho}}}
		\left\langle{ q' {\bf{b}}'_{\rm{c}}} \right\rangle \cdot \nabla \overline{\rho}
\end{equation}
plays a dominant role in producing the negative cross helicity in the shallow region, where the large mean density variation is present. In the same region, as we see in Figure~\ref{fig:sunspot_flow_Pw}, the inhomogeneity of the turbulent energy along the mean magnetic field [Eq.~(\ref{eq:comp_W_inhomo})]:
\begin{equation}
	P_{W\nabla K}
	= {\bf{B}}_{\rm{c}} \cdot \nabla \left\langle {
		\frac{1}{2} {\bf{u}}'{}^2
	} \right\rangle
\end{equation}
contributes to production of a positive cross helicity. But the magnitude of production is small compared with the density stratification-related negative production except for the 1500 G case.

\section{Concluding remarks\label{sec:conclusions}}
	The effects of cross helicity in turbulent dynamo were investigated. If the symmetry along the directions parallel and antiparallel to the magnetic field is broken, we can expect finite cross helicity in turbulence. Since the cross helicity is a pseudoscalar (and not positive-definite), it can be locally distributed positive and negative in space even if the total amount of the cross helicity integrated over the volume is zero. If a finite cross helicity exists locally in turbulence, it couples with the mean vorticity in the mean induction equation, and with the mean magnetic strain in the momentum equation, which may reduce the effects of turbulent magnetic diffusivity $\beta$ and eddy viscosity $\nu_{\rm{K}}$, in the respective case.
	
	In the dynamo equation, the cross-helicity effect coupled with the mean vortical motion contributes to the turbulent electromotive force. This effect paves the way for extending the scope of turbulent dynamos. The limitation of mean-field dynamo theory and model related to the point (iii), ``generic'' form of the turbulent electromotive form, is broken through, and point (iv), physics of main dynamo process, is drastically changed.
	
	Another important point is related to the cross-helicity effect in the momentum equation. As we have seen in particular in \S\ref{sec:flow_generation}, the turbulent cross helicity coupled with the inhomogeneity of the mean magnetic field gives rise to flow generation. This feature is entirely novel since the usual helicity or $\alpha$ effect, which leads to the force-free configuration, never contributes to the momentum equation through the mean-field Lorentz force.
	
	Considering all these points, we can conclude that the cross-helicity effects deserve to be paid much more attention in the turbulent dynamo study in the future.

\section*{Acknowledgments}
	I would like to acknowledge Akira Yoshizawa for his everlasting encouragements for my work on turbulence and dynamos. I owe most of the contents in this paper to discussions with him. I would like to thank Karl-Heinz R\"{a}dler, Paul H. Roberts, and Gene Parker for stimulating discussions on the cross-helicity effects. My thanks are also due to Simon Candelaresi, who carefully read the whole manuscript and contributed to improving the presentation of the paper. Part of this work was performed during my stay at NORDITA in February and July-August, 2011.


\appendices
\markboth{N. Yokoi}{Cross helicity and related dynamo}
\section{\vspace{12pt}\\Outline of the present two-scale direct-interaction approximation (TSDIA) analysis\label{appndx:A}}
	In the turbulent dynamo study it is of central importance to obtain expressions for the Reynolds stress $\mbox{\boldmath${\cal{R}}$}$ [Eq.~(\ref{eq:Re_strss_def})] and the turbulent electromotive force ${\bf{E}}_{\rm{M}}$ [Eq.~(\ref{eq:E_M_def})]. In this paper we present the expressions [Eqs.~(\ref{eq:Re_strss_exp}) and (\ref{eq:E_M_exp})] obtained from the two-scale direct-interaction approximation (TSDIA) by \citet{yos1990}. The TSDIA is a statistical analytical theory for inhomogeneous turbulence \citep{yos1984} constituted of a combination of the multiple-scale analysis and the direct-interaction approximation (DIA), an elaborated closure scheme for homogenous isotropic turbulence by \citet{kra1957}.

	In this Appendix, we present the outline of the TSDIA procedure that leads to Eqs.~(\ref{eq:Re_strss_exp}) and (\ref{eq:E_M_exp}). For detailed derivation, the reader is referred to \citet{yos1990,yos1998}.

\subsection{Fundamental equations and Elsasser-variable formulation}
	We apply the TSDIA formulation to the incompressible magnetohydrodynamic (MHD) turbulence. In the TSDIA formalism, by way of the differential expansion and the external-field expansion, the effects of inhomogeneity, external field, and rotation appear in the higher-order analysis. If we perform an analysis in a rotating frame, we can selectively derive the expression of the inhomogeneous helicity effects coupled with the rotation and equivalently the mean vorticity in the lower-order calculations. This makes the calculation much simpler. So, here in Appendix we present the calculation in a rotating frame.

	An incompressible magnetohydrodynamic plasma in a coordinate system
rotating with the angular velocity $\mbox{\boldmath$\omega$}_{\rm F}$  
obeys
\begin{equation}
		\frac{\partial {\bf{u}}}{\partial t}
		+ \left({ {\bf{u}} \cdot \nabla }\right){\bf{u}}
		- \left({ {\bf{b}} \cdot \nabla }\right){\bf{b}}
		= - \nabla p_{\rm M}
		- 2{\mbox{\boldmath$\omega$}}_{\rm F} \times {\bf{u}}
		+ \nu \nabla^2 {\bf{u}},
	\label{eq:u-eq}
\end{equation}
\begin{equation}
		\frac{\partial {\bf{b}}}{\partial t}
		+ \left({{\bf{u}} \cdot \nabla}\right){\bf{b}}
		- \left({{\bf{b}} \cdot \nabla}\right){\bf{u}}
		= \eta \nabla^2 {\bf{b}},
	\label{eq:b-eq}
\end{equation}
with the solenoidal conditions for the velocity ${\bf{u}}$ and the magnetic field ${\bf{b}}$:
\begin{equation}
		\nabla \cdot {\bf{u}} = \nabla \cdot {\bf{b}} = 0.
	\label{eq:slndl-ub}
\end{equation}
Here, $p_{\rm{M}} (= p + {\bf{b}}^2/2)$ is the MHD pressure, $\nu$ is the kinematic viscosity, and $\eta$ is the magnetic diffusivity. We express ${\bf{b}}$ etc.\ in Alfv\'{e}n-speed units as in Eq.~(\ref{eq:alfven_unit}). In assuming incompressibility, we do not deny the importance of compressibility in MHD turbulence. With this understanding, the present work should be regarded as an attempt to consider some basic properties of the turbulent dynamo in the framework of incompressible MHD turbulence theory.

	For the sake of clarity, we introduce the Elsasser variables:
\begin{equation}
	\mbox{\boldmath$\phi$} = {\bf{u}} + {\bf{b}},\
	\mbox{\boldmath$\psi$} = {\bf{u}} - {\bf{b}}.
	\label{elsasser_def}
\end{equation}
and rewrite Eqs.~(\ref{eq:u-eq})-(\ref{eq:slndl-ub}), neglecting the difference between $\nu$ and $\eta$. Then we have
\begin{equation}
		\frac{\partial {\mbox{\boldmath$\phi$}}}{\partial t}
		+ \left( {{\mbox{\boldmath$\psi$}}\cdot\nabla} \right){\mbox 
{\boldmath$\phi$}}
		= - \nabla p_{\rm M}
		- {\mbox{\boldmath$\omega$}}_{\rm F} \times
			\left( {{\mbox{\boldmath$\phi$}} + {\mbox{\boldmath$\psi$}}} \right)
		+ \frac{\nu + \eta}{2} \nabla^2 {\mbox{\boldmath$\phi$}}
		+ \frac{\nu - \eta}{2}\nabla^2 {\mbox{\boldmath$\psi$}},
	\label{eq:phi-eq}
\end{equation}
\begin{equation}
		\frac{\partial {\mbox{\boldmath$\psi$}}}{\partial t}
		+ \left( {{\mbox{\boldmath$\phi$}}\cdot\nabla} \right) {\mbox 
{\boldmath$\psi$}}
		= - \nabla p_{\rm M}
		- {\mbox{\boldmath$\omega$}}_{\rm F} \times
			\left( {{\mbox{\boldmath$\psi$}}+{\mbox{\boldmath$\phi$}}} \right)
		+ \frac{\nu + \eta}{2}\nabla^2 {\mbox{\boldmath$\psi$}}
		+ \frac{\nu - \eta}{2}\nabla^2 {\mbox{\boldmath$\phi$}},
	\label{eq:psi-eq}
\end{equation}
\begin{equation}
		\nabla \cdot {\mbox{\boldmath$\phi$}}
		= \nabla \cdot {\mbox{\boldmath$\psi$}} = 0.
	\label{eq:slndl-phipsi}
\end{equation}
Note that in Eqs.~(\ref{eq:phi-eq}) and (\ref{eq:psi-eq}) the nonlinearity can be expressed in terms of $\mbox{\boldmath$\phi$}$ and $\mbox{\boldmath$\psi$}$ only. Equations~(\ref{eq:phi-eq})-(\ref{eq:slndl-phipsi}) have a highly symmetric form; the interchange of $\mbox{\boldmath$\phi$}$ with $\mbox{\boldmath$\psi$}$ does not change the system of equations at all. We fully utilize this property in the following calculations.

	If we further assume that the differnce between $\nu$ and $\eta$ is not so critical, Eqs.~(\ref{eq:psi-eq}) and (\ref{eq:slndl-phipsi}) are reduced to the simplest possible form as
\begin{equation}
		\frac{\partial {\mbox{\boldmath$\phi$}}}{\partial t}
		+ \left( {{\mbox{\boldmath$\psi$}}\cdot\nabla} \right){\mbox 
{\boldmath$\phi$}}
		= - \nabla p_{\rm M}
		- {\mbox{\boldmath$\omega$}}_{\rm F} \times
			\left( {{\mbox{\boldmath$\phi$}} + {\mbox{\boldmath$\psi$}}} \right)
		+ \frac{\nu + \eta}{2}\nabla^2 {\mbox{\boldmath$\phi$}},
	\label{eq:phi-eq_simple}
\end{equation}
\begin{equation}
		\frac{\partial {\mbox{\boldmath$\psi$}}}{\partial t}
		+ \left( {{\mbox{\boldmath$\phi$}}\cdot\nabla} \right) {\mbox 
{\boldmath$\psi$}}
		= - \nabla p_{\rm M}
		- {\mbox{\boldmath$\omega$}}_{\rm F} \times
			\left( {{\mbox{\boldmath$\psi$}}+{\mbox{\boldmath$\phi$}}} \right)
		+ \frac{\nu + \eta}{2}\nabla^2 {\mbox{\boldmath$\psi$}}.
	\label{eq:psi-eq_simple}
\end{equation}

Note that the Elsasser formulation can be also applied to the compressible case \citep{mar1987}. As for an application to the compressible MHD turbulence, the reader is referred to \citet{yok2007}.

	We divide the Elsasser variables into the mean and fluctuation around it:
\begin{equation}
	\mbox{\boldmath$\phi$} 
	= \mbox{\boldmath$\Phi$} + \mbox{\boldmath$\phi$}',\;\;
	\mbox{\boldmath$\psi$} 
	= \mbox{\boldmath$\Psi$} + \mbox{\boldmath$\psi$}'.
	\label{eq:elsasser_rey_decomp}
\end{equation}
Using these variables, the Reynolds stress $\mbox{\boldmath${\cal{R}}$}$ [Eq.~(\ref{eq:Re_strss_def})] and the turbulent electromotive force ${\bf{E}}_{\rm{M}}$ [Eq.~(\ref{eq:E_M_def})] are expressed as
\begin{equation}
	E_{\rm{M}}^{\alpha}
	= - \frac{1}{2} \epsilon^{\alpha ab} {\cal{R}}_{\rm{E}}^{ab},
	\label{eq:elsasser_emf}
\end{equation}
\begin{equation}
	{\cal{R}}^{\alpha\beta}
	= \frac{1}{2} \left( {
		{\cal{R}}_{\rm{E}}^{\alpha\beta} + {\cal{R}}_{\rm{E}}^{\beta\alpha}
	} \right)
	\label{eq:elsasser_R}
\end{equation}
with $\mbox{\boldmath${\cal{R}}$}_{\rm{E}}$ being defined as
\begin{equation}
	{\cal{R}}_{\rm{E}}^{\alpha\beta}
	= \left\langle { \phi'{}^\alpha \psi'{}^\beta } \right\rangle,
	\label{eq:elsasser_Rey_strss_def}
\end{equation}
which may be called the Elsasser Reynolds stress.

	We apply the Reynolds decomposition (\ref{eq:elsasser_rey_decomp}) into Eq.~(\ref{eq:phi-eq}). In a frame rotating with the angular velocity $\mbox{\boldmath$\omega$}_{\rm F}$, $\mbox{\boldmath$\phi$}'$ obeys
\begin{eqnarray}
	\frac{\partial \phi'{}^\alpha}{\partial t}
	+& & \Psi^a \frac{\partial \phi'{}^\alpha}{\partial x^a}
	+ \frac{\partial}{\partial x^a} \left( {
		\psi'{}^a \phi'{}^\alpha - {\cal{R}}_{\rm{E}}^{\alpha a}
	} \right)
	+ \frac{\partial p'_{\rm{M}}}{\partial x^\alpha}
	- \nu \nabla^2 \phi'{}^\alpha
	\nonumber\\
	& & = - \epsilon^{\alpha ab} \omega_{{\rm F}}^a \left( {
		\phi'{}^b + \psi'{}^b 
	} \right)
	- \psi'{}^a \frac{\partial \Phi^\alpha}{\partial x^a}
	\label{eq:elsasser_fluct_eq}
\end{eqnarray}
with the solenoidal condition:
\begin{equation}
	\nabla \cdot \mbox{\boldmath$\phi$}' = 0.
	\label{eq:elsasser_sol_cond}
\end{equation}
The counterparts for $\mbox{\boldmath$\psi$}'$ are obtained by the exchange of variables
\begin{equation}
	\mbox{\boldmath$\phi$}' \to \mbox{\boldmath$\psi$}',\;\;
	\mbox{\boldmath$\psi$}' \to \mbox{\boldmath$\phi$}',\;\;
	\mbox{\boldmath$\Phi$} \to \mbox{\boldmath$\Psi$},\;\;
	\mbox{\boldmath$\Psi$} \to \mbox{\boldmath$\Phi$},\;\;
	{\cal{R}}_{\rm{E}}^{\alpha\beta} \to {\cal{R}}_{\rm{E}}^{\beta\alpha}.
	\label{eq:elsasser_transform}
\end{equation}
As will be seen later, in the Reynolds stress expression, the helicity effect occurs in a combination of the mean vorticity and the gradient of the turbulent kinetic helicity. Such a term appears at the $O(\delta^2)$ calculation in the TSDIA since both the mean vorticity and helicity gradient are the quantities of $O(\delta)$. If we consider the Reynolds stress in a rotating frame, such a combination appears in the $O(\delta)$ calculation. This is the reason why we adopt a frame rotating with $\mbox{\boldmath$\omega$}_{\rm{F}}$ in Eq.~(\ref{eq:elsasser_fluct_eq}).

\subsection{TSDIA procedure}
	The formal procedure of the TSDIA may be summarized as
\begin{enumerate}
\item Introduction of two scales;
\item Fourier representation of the rapid variables;
\item Scale-parameter expansions;
\item Calculation using the Green's functions;
\item Statistical properties for the basic field;
\item Calculation of the correlation functions using the DIA.
\end{enumerate}

	Through the steps listed above, effects of mean-field inhomogeneity, rotation, magnetic field, etc.\ are incorporated in a perturbation manner into the closure scheme of turbulence, which was originally applicable only to homogeneous isotropic turbulence.
	
	In the following, we briefly explain each step.

\subsubsection{Introduction of two scales}
			Using a scale parameter $\delta$, we introduce the slow and
rapid variables:
\begin{equation}
		\mbox{\boldmath$\xi$} = {\bf x},\ {\bf{X}} = \delta{\bf x};\
		\tau = t,\  T = \delta t.
	\label{eq:fast_slow_vars}
\end{equation}
This parameter is not necessarily small. If $\delta$ is small, the ${\bf{X}}$ and $T$ in Eq.~(\ref{eq:fast_slow_vars}) represent the slow variables. The slow variables $({\bf{X}}, T)$ provide long spatial and temporal scales since their changes are not negligible only when $\bf x$ and $t$ are large. On the other hand, the rapid
variables $({\mbox{\boldmath$\xi$}}, \tau)$ are appropriate for
describing the fine spatiotemporal motions. With these two-scale
variables, the spatial and temporal derivatives are expressed as
\begin{equation}
		\nabla = \nabla_{\xi} + \delta\nabla_{\bf{X}},\;\;
		\frac{\partial}{\partial t}
		= \frac{\partial}{\partial \tau} + \delta \frac{\partial}{\partial T},
	\label{eq:two-scale_deriv}
\end{equation}
and the field quantities $f$ are divided into $F$ and $f'$ as
\begin{equation}
		f = F({\bf{X}};T)
		+ f'({\mbox{\boldmath$\xi$}},{\bf{X}};\tau,T).
	\label{eq:mean_fluct}
\end{equation}
The expansion parameter $\delta$ is not an actual parameter but is an artificial one for implementing the effect of slowly varying quantities on the fast varying quantities. This parameter appears if we have differentiations with respect to the slow variables Eq.~(\ref{eq:two-scale_deriv}). This parameter automatically disappears in the final results through the replacement of ${\bf{X}} \to \delta{\bf{x}}$ and $T \to \delta t$.

\subsubsection{Fourier representations}
			We perform the Fourier transform with respect to the rapid
variable $\mbox{\boldmath$\xi$}$ as
\begin{equation}
		f(\mbox{\boldmath$\xi$},{\bf{X}};\tau,T)
		= \int{
			f({\bf{k}},{\bf{X}};\tau,T)
			\exp[-i{\bf{k}}\cdot(\mbox{\boldmath$\xi$} - {\bf{U}}\tau)]
		} d{\bf{k}},
	\label{eq:convFourier}
\end{equation}
and express the governing equations in wave-number space. The
factor $\exp[-i{\bf{k}}\cdot(\mbox{\boldmath$\xi - {\bf{U}}\tau$})]$ on the
RHS of Eq.~(\ref{eq:convFourier}) expresses that the transform is
performed in the frame moving with the large-scale flow ${\bf{U}}$.
For instance, the equation for
$\mbox{\boldmath$\phi$}'$ is expressed as
\begin{eqnarray}
		\lefteqn{
		\frac{\partial {\phi'}^\alpha ({\bf k};\tau)}{\partial t}
		+ \nu k^2 {\phi'}^\alpha ({\bf k};\tau)
		- ik^\alpha p'_{\rm M}({\bf k};\tau)
		}\nonumber\\
		&& -ik^b\int\!\!\!\int {\delta ({\bf{k-p-q}})
			d{\bf p}d{\bf q}
			{\psi'}^b ({\bf p};\tau)
			{\phi'}^\alpha ({\bf q};\tau)}
		\nonumber\\
		&& = i({\bf k}\cdot{\bf B}){\phi'}^\alpha({\bf k};\tau)
			- \epsilon^{\alpha ab} \omega_{{\rm F}}^a
			\left({
				{\phi'}^b({\bf k};\tau) + {\psi'}^b({\bf k};\tau)
		}\right)
		\nonumber\\
		&& + \delta \left( {
			-{\psi'}^a({\bf k};\tau)
			\frac{\partial\Phi^\alpha}{\partial X^a}
			- \frac{D\phi'({\bf k};\tau)}{DT_{\rm I}}
			\rule{0in}{4.0ex}
		}\right.
		+ B^a \frac{\partial {\phi'}^\alpha({\bf k};\tau)}{\partial X_{\rm I}^a}
			- \frac{\partial p'_{\rm M}({\bf k};\tau)}{\partial X_{\rm I}^\alpha}
		\nonumber\\
		&& \hspace{30pt} \left.\rule{0in}{4.0ex}{
			-\frac{\partial}{\partial X_{\rm I}^a}\!\!\int\!\!\!
			\int\!\!
{\delta({\bf{k-p-q}})d{\bf{p}}d{\bf{q}}{{\mbox{\boldmath$\psi$}}'}^a
		(p;\tau) {\phi'}^\alpha(q;\tau)}
		} \!\right)\!\!,
		\label{eq:phi-Fourier-eq}
\end{eqnarray}
\begin{equation}
		{\bf k}\cdot\mbox{\boldmath$\phi$}'_{\rm S}({\bf k};\tau) = 0,
	\label{eq:FourierSlndl}
\end{equation}
where the solenoidal fluctuation $\mbox{\boldmath$\phi$}'_{\rm{S}}$ is defined by
\begin{equation}
		{\mbox{\boldmath$\phi$}}'_{\rm S}({\bf k};\tau)
		= {\mbox{\boldmath$\phi$}}'({\bf k};\tau)
		+ \delta \left({
			i\frac{\bf k}{k^2}
			\frac{\partial{\phi'}^a({\bf k};t)}{\partial X_{\rm I}^a}
			}\right).
	\label{eq:phisDef}
\end{equation}
In Eq.~(\ref{eq:phi-Fourier-eq}), $\delta({\bf k} - {\bf q}-{\bf r})$ denotes the delta function which vanishes unless the wave-vector relation ${\bf k} = {\bf q}
+ {\bf r}$ is satisfied. Here and hereafter in the argument notation
dependence on the slow variables ${\bf{X}}$ and $T$ is suppressed.

\subsubsection{Scale-parameter expansion}
	We expand the field quantities ${\mbox{\boldmath$\vartheta$}}' =
({\mbox{\boldmath$\phi$}}', {\mbox{\boldmath$\psi$}}')$ in the scale  
parameter
$\delta$:
\begin{equation}
		{\mbox{\boldmath$\vartheta$}}'
		= {\mbox{\boldmath$\vartheta$}}'_0
		+ \delta {\mbox{\boldmath$\vartheta$}}'_1
		+ \delta^2 {\mbox{\boldmath$\vartheta$}}'_2
		+ \cdots,
	\label{eq:scl-prmtr-expnd}
\end{equation}
where ${\mbox{\boldmath$\vartheta$}}_0$ is the field without the mean  
field.
We further expand this in the external-field parameter such as the
mean magnetic field $\bf B$, the angular velocity
${\mbox{\boldmath$\omega$}}_{\rm F}$, etc.:
\begin{equation}
		{\mbox{\boldmath$\vartheta$}}'
		= {\mbox{\boldmath$\vartheta$}}'_{\rm{B}}
		+ {\mbox{\boldmath$\vartheta$}}'_{01}
		+ {\mbox{\boldmath$\vartheta$}}'_{02} + \cdots
		+ {\mbox{\boldmath$\vartheta$}}'_{1}
		+ {\mbox{\boldmath$\vartheta$}}'_{2} + \cdots.
	\label{eq:ext-prmtr-expnd}
\end{equation}
Here, ${\mbox{\boldmath$\vartheta$}}'_{\rm{B}}$ is the basic field
corresponding to the homogeneous isotropic turbulence. For
instance, the equation for ${\mbox{\boldmath$\phi$}}'_{\rm{B}}$ is  
written as
\begin{equation}
	\frac{\partial \phi'_{\rm{B}}{}^\alpha ({\bf k};\tau)}{\partial \tau}
	+ \nu k^2 \phi'_{\rm{B}}{}^\alpha
	- i Z^{\alpha ab}({\bf k})
	\iint {\delta({\bf k}-{\bf p}-{\bf q})} d{\bf p} d{\bf q}
	\psi'_{\rm{B}}{}^a({\bf p};\tau) \phi'_{\rm{B}}{}^b({\bf q};\tau)
   	=0,
	\label{eq:phiBeq}
\end{equation}
where $Z^{\alpha ab}({\bf{k}}) = k^a D^{\alpha b}({\bf{k}})$ [$D^{\alpha\beta} (= \delta^{\alpha\beta} - k^\alpha k^\beta/k^2)$ is the projection operator in the wave-number space]. Note that the basic field is free from the effect of mean shear, frame rotation, and mean magnetic field. This equation is considered to be the equation for the homogeneous isotropic turbulence except for the implicit dependence on the slow variables $\bf X$ and $T$.

	The $O(\delta^1)$ field $\mbox{\boldmath$\phi$}_1$ obeys
\begin{eqnarray}
	\frac
	{\partial \phi'{}_{{\rm{S}}1}^\alpha \left( {{\bf{k}};\tau} \right)}
	{\partial \tau}
	&+& \nu k^2 \phi'{}_{{\rm{S}}1}^\alpha \left( {{\bf{k}};\tau} \right)
	- i Z^{\alpha ab}\left( {\bf{k}} \right) 
	\iint \delta ({\bf{k}} - {\bf{p}} -{\bf{q}}) d{\bf{p}} d{\bf{q}}\ 
	\psi'{}_0^a({\bf{p}};\tau) \phi'{}_{{\rm{S}}1}^b({\bf{q}};\tau)
	\nonumber\\
	&=& - D^{\alpha b}({\bf{k}}) \psi'{}_0^a \left( {{\bf{k}};\tau} \right)
	\frac{\partial \Phi^b}{\partial X^a}
	- \frac{D \phi'{}_0^\alpha \left( {{\bf{k}};\tau} \right)}{DT_{\rm{I}}}
	+ B^a 
	\frac{\partial \phi'{}_0^\alpha \left( {{\bf{k}};\tau} \right)}{\partial X_{\rm{I}}^a}
	\nonumber\\
	& & + i \epsilon^{cab} \omega_{{\rm F}}^a  \frac{k^b}{k^2} D^{\alpha c}({\bf{k}}) 
		\frac{\partial}{\partial X_{\rm{I}}^d} \left( {
			\phi'{}_0^d({\bf{k}};\tau) + \psi'{}_0^d({\bf{k}};\tau)
		} \right)
	\nonumber\\
	& & + i \epsilon^{cab} \omega_{{\rm F}}^a  \frac{k^c}{k^2} D^{\alpha d}({\bf{k}}) 
		\frac{\partial}{\partial X_{\rm{I}}^d} \left( {
			\phi'{}_0^b({\bf{k}};\tau) + \psi'{}_0^b({\bf{k}};\tau)
		} \right)
	\nonumber\\
	& & -i ({\bf{k}} \cdot {\bf{B}}) \phi'{}_{{\rm{S}}1}^\alpha({\bf{k}};\tau)
	- \epsilon^{cab} \omega_{{\rm F}}^a D^{\alpha c}({\bf{k}}) \left( {
		\phi'{}_{{\rm{S}}1}^b({\bf{k}};\tau) + \psi'{}_{{\rm{S}}1}^b({\bf{k}};\tau)
	} \right),
	\label{eq:1st-fld_eq_TSDIA}
\end{eqnarray}
with the solenoidal condition:
\begin{equation}
	{\bf{k}} \cdot \mbox{\boldmath$\phi$}'_{{\rm{S}}1}({\bf{k}};\tau) = 0,
	\label{eq:sol_cond_phi_S1}
\end{equation}
for $\mbox{\boldmath$\phi$}_{{\rm{S}}1}$ defined by
\begin{equation}
	\mbox{\boldmath$\phi$}'_{{\rm{S}}1}({\bf{k}};\tau)
	= \mbox{\boldmath$\phi$}'_1({\bf{k}};\tau)
	+ i \frac{{\bf{k}}}{k^2}
	\frac{\partial \phi'{}_{\rm{B}}^a({\bf{k}};\tau)}{\partial X_{\rm{I}}^a}.
	\label{eq:1_1st_u_sol}
\end{equation}
In Eq.~(\ref{eq:1st-fld_eq_TSDIA}) we have eliminated the MHD pressure $p'_{\rm{M}}$ using the solenoidal condition (\ref{eq:sol_cond_phi_S1}).

\subsubsection{Calculation using the Green's functions}
			We define the Green's functions for
$\mbox{\boldmath$\phi$}'_{\rm{B}}$,
$\mbox{\boldmath$\phi$}'_{01}$, etc. For example, the one for
$\mbox{\boldmath$\phi$}'_{\rm{B}}$,
$G'_\phi{}^{\alpha\beta}({\bf k};\tau,\tau')$, is defined by
\begin{eqnarray}
	\frac{\partial G'_\phi{}^{\alpha\beta}({\bf k};\tau,\tau')}{\partial \tau}
	& & + \nu k^2 G'_\phi{}^{\alpha\beta}({\bf k};\tau,\tau')
	\nonumber\\
	& & \hspace{-30pt} - i Z^{\alpha ab}({\bf k})
		\iint {
			\!\!\delta({\bf k}-{\bf p}-{\bf q})}
			d{\bf p} d{\bf q}
		 \psi'_{\rm{B}}{}^a({\bf p};\tau)
					G'_\phi{}^{b\beta}({\bf q};\tau,\tau')
	= \delta^{\alpha\beta}\delta(\tau - \tau').\hspace{0pt}
	\label{eq:GreenFnDef}
\end{eqnarray}
Using these Green's functions, we formally solve
${\mbox{\boldmath$\vartheta$}}'_{01}$ and ${\mbox{\boldmath$\vartheta 
$}}'_{1}$.
\begin{eqnarray}
	\phi'{}_{{\rm{S}}1}^\alpha \left( {{\bf{k}};\tau} \right)
	&=& - \frac{\partial \Phi^b}{\partial X^a}
		D^{\alpha b}({\bf{k}}) \int_{-\infty}^\tau {d\tau_1} 
		G'{}_\phi^{\alpha c}\left( {{\bf{k}};\tau,\tau_1} \right) 
		\psi'{}_0^a \left( {{\bf{k}};\tau} \right)
	\nonumber\\
	& & -  \int_{-\infty}^\tau {d\tau_1} 
		G'{}_\phi^{\alpha a}\left( {{\bf{k}};\tau,\tau_1} \right) 
		\frac{D \phi'{}_0^a \left( {{\bf{k}};\tau} \right)}{DT_{\rm{I}}}
	\nonumber\\
	& & + B^a  \int_{-\infty}^\tau {d\tau_1} 
		G'{}_\phi^{\alpha b}\left( {{\bf{k}};\tau,\tau_1} \right) 
		\frac{\partial \phi'{}_0^b \left( {{\bf{k}};\tau} \right)}{\partial X_{\rm{I}}^a}
	\nonumber\\
	& & + i \epsilon^{cab} \omega_{{\rm F}}^a  \frac{k^b}{k^2} D^{\alpha c}({\bf{k}}) 
		 \int_{-\infty}^\tau {d\tau_1} G'{}_\phi^{\alpha d}\left( {{\bf{k}};\tau,\tau_1} \right)\frac{\partial}{\partial X_{\rm{I}}^e} \left( {
			\phi'{}_0^e({\bf{k}};\tau) + \psi'{}_0^d({\bf{k}};\tau)
		} \right)
	\nonumber\\
	& & + i \epsilon^{cab} \omega_{{\rm F}}^a  \frac{k^c}{k^2} D^{de}({\bf{k}}) 
		 \int_{-\infty}^\tau {d\tau_1} 
		 G'{}_\phi^{\alpha d}\left( {{\bf{k}};\tau,\tau_1} \right)
		 	\frac{\partial}{\partial X_{\rm{I}}^e} \left( {\phi'{}_0^b({\bf{k}};\tau) 
			+ \psi'{}_0^b({\bf{k}};\tau)
		} \right)
	\nonumber\\
	& & -i ({\bf{k}} \cdot {\bf{B}})  \int_{-\infty}^\tau {d\tau_1} G'{}_\phi^{\alpha a}\left( {{\bf{k}};\tau,\tau_1} \right)\phi'{}_{{\rm{S}}1}^a({\bf{k}};\tau)
	\nonumber\\
	& & - \epsilon^{cab} \omega_{{\rm F}}^a D^{cd}({\bf{k}})  \int_{-\infty}^\tau {d\tau_1} G'{}_\phi^{\alpha d}\left( {{\bf{k}};\tau,\tau_1} \right)\left( {
		\phi'{}_{{\rm{S}}1}^b({\bf{k}};\tau) + \psi'{}_{{\rm{S}}1}^b({\bf{k}};\tau)
	} \right).
	\label{eq:1st-fld_sol_TSDIA}
	\label{eq:phi_S1_sol}
\end{eqnarray}
Here the RHS still contains $\mbox{\boldmath$\phi$}_{{\rm{S}}1}$ and $\mbox{\boldmath$\psi$}_{{\rm{S}}1}$. By the iteration method, we get the leading expression for $\mbox{\boldmath$\phi$}_{{\rm{S}}1}$, which is the same as Eq.(\ref{eq:phi_S1_sol}) with the last two terms of the RHS dropped.

\subsubsection{Statistical properties for the basic fields}
			Since the basic fields are homogeneous and isotropic, we assume
the statistical properties for them in the form:
\begin{equation}
		{{\left\langle {
				\vartheta_{\rm{B}}^\alpha({\bf k};\tau)
				\chi_{\rm{B}}^\beta({\bf k}';\tau')
			} \right\rangle }
			\over
			{\delta({\bf k}+{\bf k'})}
		}
		= D^{\alpha\beta}({\bf k})
				Q_{\vartheta\chi}({\bf k};\tau,\tau')
		+ \frac{i}{2} \frac{k^a}{k^2}
			\epsilon^{\alpha \beta a}
			H_{\vartheta\chi}({\bf k};\tau ,\tau'),
	\label{eq:HomoIsoFlds}
\end{equation}
\begin{equation}
		\left\langle {
			G'_\vartheta{}^{\alpha\beta}({\bf k};\tau,\tau')
		} \right\rangle
		= \delta^{\alpha\beta} G_\vartheta({\bf k};\tau,\tau'),
	\label{eq:HomoIsoGreen}
\end{equation}
where $\mbox{\boldmath$\vartheta$}$ and $\mbox{\boldmath$\chi$}$ denote
$\mbox{\boldmath$\phi$}$ and/or $\mbox{\boldmath$\psi$}$. For later  
convenience, we introduce the symmetric and anti-symmetric parts of the Green's
functions as
\begin{eqnarray}
		G_{\rm{S}}({\bf k};\tau,\tau')
		&=& \!\frac{1}{2} \left({
			G_\phi({\bf k};\tau,\tau') + G_\psi({\bf k};\tau,\tau')
		}\right),
	\label{eq:symmetricGreenFn}\\
		G_{\rm{A}}({\bf k};\tau,\tau')
		&=& \!\frac{1}{2} \left({
			G_\phi({\bf k};\tau,\tau') - G_\psi({\bf k};\tau,\tau')
		}\right).
	\label{eq:antisymGreenFn}
\end{eqnarray}

\subsubsection{Calculation of the correlation functions}
			Following the above procedure, we calculate the correlation
functions. Especially, the Elsasser Reynolds stress $\mbox{\boldmath${\cal{R}}$}$ [Eq.~(\ref{eq:elsasser_Rey_strss_def})] is caluculated as
\begin{eqnarray}
	\left\langle{{\phi'}^\alpha {\psi'}^\beta}\right\rangle
	&=&\left\langle{
			\phi'_{\rm{B}}{}^\alpha \psi'_{\rm{B}}{}^\beta
		}\right\rangle
	+ \left\langle{
			\phi'_{\rm{B}}{}^\alpha \psi'_{01}{}^\beta
		}\right\rangle
	+ \left\langle{
			\phi'_{01}{}^\alpha \psi'_{\rm{B}}{}^\beta
		}\right\rangle
	+ \cdots
	\nonumber\\
	&& \hspace{10pt} +\left\langle{
			\phi'_{\rm{B}}{}^\alpha \psi'_{1}{}^\beta
		}\right\rangle
	+ \left\langle{
			\phi'_{1}{}^\alpha \psi'_{\rm{B}}{}^\beta
		}\right\rangle
		+ \cdots,
\end{eqnarray}
with the renormalization of the propagators:
\begin{subequations}
\begin{equation}
	Q_{\rm{B}}^{\alpha\beta}({\bf{k}};\tau,\tau') 
	\mapsto Q^{\alpha\beta}({\bf{k}};\tau,\tau'),
	\label{eq:Q_renormalization}
\end{equation}
\begin{equation}
	G_{\rm{B}}^{\alpha\beta}({\bf{k}};\tau,\tau') 
	\mapsto G^{\alpha\beta}({\bf{k}};\tau,\tau'),
	\label{eq:G_renormalization}
\end{equation}
\end{subequations}
where $Q_{\rm{B}}^{\alpha\beta}$ and $G_{\rm{B}}^{\alpha\beta}$ are the lowest-order propagators whereas $Q^{\alpha\beta}$ and $G^{\alpha\beta}$ are their exact counterparts.

\subsection{Results}
	With the abbreviate expressions for the spectral and time integrals:
\begin{subequations}
\begin{equation}
	I_n\{ {A} \}
	= \int k^{2n} A(k,{\bf{x}};\tau,\tau,t) d{\bf{k}},
\end{equation}
\begin{equation}
	I_n\{ {A, B} \}
	= \int k^{2n} d{\bf{k}} \int_{-\infty}^{\tau} \!\! d\tau_1 
		A(k,{\bf{x}};\tau,\tau_1,t) B(k,{\bf{x}};\tau,\tau_1,t),
\end{equation}
\end{subequations}
the main results of the TSDIA analysis is given as follows.

\paragraph{Turbulent electromotive force}
	The turbulent electromotive force ${\bf{E}}_{\rm{M}}$ is obtained from Eq.~(\ref{eq:elsasser_emf}) as
\begin{equation}
	{\bf{E}}_{\rm{M}}
	= \alpha {\bf{B}}
	- \beta {\bf{J}}
	+ \gamma \mbox{\boldmath$\Omega$}
	+ 2 \gamma_{\rm{F}} \mbox{\boldmath$\omega$}_{\rm{F}}.
\end{equation}
The transport coefficients $\alpha$, $\beta$, $\gamma$, and $\gamma_{\rm{F}}$ are expressed as
\begin{equation}
	\alpha = \frac{1}{3} \left( {
	I_0 \left\{ {G_{\rm{S}}, -H_{uu} + H_{bb}} \right\}
	- I_0 \left\{ {G_{\rm{A}}, -H_{ub} + H_{bu}} \right\}
	} \right),
	\label{eq:alpha_spec_exp_full}
\end{equation}
\begin{equation}
	\beta = \frac{1}{3} \left( {
	I_0 \left\{ {G_{\rm{S}}, Q_{uu} + Q_{bb}} \right\}
	- I_0 \left\{ {G_{\rm{A}}, Q_{ub} + Q_{bu}} \right\}
	} \right),
	\label{eq:beta_spec_exp_full}
\end{equation}
\begin{equation}
	\gamma = \frac{1}{3} \left( {
	I_0 \left\{ {G_{\rm{S}}, Q_{ub} + Q_{bu}} \right\}
	- I_0 \left\{ {G_{\rm{A}}, Q_{uu} + Q_{bb}} \right\}
	} \right),
	\label{eq:gamma_spec_exp_full}
\end{equation}
\begin{equation}
	\gamma_{\rm{F}} = \frac{2}{3} \left( {
	I_0 \left\{ {G_{\rm{S}}, Q_{bu}} \right\}
	- I_0 \left\{ {G_{\rm{A}}, Q_{uu}} \right\}
	} \right).
	\label{eq:gammaF_spec_exp_full}
\end{equation}
Here, $Q_{uu}$, $Q_{bb}$, $Q_{ub}$, $H_{uu}$, $H_{bb}$, etc.\ are the spectral functions of the kinetic energy, magnetic energy, cross helicity, kinetic helicity, current helicity, etc. for the basic field, respectively. They are written as
\begin{equation}
	\frac{1}{2} \left\langle {
		{\bf{u}}'_{\rm{B}}{}^2 + {\bf{b}}'_{\rm{B}}{}^2
	} \right\rangle
	= \int \left( {Q_{uu}(k;\tau,\tau) + Q_{bb}(k;\tau,\tau)} \right) d{\bf{k}},
\end{equation}
\begin{equation}
	\left\langle {
		{\bf{u}}'_{\rm{B}} \cdot {\bf{b}}'_{\rm{B}}
	} \right\rangle
	= 2 \int Q_{ub}(k;\tau,\tau) d{\bf{k}},
\end{equation}
\begin{equation}
	\left\langle {
		- {\bf{u}}'_{\rm{B}} \cdot \mbox{\boldmath$\omega$}'_{\rm{B}}
		+ {\bf{b}}'_{\rm{B}} \cdot {\bf{j}}'_{\rm{B}}
	} \right\rangle
	=  \int \left( {- H_{uu}(k;\tau,\tau) + H_{bb}(k;\tau,\tau)} \right) d{\bf{k}},
	\label{eq:def_spctral_fn_Huu_Hbb}
\end{equation}
\begin{equation}
	\left\langle {
		{\bf{u}}'_{\rm{B}} \cdot {\bf{j}}'_{\rm{B}}
	} \right\rangle
	=  \int H_{ub}(k;\tau,\tau) d{\bf{k}}.
\end{equation}
If we retain only the part of Green's function with mirrorsymmetry, $G_{\rm{S}}$, we get Eqs.~(\ref{eq:alpha_spec_exp})-(\ref{eq:gamma_spec_exp}). Green's functions are closely related to the characteristic time scales of MHD turbulence. Since time scales are in general pure scalars, the mirrorsymmetric part of the Green's function is supposed to play a dominant role as compared with the anti-symmetric part of it. In such a case, the transport coefficients for $\mbox{\boldmath$\Omega$}$ and $2\mbox{\boldmath$\omega$}_{\rm{F}}$ are the same
\begin{equation}
	\gamma = \gamma_{\rm{F}}
	\label{eq:gamma_gammaF}
\end{equation}
if the velocity and magnetic-field fluctuations are statistically stationary ($Q_{ub} = Q_{bu}$). Then we have
\begin{equation}
	{\bf{E}}_{\rm{M}}
	= \alpha {\bf{B}}
	- \beta {\bf{J}}
	+ \gamma \left( {
		\mbox{\boldmath$\Omega$}
		+ 2\mbox{\boldmath$\omega$}_{\rm{F}}
	} \right).
	\label{eq:E_M_OmegaF}
\end{equation}
A TSDIA analysis using not the Lagrange derivative but the co-rotational derivative, which assures the material frame indifference of turbulent fields, exactly gives relation (\ref{eq:gamma_gammaF}) in the calculation of order up to $O(\delta^2)$ \citep{ham2008}.

\paragraph{Reynolds stress}
	The Reynolds stress is obtained from Eq.~(\ref{eq:elsasser_R}) as
\begin{eqnarray}
	{\cal{R}}^{\alpha\beta}
	= \frac{2}{3} K_{\rm{R}} \delta^{\alpha\beta}
	&-& \nu_{\rm{K}} {\cal{S}}^{\alpha\beta}
	+ \nu_{\rm{M}} {\cal{M}}^{\alpha\beta}
	\nonumber\\
	&+& (\Omega^\alpha + 2\omega_{{\rm F}}^\alpha) \Gamma^\beta
		+ (\Omega^\beta + 2\omega_{{\rm F}}^\beta) \Gamma^\alpha
		- \frac{1}{3} \delta^{\alpha\beta} \left( {
			\mbox{\boldmath$\Omega$} 
			+ 2\mbox{\boldmath$\omega$}_{\rm{F}}
		} \right) \cdot \mbox{\boldmath$\Gamma$},
	\label{eq:Re_strss_OmegaF}
\end{eqnarray}
where $K_{\rm{R}} (= \langle{{\bf{u}}'{}^2 + {\bf{b}}'{}^2}\rangle /2)$ is the turbulent MHD residual energy. The transport coefficient $\nu_{\rm{K}}$ (turbulent viscosity) and $\nu_{\rm{M}}$ are related to $\beta$ and $\gamma$ as
\begin{equation}
	\nu_{\rm{K}} = \frac{7}{5} \beta,
	\label{eq:nu_K_beta_rel}
\end{equation}
\begin{equation}
	\nu_{\rm{M}} = \frac{7}{5} \gamma.
	\label{eq:nu_M_gamma_rel}
\end{equation}
The other coupling coefficient related to the vorticity $\mbox{\boldmath$\Omega$}$ and angular velocity $\mbox{\boldmath$\omega$}_{\rm{F}}$, $\mbox{\boldmath$\Gamma$}$, is expressed as
\begin{equation}
	\mbox{\boldmath$\Gamma$}
	= \frac{1}{15} \left( {
		I_{-1} \{ {G_{\rm{S}}, \nabla H_{uu}} \} 
		- I_{-1} \{ {G_{\rm{A}}, \nabla H_{bu}} \}
	} \right).
	\label{eq:Gamma_exp}
\end{equation}
We see from the $\mbox{\boldmath$\Omega$}$ and $\mbox{\boldmath$\omega$}_{\rm{F}}$-related terms in Eq.~(\ref{eq:Re_strss_OmegaF}) with Eq.~(\ref{eq:Gamma_exp}) that the inhomogeneity of kinetic helicity coupled with the mean vortical motion contributes to the Reynolds stress. This effect is the MHD counterpart of the helicity effect in the hydrodynamic turbulence \citep{yok1993}. These terms related to the helicity gradient arise from the $O(\delta^2)$ calculation of the TSDIA analysis, so they are sometimes dropped for practical applications.

\subsection{Features of the TSDIA formulation}
	The two-scale direct-interaction approximation (TSDIA) is a combination of the direct-interaction approximation (DIA) and the multiple-scale analysis. Its procedure may appear to be complicated although each step of calculation is straightforward.  Several features of the TSDIA analysis can be indicated, including several assumptions and intrinsic restrictions of this formalism. They are divided into three classes: those intrinsic to the DIA formalism; those arising from multiple-scale treatment; and those related to the system of basic equations to be treated.

\subsubsection{Features of the DIA}
	The direct-interaction approximation (DIA), a second-order renormalized perturbation theory, is a modern closure scheme for homogeneous isotropic turbulence \citep{kra1957}. In the earlier studies of homogeneous isotropic turbulence, closure schemes for the correlation functions have been intensively explored \citep{bat1953}. If the homogeneity of the fluctuating field is presumed, the Fourier representation provides a powerful tool for describing the properties of fluctuating quantities. In addition to the correlation functions, \citet{kra1957} introduced the notion of response function into the study of turbulence. Propagators (correlation and Green's function) are introduced in the wave-number space. Through the Green's functions, velocity fluctuations are related to the steering force or noise. The spirit of the DIA approach is embodied in Eq.~(\ref{eq:GreenFnDef}) for the Green's function: The dynamics of the Green's function is explored with the nonlinear interactions among the modes. Using the propagator renormalization (so-called line renormalization, no vertex renormalization), a particular sort of interaction between the modes ${\bf{k}}$ , ${\bf{p}}$, and ${\bf{q}}$ (called the direct interaction) are calculated up to the infinite order. In the Green's function equation (\ref{eq:GreenFnDef}), the nonlinear mode coupling term is the most important part, and the molecular viscosity plays only a minor role. In this sense, this approach is suitable for treating fully developed turbulence at high Reynolds number. This is one of the most prominent features of the DIA approach as compared with the quasi-linear or first-order smoothing approximation. 
	
	Its Lagrangian version succeeded in reproducing turbulence statistics, which includes the Kolmogorov's scaling law, from the Navier--Stokes equations without putting any Ansatz for the first time in the history.

\subsubsection{Features of the two-scale analysis coupled with the DIA}
Following the DIA formulation, the Green's functions are introduced in the wave-number space in the TSDIA. We use the average of the Green's functions with the isotropic assumption [Eq.~(\ref{eq:HomoIsoGreen})]. 

	The first-order field $\mbox{\boldmath$\phi$}'_{{\rm{S}}1}$ obeys Eq.~(\ref{eq:1st-fld_eq_TSDIA}). The left-hand side (LHS) of this equation is essentially the same as that of the basic field Eq.~(\ref{eq:phiBeq}). Terms directly related to the mean-field shear, rotation, and magnetic field appear on the right-hand side (RHS). We regard these terms on the RHS as the force terms. Then we can formally solve Eq.~(\ref{eq:1st-fld_eq_TSDIA}) with the aid of the Green's function [Eq.~(\ref{eq:GreenFnDef})]. In this sense, effects of mean-field inhomogeneity etc.\ are incorporated as a perturbation for the basic field, homogeneous isotropic turbulence.
	
\paragraph{Homogeneous isotropic basic field}
	In the TSDIA, effects of the mean-field inhomogeneity are incorporated through the differential expansion (\ref{eq:two-scale_deriv}) in a perturbational manner (\ref{eq:scl-prmtr-expnd}). As for the basic field, a homogeneous and isotropic field is assumed following the DIA approach. Spectral function and the Green's function for the homogeneous isotropic turbulence are expressed in very simple forms, Eqs.~(\ref{eq:HomoIsoFlds}) and (\ref{eq:HomoIsoGreen}). Due to this choice, this approach is not valid in the case that the higher-order derivatives of the mean field play more important role than the lower-order derivatives. For the hydrodynamic turbulence, this corresponds to situations where the eddy-viscosity expression of the Reynolds stress itself is not valid.
	
	A measure of mean-field inhomogeneity may be the relative magnitude of the turbulence time scale $\tau_{\rm{turb}}$ to the mean inhomogeneity counterpart $\tau_{\rm{mean}}$. If $\tau_{\rm{turb}}$ is similar to or less than $\tau_{\rm{mean}}$:
\begin{equation}
	\frac{\tau_{\rm{turb}}}{\tau_{\rm{mean}}}
	\sim \frac{KS}{\varepsilon} \lesssim 1,
\end{equation}
the differential expansion is not bad. Here, $K$ is the turbulent energy and $\varepsilon$ is its dissipation rate. The characteristic time scale of turbulence, $\tau_{\rm{turb}}$, is given by the energy cascade time $K / \varepsilon$. On the other hand, the characteristic time scale of the mean-field inhomogeneity $\tau_{\rm{mean}}$ is given by the reciprocal of the mean shear rate $S$ (for example, $S= \sqrt{({\cal{S}}^{ab} )^{2}/2}$, with $\mbox{\boldmath${\cal{S}}$}$ [Eq.~(\ref{eq:mean_vel_strain})]). In a typical turbulent channel flow, $KS / \varepsilon \sim 3$.

	On the other hand, if $\tau_{\rm{mean}}$ is much shorter than $\tau_{\rm{turb}}$:
\begin{equation}
	\frac{\tau_{\rm{turb}}}{\tau_{\rm{mean}}}
	\sim \frac{KS}{\varepsilon} \gg 1,
\end{equation}
the differential expansion from the homogeneous isotropic field is not good. In this case, other approach such as the rapid distortion theory (RDT) might be more appropriate, although its applicability to the fully developed nonlinear stage is open.

\paragraph{External-field expansion}
	In order to incorporate the effects of frame rotation and/or large-scale magnetic field, we invoke the external-field expansion (\ref{eq:ext-prmtr-expnd}). This is basically appropriate if the external fields (frame rotation, magnetic field, etc.) are not so strong. As is well known, an external field often makes turbulence anisotropic. In the case of a strong external field, the expansion from the homogeneous isotropic basic field is considered to be inappropriate. In such a situation, assuming a homogeneous but anisotropic basic field may be appropriate. The isotropic form of the Green's function Eq.~(\ref{eq:HomoIsoGreen}) should be also reappraised. In Eqs.~(\ref{eq:HomoIsoFlds}) and (\ref{eq:HomoIsoGreen}) we assume that the basic fields (lowest-order fields) are isotropic. However this is just for the sake of simplicity of calculation, not the essential ingredients of this approach. Introduction of anisotropy is a very important point in the research of turbulence with rotation, density stratification, magnetic field, etc. Starting with the simplest anisotropy with axisymmetry with respect to the rotation, magnetic field, etc. is one good starting point. Actually such analysis is in progress.

\paragraph{Green's function in the wave-number space}
	By introducing the Green's functions in wave-number space, we can fully treat nonlinear mode coupling of turbulence in the sense of the DIA. On the other hand, if we introduce the Green's functions in the real or configuration space, the real-space non-locality and memory effects related to the mean-field quantities can be incorporated. For example, the Reynolds stress may be expressed as
\begin{equation}
	\left\langle{ u'{}^\alpha u'{}^\beta } \right\rangle({\bf{x}};t)
	= \int_0^t ds \int d{\bf{y}} G^{\alpha a}({\bf{x}}, {\bf{y}}; t, s) 
	Q^{\beta b} \frac{\partial U^a({\bf{y}};s)}{\partial x^b}.
	\label{eq:nonlocality_space}
\end{equation}
Here, the Reynolds stress is expressed in a form non-local in space and time. In this paper we do not discuss about such approaches. In the context of the cross-helicity effect, the reader is referred to \citet{rae2010}. Their Eq.~(42) for the mean-vorticity-related coefficient should be compared with Eq.~(\ref{eq:gamma_spec_exp}) in the present paper.

\subsubsection{Simplifications}

	As shown in Eqs.~(\ref{eq:u-eq})-(\ref{eq:slndl-ub}), we consider incompressible MHD turbulence and further assume that the difference between the viscosity $\nu$ and the magnetic diffusivity $\eta$ is not so critical. This makes the equations of $\mbox{\boldmath$\phi$}$ and $\mbox{\boldmath$\psi$}$ very simple.

	One justification for this treatment may be as follows. In geo/astrophysical magnetic phenomena, both the Reynolds and magnetic Reynolds numbers are usually huge. Unless the difference between them is so critical, we can assume that the difference of viscosity ($\nu$) and the magnetic diffusivity ($\eta$) can be negligible as compared with the magnitudes of $\nu$ and $\eta$ themselves. If we use an approach with the assumption that the Reynolds numbers are small, the results may depend on the relative values of $\nu$ and $\eta$. However, the present approach is most suitable for the case with infinite Reynolds numbers. In this sense, this assumption is not so critical as compared with the case in the latter approach.

\paragraph{Symmetry of Green's functions}
	Owing to the coupling between the velocity and magnetic-field fluctuations in the analysis of MHD turbulence, we have to examine the velocity fluctuation responses to the infinitesimal disturbances both on the velocity and magnetic-field evolutions, and the counterparts of magnetic fluctuation responses. This means that  in order to treat MHD turbulence strictly we have at least four Green's functions, which may be schematically denoted as $G_{uu}$, $G_{ub}$, $G_{bu}$, and $G_{bb}$. In this work, we adopt the Elsasser variables $\mbox{\boldmath$\phi$}$ and $\mbox{\boldmath$\psi$}$. Equivalently, in the Elsasser formulation, we in general have four Green's functions, which may be denoted as $G_{\phi\phi}$, $G_{\phi\psi}$, $G_{\psi\phi}$, and $G_{\psi\psi}$. By putting $\nu = \eta$, we drop the cross diffusion term [the last terms in Eq.~(\ref{eq:phi-eq}) and (\ref{eq:psi-eq})] to have Eqs.~(\ref{eq:phi-eq_simple}) and (\ref{eq:psi-eq_simple}). With this treatment, we assume that $G_{\phi\psi}$ and $G_{\psi\phi}$ are negligibly small ($G_{\phi\psi} = G_{\psi\phi} = 0$) as compared with $G_{\phi\phi} (\equiv G_\phi)$ and $G_{\psi\psi} (\equiv G_\psi)$.

	In addition, we assume that the antisymmetric part of the Green's function, $G_{\rm{A}}$ [Eq.~(\ref{eq:antisymGreenFn})], is negligible as compared with the symmetric part $G_{\rm{S}}$ [Eq.~(\ref{eq:symmetricGreenFn})]. This treatment corresponds to the situation where the time scales associated with $\mbox{\boldmath$\phi$}$ and $\mbox{\boldmath$\psi$}$ are the same. The turbulent time scales associated with the motions parallel and anti-parallel to the magnetic field are the same.
		
	Under these considerations, we assumed that the responses of fluctuating fields are represented by only one Green's function.

\vspace{12pt}	
To summarize:\\

	Effects of inhomogeneities are treated in the differential expansion from the basic field. Nonlinear interactions of the fluctuation field is fully taken into account through the introduction of the Green's function equations and the renormalization of the propagators. This is a ``partial sum'' of the direct interactions. The summation is partial but to the infinite order. This approach is considered to be most suitable for fully developed turbulence with very high Reynolds numbers.

\begin{itemize}
\item The basic field of turbulence is homogeneous isotropic;
	\begin{itemize}
	\item The spectrum of the velocity correlation and the average of the Green's functions for the basic fields are isotropic.
	\end{itemize}

\item Responses of the velocity and magnetic-field fluctuations to the steering force or noise are treated by introducing the Green's functions;

	\begin{itemize}
	\item Green's functions are introduced in the wave-number space;
	\item Nonlinear dynamics of fluctuation are fully considered with the mode couplings in the Green's function equation;
	\item In order to treat the mean-field non-locality in space, we have to  introduce the Green's function in the configuration space.
	\end{itemize}

\item For the incompressible magnetohydrodynamic turbulence, we have four Green's functions;
	\begin{itemize}
	\item By introducing the Elsasser formalism with some symmetries, we assume that dynamics of MHD turbulence can be described by only one Green's function.
		\begin{itemize}
		\item $G_{\phi\phi}$, $G_{\phi\psi}$, $G_{\psi\phi}$, $G_{\psi\psi}$
			in the Elsasser veriable formulation;
			\begin{itemize}
			\item $G_{\phi\psi} = G_{\psi\phi} =0,\; 
				G_{\phi\phi} (\equiv G_{\phi}),\; G_{\psi\psi} (\equiv G_{\psi})$;
			\item $G_{\phi} = G_{\psi},\;  
				G_{\rm{S}} [= (G_{\phi} + G_{\psi})/2] = G,\; 
				G_{\rm{A}} [= (G_{\phi} - G_{\psi})/2] =0.$
			\end{itemize}
		\end{itemize}
	\end{itemize}

\item Inhomogeneities of the mean fields are incorporated through the differential expansion from the basic fields;
	\begin{itemize}
	\item Higher-order derivatives ($\nabla^2 {\bf{U}}$, $\nabla^2 {\bf{B}}$, $\cdots$) and nonlinear terms of inhomogeneity [$(\nabla {\bf{U}})^2$, $(\nabla{\bf{B}})^2$,$(\nabla{\bf{U}})(\nabla{\bf{B}})$, $\dots$] occur in higher-order in the expansion.
	\end{itemize}

\item External-field effects such as the system rotation $\mbox{\boldmath$\omega$}_{\rm{F}}$, mean magnetic field ${\bf{B}}$, etc. are taken into account by way of the external-field expansion;
	\begin{itemize}
	\item This is the so-called weak-field expansion; 
	\item Nonlinear terms like $\mbox{\boldmath$\omega$}_{\rm{F}}^2$, ${\bf{B}}^2$, etc.\ enter in the higher-order term.
	\end{itemize}

\end{itemize}


\label{lastpage}

\end{document}